\documentclass[12pt]{article}
\usepackage[margin=1.25in]{geometry}
\usepackage{amsmath,amsfonts}
\newcommand{\Kappa}{\mathcal{K}}
\usepackage{array}
\usepackage{tikz}
\usepackage{multirow}
\usepackage{tablefootnote}
\usetikzlibrary{shapes.geometric, arrows.meta, positioning, fit, backgrounds}
\usepackage{booktabs}
\usepackage[caption=false,font=normalsize,labelfont=sf,textfont=sf]{subfig}
\usepackage{textcomp}
\usepackage[dvipsnames]{xcolor}
\usepackage{amssymb}
\usepackage{mathrsfs}
\usepackage{stfloats}
\usepackage{blkarray}

\usepackage{tikz}
\usepackage{url}
\usepackage{verbatim}
\usepackage{amsthm}
\usepackage{subcaption}
\usepackage[normalem]{ulem}
\usepackage{graphicx}
\usepackage{arydshln}
\usepackage{makecell}
\usepackage{mathrsfs}
\usepackage{enumerate} 
\newtheorem{theorem}{Theorem}
\DeclareMathOperator*{\argmin}{arg\,min}
\usepackage{adjustbox}
\usepackage{enumitem}
\usepackage{algorithm}
\usepackage{algpseudocode}
\newtheorem{proposition}{Proposition}

\newtheorem{definition}{Definition}
\usepackage{float}
\newtheorem{lemma}{Lemma}
\usepackage{hyperref}

\usepackage{authblk}

\title{A structural equation formulation for general quasi-periodic Gaussian processes}

\author[1,2,3]{Unnati Nigam\thanks{Corresponding author: \texttt{unnati.nigam@iitb.ac.in}}}
\author[2]{Radhendushka Srivastava\thanks{\texttt{rsrivastava@iitb.ac.in}}}
\author[3]{Faezeh Marzbanrad\thanks{\texttt{faezeh.marzbanrad@monash.edu}}}
\author[3]{Michael Burke\thanks{\texttt{michael.g.burke@monash.edu}}}

\affil[1]{IITB-Monash Research Academy, Mumbai, India}
\affil[2]{Department of Mathematics, Indian Institute of Technology Bombay, Mumbai 400076, Maharashtra, India}
\affil[3]{Department of Electrical and Computer Systems Engineering, Monash University, Clayton 3168, Victoria, Australia}

\date{}

\begin{document}

\maketitle

\begin{abstract}
This paper introduces a structural equation formulation that gives rise to a new family of quasi-periodic Gaussian processes, useful to process a broad class of natural and physiological signals. The proposed formulation simplifies generation and forecasting, and provides hyperparameter estimates, which we exploit in a convergent and consistent iterative estimation algorithm. A bootstrap approach for standard error estimation and confidence intervals is also provided. We demonstrate the computational and scaling benefits of the proposed approach on a broad class of problems, including water level tidal analysis, CO\textsubscript{2} emission data, and sunspot numbers data. By leveraging the structural equations, our method reduces the cost of likelihood evaluations and predictions from $\mathcal{O}(k^2 p^2)$ to $\mathcal{O}(p^2)$, significantly improving scalability.
\end{abstract}

\section{Introduction}
Periodic signals are prevalent in fields like robotics, physiology, astronomy, and communication systems. However, random noise and unmodeled disturbances often disrupt these regular periodic patterns, leading to signals that display quasi-periodic or pseudo-periodic characteristics. The task of identifying the periodic components and reconstructing the original signal from such quasi-periodic data is a well-established challenge in the field of signal processing.

A range of methods have been proposed for analyzing strictly periodic signals. Among the most commonly used techniques are correlation-based approaches, such as those described in \cite{Fan2018,Li2021a} and \cite{Rabiner1977}. Regression analysis is another widely applied tool for modeling periodic signals, as noted in \cite{Quinn1991}. To improve computational efficiency, \cite{Nielsen2017} introduced a non-linear regression approach for modeling periodic signals. The maximum likelihood estimation technique has also been extensively used to estimate periodic structures, with key contributions in \cite{Clarkson2008} and \cite{Wise1976}. Furthermore, Bayesian methods have been employed to model quasi-periodic signals, with \cite{mackay_paper} using MacKay’s kernel and prior covariance information to enhance signal modeling. 

The periodic Gaussian process, as introduced in \cite{mackay_paper}, is a widely adopted noise model for quasi-periodic signals in various applications. It has been used in scenarios such as fault vibration detection in mechanical systems \cite{pgp}, pitch estimation in speech signals \cite{pgp}, analysis of climatological data like rainfall and famine trends \cite{hajighassemi2014analytic}, modeling joint angles of rotating robotic arms \cite{hajighassemi2014analytic}, and traffic pollution analysis \cite{bandit}. Simulation results from \cite{pgp} demonstrate that periodic Gaussian processes excel in period estimation, particularly under low signal-to-noise ratios, outperforming traditional methods. Periodic Gaussian processes have also been used to detect the multiple unknown seasonal components as well as estimation of respective periods sequentially \cite{sstp}. However, while periodic Gaussian processes effectively capture correlations within periods, they do not model dependencies across different periods, as highlighted in \cite{quasiperiodic}.  

In addition to periodic Gaussian processes, Gaussian processes with covariance functions formed by the product of an exponential kernel and MacKay’s periodic kernel have also been explored to model quasi-periodic signals (see \cite{rasmussen, Chandola2011, qpgp_def, bandit}). More recently, \cite{quasiperiodic} introduced the Quasi-Periodic Gaussian Process (QPGP), with a covariance function defined as the product of a geometrically decaying kernel and MacKay’s periodic covariance kernel (see \eqref{TSP_eqn}), specifically designed to model quasi-periodic signals. This approach effectively captures both intra-period (within-period) and inter-period (between-period) correlations, and allows a maximum likelihood estimation algorithm for model parameters by utilizing the structural properties of MacKay’s kernel to enhance computational efficiency. Simulation studies have demonstrated the improved performance of the QPGP compared to alternative models.  While standard QPGPs often require computationally intensive likelihood evaluations, the algorithm presented in \cite{quasiperiodic} offers a significant reduction in complexity. However, this method is limited to MacKay’s kernel and cannot be extended to other commonly used periodic covariance kernels. In another approach, \cite{sGP_paper} considered the seasonal Gaussian process which  is derived through stochastic differential equations and used B-spline approximations for scalable modelling of large irregular quasi-periodic signals.

In this article\footnote{ The preliminary idea of a dynamical system based QPGP was presented by the authors in \cite{our_paper}.}, we propose a new dynamical equation system, based on a Gaussian linear state-space model, that gives rise to a broad family of QPGPs (Section~\ref{s3}). This new family of QPGP allows extensive selection of periodic kernels to model the variation within periodic blocks along with diminishing variation between the elements of different blocks. In Section~\ref{s4}, a likelihood-based computationally inexpensive algorithm is presented for the estimation of model parameters and one-step prediction. The dynamical equations lead to the rapid generation of bootstrap resamples. In light of this, we present a bootstrap procedure to estimate the standard errors of the parameter estimates along with the 95\% bootstrap confidence intervals for the parameter estimates. A finite sample performance of the proposed estimation methodology is illustrated in Section~\ref{s5}. We apply the proposed QPGP model to quasi-periodic signals on carbon dioxide data, sunspot number data, and water level tidal signals in Section~\ref{s6}. We make the concluding remarks in section~\ref{s7}. 
Proofs of theoretical results are given in the supplementary material. MATLAB codes for parameter estimation, along with bootstrap standard errors and their confidence intervals, are available on GitHub\footnote{ \url{https://github.com/unnati-nigam/quasiperiodicGP}.}. 

In summary, the core contributions of this work are:
\begin{itemize}
    \item A structural formulation of quasi-periodic Gaussian processes that allows flexible specification of the within-period covariance structure, enabling richer periodic dependence modeling.
    \item A computationally efficient framework for QPGPs that enables fast simulation, likelihood evaluation, parameter estimation, and prediction, achieving significant runtime reduction, particularly for large sample sizes.
    \item A scalable inference strategy tailored for large-scale and streaming data settings, making the proposed methodology suitable for large-scale time series applications.
    \item A principled approach for quantifying uncertainty in parameter estimates, providing statistically meaningful measures of estimation reliability.
\end{itemize}
We now review the literature on quasi-periodic Gaussian processes.

\section{Quasi-Periodic Gaussian Processes}\label{s2}

A zero-mean stationary Gaussian process $\{X(t), t\in\mathbb{Z}\}$ is completely specified by its covariance kernel, $\kappa(t)\triangleq E(X(0)X(t)) \   \forall \  t\in\mathbb{Z}$. 
A Gaussian process  $\{X(t), t\in\mathbb{Z}\}$ is referred to as a Periodic Gaussian Process with period $p$ if its covariance kernel $\kappa_p$ is a periodic function with period $p$ \cite{rasmussen}, i.e., 
\begin{equation}\label{periodic_kernel_GP}
\kappa_p(t+p)=\kappa_p(t) \ \ \forall t\in \mathbb{Z}.   \end{equation}
Using the fact that the covariance kernel of a stationary process is an even function, i.e., $\kappa_p(-t) = \kappa_p(t)$, and \eqref{periodic_kernel_GP}, specifying $\kappa_p(t)$ for $t = 0, 1, \dots, T$ where $T$ is the maximum lag needed to identify $\kappa_p$, (with $T = p/2$ if $p$ is even and $T = (p-1)/2$ if $p$ is odd) completely determines the periodic covariance kernel $\kappa_p$.

We list below some of the popular periodic families of covariance kernels with period $p$ used in various applications for modelling periodic signals. 
\begin{enumerate}
    \item MacKay's Kernel \cite{mackay_paper}: Given $ \theta>0$, $\sigma^2>0$,
        \begin{equation}\label{McKer}
    \kappa_p(t)=\sigma^2\exp\left(-\theta^2 \sin^2\left(\pi |t|/p\right)\right).    
    \end{equation}
   Here, the parameter $\theta$ represents the inverse of characteristic length-scale (see pp. 14 in \cite{rasmussen}). 
   \item Periodic Mat\'ern Kernel: Given $\nu>0$, $\theta>0$, $\sigma^2>0$, 
    \begin{equation}
    \kappa_p(t)=\sigma^2\ \frac{2^{1-\nu}}{\Gamma(\nu)} (\phi(|t|))^\nu K_\nu (\phi(|t|)), \label{MatKer}
    \end{equation}
    where  $\phi(t)= \dfrac{2}{\theta} \sqrt{2\nu\sin^2 \left( \pi t/p\right)}$
     and $K_\nu(\cdot)$ is a modified Bessel function of second type \cite{abramowitz1965handbook}. This periodic covariance function is formed by warping the Mat\'ern kernel \cite{stein1999interpolation}. The warping of an aperiodic kernel is a general technique to construct a periodic kernel. The parameters $\nu$ and  $\theta$ represent the degree of smoothness and characteristic length scale, respectively. In particular, if $\nu=1.5$, then the corresponding periodic Gaussian process is differentiable in the mean square sense (see pp. 81--84 in \cite{rasmussen}).

     \item Cosine Kernel \cite{qpc}: Given $\sigma^2>0$ and $\iota\in\mathbb{Z}^+$,
    \begin{equation}\label{CosKer}
        \kappa_p(t)=\sigma^2\cos \left(2\pi \iota |t|/p\right).
    \end{equation}
The family of cosine kernels constitutes a basis of periodic covariance kernels. 
A periodic covariance kernel $\kappa_p$ can be pointwise approximated by a linear combination of elements of the cosine family.  
\end{enumerate}
Moreover, note that the parameter $\sigma^2(= \kappa_p(0))$, for above above-listed kernels, represents the variance of the periodic Gaussian process.

The periodic behavior of the covariance kernel $\kappa_p$ only leads to a periodic sample path from the Gaussian process. This is a major limitation of periodic Gaussian processes in modelling a quasi-periodic signal. 
To better reflect quasi-periodicity, the periodic pattern of  the covariance kernel of a periodic Gaussian process can be adjusted using non-periodic covariance kernels to form quasi-periodic kernels. For example, \cite{rasmussen} studied a stationary quasi-periodic covariance kernel
$$\kappa(t)=\sigma^2  e^{-t^2} \exp(-\theta^2 \sin^2(\pi t/p)),$$
where $\theta>0$. This kernel is a product of a squared-exponential covariance  kernel and periodic MacKay's kernel. 
This kernel has been used to model quasi-periodic Gaussian noise in ECG and Pulsatile physiological signals, crop biomass data \cite{Chandola2011} and the stellar activity of stars \cite{qpgp_def}. 

In a study of astrophysical phenomena, \cite{qpc} used a different stationary quasi-periodic covariance kernel
$$
\kappa(t) = \sigma^2  e^{-t^2} \left( \cos\left(2\pi t/p\right) + \exp\left(-\theta^2 \sin^2\left(\pi t/p\right)\right) \right).
$$
For carbon dioxide concentration data, \cite{rasmussen} used a quasi-periodic kernel that is a linear combination of different non-periodic covariance kernels and a periodic Mackay's kernel. \cite{Foreman-Mackey_2017} considered a quasi-periodic kernel (termed a Simple Harmonic Oscillator kernel) for astronomical quasi-periodic data.

These stationary quasi-periodic covariance kernels have been used to model quasi-periodic signals in various applications. However, these kernels do not explicitly model the correlation between the blocks of successive periodic patterns in the process.  
To address the limitations of standard periodic Gaussian processes, \cite{quasiperiodic} explored a non-stationary covariance kernel that models both the between- and within-period correlation of the quasi-periodic signal. They referred to a Gaussian process $\{X_t, t\in\mathbb{Z}\}$ as a Quasi-Periodic Gaussian Processes with period $p$ if the covariance between $X_t$ and $X_s$ for $t,s\in\mathbb{Z}$ is given by 
\begin{equation}
\mbox{Cov}(X_t,X_s)= \omega^{\left\lvert\left\lceil t/p\right\rceil-\left\lceil s/p\right\rceil\right\rvert} \sigma^2\exp \left(-\theta^2\sin^2\left(\pi(t-s)/p\right)\right),\label{TSP_eqn}
\end{equation}
where $\lceil \cdot \rceil$ denotes the ceiling function, $\sigma^2>0$ is the variance, $\omega\in(-1,1)$ denotes the between-period correlation,  and $\theta>0$ denotes the within-period correlation.
The second term on the RHS of \eqref{TSP_eqn} corresponds to MacKay's covariance kernel, which determines the periodic pattern of the process, while the first term is a geometrically decaying function, i.e. the covariance decreases as the difference between $\lceil t/p\rceil$ and $\lceil s/p \rceil$ increases. The sample paths of a QPGP with covariance as in \eqref{TSP_eqn} exhibit quasi-periodic patterns. Note that, for $t, s\in\mathbb{Z}$ such that $\left\lvert\left\lceil t/p\right\rceil-\left\lceil s/p\right\rceil\right\rvert=0$, $X_t$ and $X_s$ are in the same periodic block of the QPGP and the covariance between them coincides with MacKay's kernel. 
Further, when $t, s\in\mathbb{Z}$ such that $\left\lvert\left\lceil t/p\right\rceil-\left\lceil s/p\right\rceil\right\rvert=k$, then $X_t$ and $X_s$ belong to periodic blocks that are successively $k$ apart. In this case, the covariance between $X_t$ and $X_s$ decays as $\omega^k$.  
Although, the covariance structure of QPGP given in \eqref{TSP_eqn} is non-stationary, it models the correlation between and within block elements explicitly.

Given $n$ samples $X_1,X_2,\ldots,X_n$ of the QPGP, the likelihood function involves the computation of determinant and inverse of a covariance matrix of order $n$ with elements as in \eqref{TSP_eqn}, which is computationally expensive. The high computational complexity of Gaussian processes, $\mathcal{O}(n^3)$, presents a major barrier when modeling long-duration periodic and quasi-periodic time series. To address this bottleneck, \cite{pmlr-v33-solin14} reformulated periodic GPs into continuous-time state-space models by mapping covariance functions to linear time-invariant stochastic differential equations via Fourier series expansions, enabling linear-time inference through Kalman filtering. However, their approach relies on truncating infinite Fourier series. In contrast, our work bypasses continuous state-space transformations and spectral approximations entirely by formulating a discrete-time structural equation framework. By partitioning the series into $p$-dimensional periodic blocks and capturing between-period correlation via a vector autoregressive recursion, we reduce the computational burden to scale directly with period length, rather than expanded state dimensions while retaining exact base kernel representations.

\cite{quasiperiodic} expressed  the covariance  matrix, given by \eqref{TSP_eqn} as $\sigma^2$ times a Kronecker product of a Kac-Murdock-Szego (KMS) matrix (\cite{Kac1953}) and a symmetric-circulant matrix. \cite{quasiperiodic} developed an efficient algorithm for fast likelihood computation by exploiting the structure of special matrices. Specifically, the method leverages properties of the Kronecker product inverse and utilizes the factorization of the KMS matrix alongside a symmetric circulant matrix, enabling the use of the Fast Fourier Transform (FFT). However, such a computation restricts the scope of generalization of the within- and between-period correlation structures of a covariance function. 

Below, we introduce a more general QPGP formulation.

\section{Dynamical Equation Model for QPGP}\label{s3}
The covariance structure of the proposed QPGP  mainly consists of two parts: (a) a component that models the covariance between the elements of different periodic blocks, and (b) a component that models the covariance between elements within the same periodic block. Our proposed QPGP uses a dynamical equation system, a Gaussian linear state-space model \cite{linear_ssm}, to permit the modeling of the within-period correlations for arbitrary choices of the periodic covariance kernel, while providing the flexibility to adjust the correlation between elements of successive periods. We now formally define this new family of Quasi-Periodic Gaussian Processes.

\medskip
\begin{definition}[QPGP] \label{def:QPGP}
    A Gaussian process $\{Y_t, \ t\in \mathbb{Z}^{+}\}$ is said to be a zero-mean Quasi-Periodic Gaussian process with period $p$, periodic covariance kernel $\kappa_p$ and between-period correlation $\omega$ if 
    \begin{equation}
        E(Y_t)=0, \mbox{ for all } t\in \mathbb{Z}^+,
    \end{equation}
    and the $p$-dimensional random vectors, $$\boldsymbol{\mathsf{Y}}_{i+1}\triangleq[Y_{ip+1}, Y_{ip+2},\dots,Y_{ip+p}]^{\top}, \text{ for }i=0,1,2\ldots$$ satisfy the recursion
\begin{equation}  
    \boldsymbol{\mathsf{Y}}_{i+1} = \omega \boldsymbol{\mathsf{Y}}_{i} + \boldsymbol{\mathsf{Z}}_{i+1}, \mbox{ for } i \ge  1 ,\label{eq:recursion}
\end{equation}  
where $\{\boldsymbol{\mathsf{Z}}_{i+1}\}_{i\ge 1}$ is a sequence of independently and identically distributed  $p$-dimensional zero-mean Gaussian random vectors with covariance matrix 
\begin{equation}\label{def_bloc_cov}
    \boldsymbol{\Kappa}\ \triangleq(\kappa_p(i-j))_{1\le i,j\le p},
\end{equation} which is independent of initial vector $\boldsymbol{\mathsf{Y}}_1$.  
\end{definition} 

\medskip
Note that the random vectors $\boldsymbol{\mathsf{Y}}_i$'s represent the periodic blocks of the QPGP. The recursion given in \eqref{eq:recursion} shows that the correlation between successive periods of the QPGP is described by the parameter $\omega$. Since the random vectors $\boldsymbol{\mathsf{Z}}_i$'s are independent copies of a zero-mean periodic Gaussian process with covariance kernel $\kappa_p$, the recursion given in \eqref{eq:recursion} also shows that the within-period correlation of the QPGP is described by $\kappa_p$. We refer to the random vectors $\boldsymbol{\mathsf{Z}}_{i}$'s as the periodic building-blocks of the QPGP. The rapid generation of periodic building blocks and the dynamical equation in \eqref{eq:recursion} leads to the rapid generation of sample paths from a QPGP.
Theorem~\ref{thm:cov_QPGP}, given below, provides an expression for the covariance structure of the QPGP. Note that Eqn. \eqref{eq:recursion} is essentially a vector autoregressive \cite{Tiao01121981} model of first-order with an isotropic transition operator $\omega$.

The decomposition of periodic and quasi-periodic functions using state-space models has previously been proposed in \cite{pmlr-v33-solin14}. While state-space methods effectively mitigate scaling along the length of the time series $n$, our approach differs fundamentally in both representation and execution. Rather than approximating continuous-time dynamics via SDEs, we introduce a discrete-time structural equation framework, given by \eqref{eq:recursion}. While the state-space model in \cite{pmlr-v33-solin14} relies on truncating infinite Fourier expansions to finite harmonics, our structural framework accommodates arbitrary periodic building blocks without continuous state-space transformations or spectral truncation errors. Further, the filtering complexity of \cite{pmlr-v33-solin14} depends on the expanded state-space dimension, our method reduces the primary computational complexity, scaling with period length.

\bigskip
\begin{theorem}\label{thm:cov_QPGP}
    Let $\{Y_t, \ t\in\mathbb{Z}^+\}$ be a zero mean Quasi-Periodic Gaussian Process with parameters $p, \kappa_p$ and $\omega$. Then, for $s\le t  \in \mathbb{Z}^+$, we have
    \begin{equation}
        Cov(Y_t, Y_s)=\omega^{\left\lvert\left\lceil\frac{t}{p}\right\rceil- \left\lceil \frac{s}{p}\right\rceil\right \rvert}  
        \Bigg( \kappa_p(t-s)\left[\frac{1-\omega^{2\left\lfloor \frac{s}{p}\right\rfloor}}{1-\omega^2}\right]+ \omega^{2\left\lfloor \frac{s}{p}\right\rfloor}Cov\left(Y_{l(t)},Y_{l(s)}\right)\Bigg)\label{quasi_cov}
    \end{equation}
where $l(t)\triangleq t-\lfloor t/p\rfloor p$,  $l(s)\triangleq s-\lfloor s/p\rfloor p$ with $l(t),l(s)\in\{1,2,\ldots,p\}$ and $\lfloor \cdot \rfloor$ denotes the floor function.
\end{theorem}

\medskip
The second term on the RHS of~\eqref{quasi_cov} represents the effect of initial vector $\boldsymbol{\mathsf{Y}}_1$ on the covariance of QPGP, which diminishes for large~$s$. Therefore, if one burns out or discards a sufficiently large  number of initial observations of the QPGP, then \eqref{quasi_cov} can be approximated as 
\begin{align}
        \hskip-15pt Cov(Y_t, Y_s)&\approx \frac{\omega^{\left\lvert\left\lceil\frac{t}{p}\right\rceil- \left\lceil \frac{s}{p}\right\rceil\right \rvert}}{1-\omega^2}  
        \kappa_p(t-s).\label{quasi_cov_approx}
    \end{align}
The approximation error in \eqref{quasi_cov_approx} depends on the variance-covariance matrix of the initial block $\boldsymbol{\mathsf{Y}}_1$. One can obtain an upper bound on the burnout initial blocks so that this approximation error is contained in a specified tolerance limit (see supplementary material).
 Proposition~\ref{prop:quasi_cov_simple}, given below, shows that a special choice of distribution of the initial vector $\boldsymbol{\mathsf{Y}}_1$ also simplifies~\eqref{quasi_cov} to  \eqref{quasi_cov_approx} in an exact manner.

\medskip
\begin{proposition} \label{prop:quasi_cov_simple}
Let $\{Y_t, \ t\in\mathbb{Z}^+\}$ be a zero-mean Quasi-Periodic Gaussian Process with parameters $p, \kappa_p$ and $\omega$. Let the initial vector $\boldsymbol{\mathsf{Y}}_1$ be a zero mean Gaussian vector with covariance matrix $\frac1{1-\omega^2}\boldsymbol{\Kappa}$. Then, for $s,t  \in \mathbb{Z}^+$, $Cov(Y_t, Y_s)$ is given by the right-hand side of \eqref{quasi_cov_approx}.
\end{proposition}
We now define a Standard
Quasi-Periodic Gaussian Process.
\begin{definition}[Standard QPGP]
    A QPGP $\{Y_t, t\in\mathbb{Z}^+\}$ is said to be a Standard QPGP if the initial vector $\boldsymbol{\mathsf{Y}}_1$ is a Gaussian vector with mean $0$ and covariance matrix $\frac1{1-\omega^2}\boldsymbol{\Kappa}$.
\end{definition}

\medskip
The covariance structure  of the QPGP given by \cite{quasiperiodic} (see \eqref{TSP_eqn}) coincides with that of the proposed standard QPGP when $\kappa_p$ is chosen as MacKay's kernel \eqref{McKer} with a scale factor of $\frac1{1-\omega^2}$. The proposed QPGP  enables constructing a new family of QPGP with arbitrary periodic covariance kernels for modeling within-period correlation, providing far greater flexibility than the covariance function given by \eqref{TSP_eqn}.  

\section{Estimation strategy}\label{s4}
Given the $n$-dimensional data vector $\boldsymbol{y}=[y_1,y_2,\ldots,y_n]^\top$ from the standard QPGP with period $p$, periodic covariance kernel $\kappa_p$, and between-period correlation $\omega$, we consider the likelihood approach for estimation of the parameters. We begin with $n=kp$ for some $k\in\mathbb{Z}^+$ for simplicity of the likelihood expression. When $n\ne kp$, the amended estimation methodology is provided in Appendix~\ref{incomplete_block}. The negative logarithm of the likelihood function of the data vector $\boldsymbol{y}$ is given as 
\begin{align}\label{likelihood_datavector}
    \ell_n(\omega, \kappa_p)&= \frac12\log(|\boldsymbol{\Sigma}_n|)+\frac12 \boldsymbol{y}^\top \mathbf{\Sigma}_n^{-1}\boldsymbol{y}+n\log(\sqrt{2\pi}),
\end{align}
where 
\begin{equation}
(1-\omega^2)\boldsymbol{\Sigma}_n\triangleq \left(\omega^{\left\lvert\left\lceil\frac{i}{p}\right\rceil- \left\lceil \frac{j}{p}\right\rceil\right \rvert}  
        \kappa_p(i-j)\right)_{1\le i,j\le n}.\label{Sigma_def}
\end{equation}
The evaluation of \eqref{likelihood_datavector} involves a computationally expensive computation of the determinant and inverse of the covariance matrix $\boldsymbol{\Sigma}_n$. 
When $\kappa_p$ is chosen to be MacKay's kernel, then the fast algorithm developed in \cite{quasiperiodic} for the computation of the determinant and inverse of $(1-\omega^2)\boldsymbol{\Sigma}_n$ can be used for the likelihood evaluation. 

By using \eqref{eq:recursion} and the conditional distribution of the periodic blocks of the standard QPGP, the negative logarithm of the likelihood function is expressed as follows.  
\begin{align}
    \ell_n(\omega, \kappa_p)&= \frac{k-1}{2} \log(|\boldsymbol{\Kappa}|)
    +\underbrace{ \ \ \frac12\sum_{i=1}^{k-1}(\boldsymbol{\mathsf{y}}_{k-i+1}-\omega \boldsymbol{\mathsf{y}}_{k-i})^{\top}\boldsymbol{\Kappa}^{-1}(\boldsymbol{\mathsf{y}}_{k-i+1}-\omega \boldsymbol{\mathsf{y}}_{k-i})}_{\text{Contribution of periodic blocks $\boldsymbol{\mathsf{Y}}_k,\ldots, \boldsymbol{\mathsf{Y}}_1$}}\nonumber\\
    &\quad+\underbrace{ \frac{1}{2}\log\left(\Big|\frac{1}{1-\omega^2}\boldsymbol{\Kappa}\Big|\right)
    +\frac{1}{2}\boldsymbol{\mathsf{y}}_1 ^\top \left(\frac{1}{1-\omega^2}\boldsymbol{\Kappa}\right)^{-1}\boldsymbol{\mathsf{y}}_1}_{\text{Marginal contribution of periodic block $\boldsymbol{\mathsf{Y}}_1$}} +c,\label{likelihood_complete}
\end{align}
where $\boldsymbol{\mathsf{y}}_1, \boldsymbol{\mathsf{y}}_2,\ldots, \boldsymbol{\mathsf{y}}_k$ are the observed periodic blocks of the QPGP data vector $\boldsymbol{y}$ and $c=n\log(\sqrt{2\pi})$. The evaluation of \eqref{likelihood_complete} requires computation of the inverse and determinant of the $p$-dimensional periodic covariance matrix $\boldsymbol{\Kappa}$. This simplification reduces the computational cost of likelihood evaluation immensely. In subsection~\ref{faster_likelihood_nuemeric}, we illustrate numerically that evaluation of the simplified likelihood expression (as given in \eqref{likelihood_complete}) is computationally faster than  using expression in \eqref{likelihood_datavector} (see Table \ref{tab:computational cost likeli}). This improvement arises because the proposed method reduces the matrix computation complexity from $\mathcal{O}(k^2p^2)$ to $\mathcal{O}(p^2)$.

We now formally describe the parameter space of the parameters $\omega$ and $\kappa_p$. Since $\omega$ represents the correlation between the elements of successive periodic blocks, we set $\omega\in (-1,1)$. Similarly, as $\kappa_p$ represents the covariance kernel of the periodic building-blocks, we set $\kappa_p\in \mathbb{K}_p$ where $\mathbb{K}_p$ denotes the set of all periodic covariance kernels of order $p$. The maximum likelihood estimator of $(\omega, \kappa_p)$ is obtained by minimizing $\ell_n(\omega,\kappa_p)$, in \eqref{likelihood_complete},  over $\omega\in(-1,1)$ and $\kappa_p\in\mathbb{K}_p$. 

The non-convexity of $\ell_n(\omega,\kappa_p)$ over $\kappa_p\in\mathbb{K}_p$ poses a major challenge in the maximum likelihood estimation  (see~\cite{nonconvex_toeplitz}). However, for a specific $\kappa_p$ indexed by parameters $(\boldsymbol{\theta},\sigma^2)$, the likelihood function $\ell_n(\omega, \kappa_p(\boldsymbol{\theta},\sigma^2))$ may be a convex function of  $(\omega, \boldsymbol{\theta}, \sigma^2)$ over the reduced parameter space. In such scenarios, the maximum likelihood estimates of $\omega, \theta$, and $\sigma^2$ can be obtained analytically or by using numerical methods such as grid search. In particular, \cite{quasiperiodic} considered $\kappa_p$ to be MacKay's kernel, and maximum likelihood estimates were obtained using the grid search method. The computational efficiency and accuracy of estimates based on the grid search algorithm depend on the size of the grid.  
 
By utilizing the structural equations approach, we now develop a fast algorithm for estimating QPGP parameters $\omega$ and $\kappa_p$ under a general setup. Note that the likelihood of the proposed QPGP differs from the standard QPGP through the marginal contribution of the initial periodic block $\boldsymbol{\mathsf{Y}}_1$ (see \eqref{likelihood_complete}). This shows that the maximum likelihood estimates depend on the marginal distribution of $\boldsymbol{\mathsf{Y}}_1$. Given this, we do not consider the marginal contribution of $\boldsymbol{\mathsf{Y}}_1$ in the likelihood function in our estimation approach. This loss of information provides flexibility in the applicability of our estimation approach for a general QPGP. Thus, we consider the function $\tilde{\ell}_n$, referred to as the \emph{scaled} negative logarithm of the \emph{reduced} likelihood function, given below for our estimation approach in the next subsection. 
\begin{equation}\tilde{\ell}_n(\omega, \boldsymbol{\Kappa})=\log (|\boldsymbol{\Kappa}|)+c +\frac{1}{k-1}\sum_{i=1}^{k-1}(\boldsymbol{\mathsf{y}}_{k-i+1}-\omega \boldsymbol{\mathsf{y}}_{k-i})^\top \boldsymbol{\Kappa}^{-1}(\boldsymbol{\mathsf{y}}_{k-i+1}-\omega \boldsymbol{\mathsf{y}}_{k-i}).\label{reduced_likelihood}
\end{equation}
\subsection{Two stage fast estimation algorithm} \label{s4.1}
 
We propose a two-stage algorithm to estimate the parameters $\omega$ and periodic covariance kernel $\kappa_p$ based on the reduced likelihood function $\tilde{\ell}_n$ given in \eqref{reduced_likelihood}. Note that the reduced likelihood function $\tilde{\ell}_n(\omega,\boldsymbol{\Kappa})$ is a  twice differentiable
function of $\omega\in[-1,1]$ and $\boldsymbol{\Kappa}\in \mathfrak{K}$ where $\mathfrak{K}$ is the set of all real-valued $p$-dimensional invertible matrices with bounded entries. 
In~Stage~I of the algorithm, we estimate $\omega$ and $\boldsymbol{\Kappa}$ by minimizing $\tilde{\ell}_n$ over $\omega\in[-1,1]$ and $\boldsymbol{\Kappa}\in\mathfrak{K}$.
Note that, by using matrix differentiation (see pp. 9--10 of \cite{matrix_cookbook}), we have
\begin{align}
    \frac{\partial \tilde{\ell}_n }{\partial \omega} &=- \frac1{k-1}\sum_{i=1}^{k-1}\boldsymbol{\mathsf{y}}_i^\top\boldsymbol{\Kappa}^{-1}\boldsymbol{\mathsf{y}}_{i+1}+\frac{\omega}{k-1}\sum_{i=1}^{k-1} \boldsymbol{\mathsf{y}}_i^\top\boldsymbol{\Kappa}^{-1}\boldsymbol{\mathsf{y}}_{i},\label{deriv_omega}\\
    \frac{\partial \tilde{\ell}_n }{\partial \boldsymbol{\Kappa}}&=- \boldsymbol{\Kappa}+\frac{1}{k-1}\sum_{i=1}^{k-1}(\boldsymbol{\mathsf{y}}_{i+1}-\omega \boldsymbol{\mathsf{y}}_i)(\boldsymbol{\mathsf{y}}_{i+1}-\omega \boldsymbol{\mathsf{y}}_i)^\top.\label{deriv_Kappa} 
\end{align}
By equating the right-hand side of \eqref{deriv_omega} and \eqref{deriv_Kappa} to $0$, the stationary points of $\tilde{\ell}_n$ satisfy the following relation.
\begin{eqnarray}
    {\omega}&=&\dfrac{\sum_{i=1}^{k-1}\boldsymbol{\mathsf{y}}_{i}^\top\boldsymbol{\Kappa}^{-1}\boldsymbol{\mathsf{y}}_{i+1}}{\sum_{i=1}^{k-1}\boldsymbol{\mathsf{y}}_{i}^\top\boldsymbol{\Kappa}^{-1}\boldsymbol{\mathsf{y}}_{i}}, \label{w_est}\\
    \boldsymbol{\Kappa}&=& 
    \frac{1}{(k-1)}\sum_{i=1}^{k-1}(\boldsymbol{\mathsf{y}}_{i+1}-\omega\boldsymbol{\mathsf{y}}_i)(\boldsymbol{\mathsf{y}}_{i+1}-\omega\boldsymbol{\mathsf{y}}_i)^\top.\label{R_est}
\end{eqnarray}
The explicit solution of the nonlinear equations given in \eqref{w_est} and \eqref{R_est} does not exist. We apply an alternate minimization of $\tilde{\ell}_n$ over $\omega$ and $\boldsymbol{\Kappa}$ iteratively to obtain the solution of these equations and denote them by $\tilde{\omega}_n$ and $\tilde{\boldsymbol{\Kappa}}_n$. Note that, for given $\omega$, the solution $\boldsymbol{\Kappa}$ as in \eqref{R_est} is the unique minimizer of $\tilde{\ell}_n$. Similarly, given $\boldsymbol{\Kappa}$, the solution $\omega$ given in \eqref{w_est} is the unique minimizer of $\tilde{\ell}_n$. Therefore, by using Proposition 2.7.1 of \cite{alternate_min_ref}, the iterative alternating minimization sequence  $\{(\tilde{\omega}_{(m)},\boldsymbol{\tilde{\Kappa}}_{(m)})\}$ in Stage I of Algorithm~\ref{alg1} converges to a stationary point of  $\tilde{\ell}_n$.

Note that the solution  $\tilde{\boldsymbol{\Kappa}}_n$ is a non-negative definite matrix, but not guaranteed to be a covariance matrix corresponding to a periodic covariance kernel. Therefore, in Stage~II of the Algorithm~\ref{alg1}, we constrain $\tilde{\boldsymbol{\Kappa}}_n$ to the set $\mathbb{K}_p$. For this purpose, we minimize the Frobenius norm of the matrix $(\tilde{\boldsymbol{\Kappa}}_n-\boldsymbol{\Kappa})$ over the periodic covariance kernels to estimate $\kappa_p$, i.e.,
\begin{equation}\label{kappa_tilde}
    \mbox{minimize}_{\boldsymbol{\Kappa}(=(\kappa_p(i-j))_{1\le i,j\le p})}\|\tilde{\boldsymbol{\Kappa}}_n-\boldsymbol{\Kappa}\|_F.
\end{equation}
The minimizer of \eqref{kappa_tilde} is given as follows.
\begin{align}
    \tilde{\kappa}_p(t)=\frac{1}{p-|t|}\sum_{j=1}^{p-|t|}\tilde{\boldsymbol{\Kappa}}_n(j,j+|t|), \text{ for } |t|<p. \label{tilde_k_p}
\end{align}
where, $\tilde{\boldsymbol{\Kappa}}_n(j,j+|t|)$  denotes to the $(j,j+|t|)^{\text{th}}$ element of $\tilde{\boldsymbol{\Kappa}}_n$.
To ensure the positive definiteness property of covariance kernel $\tilde{\kappa}_p$, we use the technique developed in \cite{peterhall}. We restrict the spectrum of $\tilde{\kappa}_p$ to be non-negative using truncation, and its inverse Fourier transform is our proposed estimates of $\kappa_p$, i.e.,  
\begin{equation}
    \hat{\kappa}_p(t)=  \int_{-\pi}^\pi e^{it\lambda} \max(\tilde{f}(\lambda),0) \ d\lambda \quad\mbox{ for } |t|<p,\label{hat_k_p}
\end{equation}
where 
\begin{equation}
 \tilde{f}(\lambda) =\frac1{2\pi} \sum_{|t|<p}\tilde{\kappa}_p(t )e^{-i t \lambda }, \mbox{ for all } \lambda\in [-\pi, \pi]. \label{tilde_f_def}
 \end{equation}
Now, our proposed estimate of $\omega$ is given as follows.
\begin{equation}
        \hat{\omega} =\dfrac{\sum_{i=1}^{k-1} \boldsymbol{\mathsf{y}}_{i}^\top \hat{\boldsymbol{\Kappa}}^{-1} \boldsymbol{\mathsf{y}}_{i+1}}{\sum_{i=1}^{k-1} \boldsymbol{\mathsf{y}}_{i}^\top \hat{\boldsymbol{\Kappa}}^{-1} \boldsymbol{\mathsf{y}}_{i}},\label{hat_w_def}
\end{equation}
where $\hat{\boldsymbol{\Kappa}}\triangleq (\hat{\kappa}_p(i-j))_{1\le i,j\le p}$. The proposed estimation algorithm is summarized in Alg.~\ref{alg1}.

Suppose the periodic covariance kernel $\kappa_p$ of the QPGP is known to belong to a parametric family of covariance kernel $\mathbb{K}_{(\boldsymbol{\theta},\sigma^2)}$ where hyper-parameters $(\boldsymbol{\theta}, \sigma^2)\in \Theta\times (0,\infty)$ (see the listed examples in section~\ref{s2}). In such a scenario, the stage~II of Algorithm~\ref{alg1} reduces significantly. In particular, we restrict $\tilde{\boldsymbol{\Kappa}}_n$, obtained from stage~I, to $\mathbb{K}_{(\boldsymbol{\theta},\sigma^2)}$ to estimate $(\boldsymbol{\theta}, \sigma^2)$. We estimate $(\boldsymbol{\theta},\sigma^2)$ by minimizing the Frobenius norm of $(\tilde{\boldsymbol{\Kappa}}_n-\boldsymbol{\Kappa}(\boldsymbol{\theta},\sigma^2))$ over $\boldsymbol{\Kappa}(\boldsymbol{\theta},\sigma^2)\in \mathbb{K}_{(\theta,\sigma^2)}$, i.e., 
\begin{equation}\label{parameteric_kappa}
( \hat{\boldsymbol{\theta}}, \hat{\sigma}^2)=\argmin_{(\boldsymbol{\theta}, \sigma^2)\in \Theta\times(0,\infty)}\|\tilde{\boldsymbol{\Kappa}}_n-\boldsymbol{\Kappa}(\boldsymbol{\theta},\sigma^2)\|_{F}.
\end{equation}
We then replace $\hat{\boldsymbol{\Kappa}}$ by $\boldsymbol{\Kappa}( \hat{{\boldsymbol{\theta}}}, \hat{\sigma}^2)$ in \eqref{hat_w_def} to get the stage~II estimate of $\omega$. 
In the absence of an analytical expression of $( \hat{\boldsymbol{\theta}}, \hat{\sigma}^2)$, a grid search minimization approach can be implemented for \eqref{parameteric_kappa}.

Note that Stage II of the estimation strategy is employed to ensure the periodic structure of the covariance kernel $\kappa_p$. In case of the known parametric family of $\kappa_p$, one can alternatively estimate the parameters of $\kappa_p$ in iterative Stage I itself (either by solving normal equations for $(\theta,\sigma^2)$ in place of \eqref{deriv_Kappa} or grid search over $(\theta, \sigma^2)$). If one must employ grid search at each iteration of Stage I (we refer to it as SIGS estimates), this would be computationally expensive. However, this alternative approach does not require Stage II of the proposed estimation algorithm due to the known parametric structure of $\kappa_p$.     
A simulation study to illustrate the estimation accuracy and predictive MSE, along with the computational cost of SIGS estimates are provided in Sec. II of  Supplementary Material.

While consistency results for a broader class of Gaussian state-space and vector autoregressive models are well-established in the literature, Theorem~\ref{thm5}, given below, establishes the consistency of the proposed estimators $\hat{\omega}$ and $\hat{\kappa}_p$. 
\begin{theorem}\label{thm5}
Let $\boldsymbol{y}=[y_1,y_2,\ldots,y_n]^\top$ be a $n$-dimensional sample path of a QPGP with period $p$ and parameters $\omega_0$ and $\kappa_{0p}$. Then, we have the following convergence results.
\begin{enumerate}
\item The reduced likelihood function $\tilde{\ell}_n(\omega, \boldsymbol{\Kappa}) \xrightarrow{P} \tilde{\ell}(\omega, \boldsymbol{\Kappa})$ continuously\footnote{\textit{Definition 3: }
A sequence of function $g_n$ defined over $\mathbb{R}^d$ converges to $g$ continuously if $u_n\to u$ then $g_n(u_n)\to g(u) $ as $n\to\infty$.   
} as $n\to\infty$ where 
\begin{align}\label{limit_tilde_l}
\hskip-20pt\tilde{\ell}(\omega, \boldsymbol{\Kappa})
=\log(|\boldsymbol{\Kappa}|)+tr (\boldsymbol{\Kappa}^{-1}\boldsymbol{\Kappa}_0) \left(1+\frac{(\omega-\omega_0)^2}{1-\omega_0^2}\right) 
\end{align}

\item  The limiting function $\tilde{\ell}(\omega,\boldsymbol{\Kappa})$, defined over $\omega\in[-1,1]$ and $\boldsymbol{\Kappa}\in\mathfrak{K}$, is a twice differentiable function with a unique minima at $(\omega_0,\boldsymbol{\Kappa}_0)$.

    \item  The estimators $\hat{\omega}$ and $ \hat{\boldsymbol{\Kappa}}$, obtained from the stage~II of Algorithm~\ref{alg1}, converge as follows.
    \begin{itemize}
    \item $\hat{\omega} \xrightarrow{P} \omega_0$ as $n\to\infty$.
    \item $\hat{\boldsymbol{\Kappa}} \xrightarrow{P} \boldsymbol{\Kappa}_0 \triangleq(\kappa_{0p}(i-j))_{1\le i,j\le p}$ as $n \to \infty$.
    \end{itemize}
\end{enumerate}
Here, $\xrightarrow{P}$ denotes the convergence in probability. 
\end{theorem}

\begin{algorithm}[htbp]
\caption{Estimation of QPGP parameters $\omega$ and $\kappa_p$}
\label{alg1}
\begin{algorithmic}[1]
\State \textbf{Input:} $\boldsymbol{y}=[y_1,\ldots,y_{n}]^\top$, $n=kp$; $p$; $\delta$ (threshold)
\State \textbf{Stage~I:}
\State Initialize: $\tilde{\boldsymbol{\Kappa}}_{(0)} = \mathbf{I}_p$
\For{$m = 1, 2, \dots$}
    \State $\displaystyle \tilde{\omega}_{(m)} = \frac{\sum_{i=1}^{k-1}\boldsymbol{\mathsf{y}}_{i}^\top \tilde{\boldsymbol{\Kappa}}_{(m-1)}^{-1} \boldsymbol{\mathsf{y}}_{i+1}}{\sum_{i=1}^{k-1}\boldsymbol{\mathsf{y}}_{i}^\top \tilde{\boldsymbol{\Kappa}}_{(m-1)}^{-1} \boldsymbol{\mathsf{y}}_{i}}$
    \State $\displaystyle \tilde{\boldsymbol{\Kappa}}_{(m)} = \frac{1}{k-1} \sum_{i=1}^{k-1} (\boldsymbol{\mathsf{y}}_{i+1} - \tilde{\omega}_{(m)} \boldsymbol{\mathsf{y}}_i) (\boldsymbol{\mathsf{y}}_{i+1} -\tilde{\omega}_{(m)} \boldsymbol{\mathsf{y}}_i)^\top$
    \If{$\displaystyle \max\Big(\Big|\frac{\partial \tilde{\ell}_n}{\partial \omega}\Big|, \Big|\frac{\partial \tilde{\ell}_n}{\partial \boldsymbol{\Kappa}}  \Big|_\infty \Big) \Big|_{\tilde{\omega}_{(m)}, \tilde{\boldsymbol{\Kappa}}_{(m)}} <\delta$}
        \State Set $\tilde{\omega}_n = \tilde{\omega}_{(m)}$, $\tilde{\boldsymbol{\Kappa}}_n = \tilde{\boldsymbol{\Kappa}}_{(m)}$
        \State \textbf{break}
    \EndIf
\EndFor
\State \textbf{Stage~II:}
\For{$|t|<p$}
    \State $\displaystyle \tilde{\kappa}_p(t)=\frac{1}{p-|t|}\sum_{j=1}^{p-|t|} \tilde{\boldsymbol{\Kappa}}_n(j,j+|t|)$
\EndFor
\State Compute \textbf{Spectrum}($\tilde{\kappa}_p$):
\State $\displaystyle \tilde{f}(\lambda) = \frac1{2\pi} \sum_{|t|<p}\tilde{\kappa}_p(t )e^{-i t \lambda }, \quad \forall \lambda\in [-\pi, \pi]$
\State \textbf{Estimates of} $\kappa_p$ and $\omega$:
\State $\displaystyle \hat{\kappa}_p(t)= \int_{-\pi}^\pi e^{it\lambda} \max(\tilde{f}(\lambda),0) d\lambda, \quad |t|<p$
\State $\displaystyle \hat{\omega} =\frac{\sum_{i=1}^{k-1} \boldsymbol{\mathsf{y}}_{i}^\top \hat{\boldsymbol{\Kappa}}^{-1} \boldsymbol{\mathsf{y}}_{i+1}}{\sum_{i=1}^{k-1} \boldsymbol{\mathsf{y}}_{i}^\top \hat{\boldsymbol{\Kappa}}^{-1} \boldsymbol{\mathsf{y}}_{i}}, \quad \hat{\boldsymbol{\Kappa}}\triangleq (\hat{\kappa}_p(i-j))_{1\le i,j\le p}$
\State \textbf{Output:} $\hat{\omega}$ and $\hat{\kappa}_p(t)$ for $t=0,1,2\ldots,p$
\end{algorithmic}
\footnotetext{Let $\boldsymbol{A}\triangleq(a(i,j))_{1\le i,j\le p}$ then  $|\boldsymbol{A}|_\infty\triangleq \max(|a(i,j)|; 1\le i,j\le p)$.}
\end{algorithm}

\subsection{Prediction of QPGP}
In this subsection, for a standard QPGP vector $\boldsymbol{Y}_t=[Y_1,\ldots, Y_t]^\top$, we first obtain the best linear predictor of $Y_t$ in terms of $\boldsymbol{Y}_{t-1}$ and subsequently describe a measure of the goodness of fit of the QPGP.  
For a Gaussian vector $\boldsymbol{Y}_t$, the best linear predictor  ($\hat{Y}_t$) of $Y_t$ given $\boldsymbol{Y_{t-1}}$ is the conditional expectation given as follows (see Definition 2.7.4 for the conditional mean of jointly Gaussian vectors on pp. 64 of \cite{BrockwellDavis2016}).
\begin{align}\label{eq:pred}
\hat{Y}_t&=E\left(Y_t| Y_{t-1}, Y_{t-2},\ldots,Y_{1}\right) =\boldsymbol{\Sigma}_{t-1,1}\boldsymbol{\Sigma}_{t-1}^{-1}\boldsymbol{Y}_{t-1} \mbox{ for } t>1,
\end{align}
where $\boldsymbol{\Sigma}_t=\text{Var} (\boldsymbol{Y}_t)$ as defined in \eqref{Sigma_def} and partitioned as 
$$
\boldsymbol{\Sigma}_t=\left[
\begin{array}{c:c}
\boldsymbol{\Sigma}_{t-1} &\boldsymbol{\Sigma}_{t-1,1} \\
\hdashline
\boldsymbol{\Sigma}_{1,t-1} &\frac{\kappa_p(0)}{1-\omega^2}
\end{array}
\right].
$$
Further, $\text{Var}(\hat{Y}_t)=\boldsymbol{\Sigma}_{t-1,1}\boldsymbol{\Sigma}_{t-1}^{-1}\boldsymbol{\Sigma}_{1,t-1}$. Similar to the evaluation of the likelihood function, the evaluation of $\hat{Y}_t$ also requires the expensive computation of the inverse of $\boldsymbol{\Sigma}_{t-1}$ for large $t$. Theorem~\ref{thm2}, given below, demonstrates that the proposed structural equation-based QPGP yields a computationally efficient formula for the best linear predictor of $Y_t$.  

\begin{theorem}
\label{thm2}
Let $\boldsymbol{Y}_t = [Y_1, Y_2, \ldots, Y_t]^\top$ be a standard QPGP vector with parameters $p$, $\omega$ and $\kappa_p$. Let $i(t)\triangleq\lfloor t/p \rfloor$, $l(t) \triangleq t - i(t)p$ and $\boldsymbol{\mathsf{Y}}_{j+1}^{(l)}\triangleq[Y_{jp+1},\ldots,Y_{jp+l}]^\top$ for $j=0,\ldots,k-1$. Then, the best linear predictor, $\hat{Y}_t$, of $Y_t$ given $\boldsymbol{Y_{t-1}}$  is as follows.
\begin{equation}\label{Y_pred_QPGP}
    \hat{Y}_t=\!\!\begin{cases}
    \boldsymbol{\Kappa}_{1,l(t)-1} \boldsymbol{\Kappa}_{l(t)-1}^{-1} \boldsymbol{\mathsf{Y}}_{1}^{(t-1)}  & \!\!\!\!\mbox{ if } 1<t\le p\\
        \omega Y_{t-p} + \boldsymbol{\Kappa}_{1,l(t)-1} \boldsymbol{\Kappa}_{l(t)-1}^{-1} \left({\boldsymbol{\mathsf{Y}}_{i(t)+1}^{l(t)-1}}-\omega \boldsymbol{\mathsf{Y}}_{i(t)}^{l(t)-1}\right)& \!\!\!\!\mbox{ if } t>p
    \end{cases},
\end{equation}
where $\boldsymbol{\Kappa}_{l(t)}\triangleq (\kappa_{p}(i-j))_{1\le i,j\le l(t)}$ and partitioned as
\begin{equation}
\boldsymbol{\Kappa}_{l(t)}=\left[
\begin{array}{c:c}
\boldsymbol{\Kappa}_{l(t)-1} &\boldsymbol{\Kappa}_{l(t)-1,1} \\
\hdashline
\boldsymbol{\Kappa}_{1,l(t)-1} &\kappa_p(0)
\end{array}
\right]. \label{kappa_part_thm3}
\end{equation}
Further,
\begin{equation}\label{Var_pred_Yt_expression}
   \text{Var}(\hat{Y}_t)=\begin{cases}
     \frac{1}{1-\omega^2} \boldsymbol{\Kappa}_{1,l(t)-1} \boldsymbol{\Kappa}_{l(t)-1}^{-1} \boldsymbol{\Kappa}_{l(t)-1,1} & \mbox{if } 1<t\le p\\
      \frac{\omega^2\kappa_p(0)}{1-\omega^2}+ \boldsymbol{\Kappa}_{1,l(t)-1}\boldsymbol{\Kappa}_{l(t)-1}^{-1} \boldsymbol{\Kappa}_{l(t)-1,1}  &\mbox{if } t>p
    \end{cases}.
\end{equation}    
\end{theorem}

Theorem~\ref{thm2} enables the fast evaluation of $\hat{Y}_t$ as it requires computation of the inverse of $\boldsymbol{\Kappa}_{l(t)-1}$ where $l(t)\in\{1,2,\ldots,p\}$ with computation complexity $\mathcal{O}(l(t)^2)\le \mathcal{O}(p^2)$. In contrast, the computation complexity of $\hat{Y}_t$ using expression \eqref{eq:pred} is $\mathcal{O}(t^2)$.  We numerically illustrate in subsection~\ref{faster_likelihood_nuemeric} that the computation of $\hat{Y}_t$ based on \eqref{Y_pred_QPGP} is computationally faster than that of based on \eqref{eq:pred} (see Table~\ref{tab:computational cost prediction}).

We now turn to describe a measure of goodness of fit of a standard QPGP. We first estimate the best linear predictor of $Y_t$ by using Theorem~\ref{thm2}. 
Given $n$-dimensional data vector $\boldsymbol{y}=[y_1,y_2,\ldots,y_n]^\top$ of the standard QPGP with period $p$, periodic covariance kernel $\kappa_p$, and between-period correlation $\omega$, we estimate the best linear predictor of $y_t$ using the plugin estimator as follows.
\begin{equation}\label{estimate_pred}
\hat{y}_t=\begin{cases}
    \hat{\boldsymbol{\Kappa}}_{1,l(t)-1} \hat{\boldsymbol{\Kappa}}_{l(t)-1}^{-1} \boldsymbol{\mathsf{y}}_{1}^{(t-1)}  & \mbox{ if } 1<t\le p\\
       \hat{\omega} y_{t-p} + \hat{\boldsymbol{\Kappa}}_{1,l(t)-1} \hat{\boldsymbol{\Kappa}}_{l(t)-1}^{-1} \left({\boldsymbol{\mathsf{y}}_{i(t)+1}^{l(t)-1}}-\hat{\omega} \boldsymbol{\mathsf{y}}_{i(t)}^{l(t)-1}\right) &\mbox{ if } t>p
       \end{cases},
\end{equation}
where $\hat{\omega}$ and $\hat{\boldsymbol{\Kappa}}$ are proposed estimators obtained from Algorithm~\ref{alg1}. Further, $\hat{\boldsymbol{\Kappa}}_{1,l(t)-1}$ and  $\hat{\boldsymbol{\Kappa}}_{l(t)-1}$ are obtained from $\hat{\boldsymbol{\Kappa}}$ by using \eqref{kappa_part_thm3}. The estimated predicted value, $\hat{y}_t$ given by \eqref{estimate_pred}, is also referred to as the fitted QPGP at time $t$. We choose root mean squared error (RMSE), defined below, as a measure of goodness of fit.
\begin{equation}
    \mbox{RMSE}=\sqrt{\frac1{n}\sum_{t=2}^n (y_t-\hat{y}_t)^2 }\label{chi}
\end{equation}
Note that $\mbox{RMSE}$ measures the square root of scaled squared Euclidean distance between the observation vector ($\boldsymbol{y}_n$) and its best linear predictor vector  ($\hat{\boldsymbol{y}}_n$). The measure $\mbox{RMSE}$ can be used to determine the covariance kernel $\kappa_p$ among a class of potential parametric families of covariance kernels (see examples in section~\ref{s2}) in a particular application. In such a scenario, we recommend choosing $\kappa_p$ that corresponds to the smallest $\mbox{RMSE}$. For a general QPGP, we discard the initial periodic block $\boldsymbol{\mathsf{Y}}_1$ in the computation of $\mbox{RMSE}$.

\subsection{Uncertainty quantification} \label{s4.3} 
In this subsection, we present a model-based bootstrap approach to quantify the uncertainty of the estimators $\hat{\omega}$ and $\hat{\kappa}_p$. The rapid generation of the proposed QPGP using the structural equations \eqref{eq:recursion} plays an advantageous role in the resampling procedure to generate the bootstrap samples. 

Given the QPGP data vector $\boldsymbol{y}=[y_1,y_2, \ldots, y_n]^\top$ and $\hat{\omega}$ and $\hat{\kappa}_p$ obtained from Algorithm~\ref{alg1}, we use the following resampling steps to get bootstrap estimates of the parameters $\omega$, $\boldsymbol{\Kappa}$ and best linear prediction of $y_t$ for $t=1,2,\ldots,n$. 
\begin{enumerate}
    \item{\it Residuals}: Compute the residuals, $\hat{\boldsymbol{\mathsf{z}}}_i= \boldsymbol{\mathsf{y}}_i-\hat{\omega}\boldsymbol{\mathsf{y}}_{i-1}$, for $i=2,3,.\dots,k$. 
    \item {\it Resampled periodic building blocks}: Generate $\boldsymbol{\mathsf{z}}_i^*$, for $i=2,\ldots,{k}$ by using simple random sampling with replacement from $\{\hat{\boldsymbol{\mathsf{z}}}_2, \hat{\boldsymbol{\mathsf{z}}}_3,\dots, \hat{\boldsymbol{\mathsf{z}}}_k\}$.
    \item {\it Initial periodic block}: Set $\boldsymbol{\mathsf{y}}_1^*$ as $\boldsymbol{\mathsf{y}}_1$.
    \item {\it Resampled QPGP}: Generate $\boldsymbol{y}^*=[y_1^*,y_2^*, \ldots, y_n^*]^\top$ by using \eqref{eq:recursion} and $\boldsymbol{\mathsf{y}}_1^*, \boldsymbol{\mathsf{z}}_2^*,\ldots, \boldsymbol{\mathsf{z}}_{k}^*$. 
    \item {\it Bootstrap estimates}: $\hat{\omega}^*$, $\hat{\kappa}_p^*$ by using Algorithm~\ref{alg1}.
 
\end{enumerate}
In step (3), the choice of the initial building block for the resampling scheme is in line with \cite{bootstrap_ar1}. We repeat the resampling and bootstrap estimation steps (1) to (5) a large number of times (say, $M$). The bootstrap standard error of $\hat{\omega}$ and $\hat{\kappa}_p$ is given by standard deviation of the $M$ bootstrap estimates of $\hat{\omega}^*$ and $\hat{\kappa}_p^*$, respectively. A $(1-\alpha)\%$ bootstrap confidence interval of $\omega$ is constructed using the empirical $\alpha/2$ and $(1-\alpha/2)$ quantiles of $M$ bootstraps estimates $\hat{\omega}^*$. Similarly, a pointwise confidence interval of $\kappa_p(\cdot)$ is also constructed using empirical quantiles $M$ bootstrap estimates $\hat{\kappa}_p^*(\cdot)$. 

Suppose $\kappa_p$ is known to belong to a parametric family of periodic covariance kernel $\mathbb{K}_{(\boldsymbol{\theta},\sigma^2)}$ with hyper-parameters $(\boldsymbol{\theta}, \sigma^2)$. Given an estimate of $\hat{\omega}$,  we modify step (5) of the bootstrap procedure in this scenario. The bootstrap estimates of parameter $\hat{\omega}^*$ and $ (\hat{\theta}^*, \hat{\sigma}^{2 *})$,
based on resampled QPGP $\boldsymbol{y}^*$, are obtained by using the estimation strategy as described in \eqref{parameteric_kappa}. As discussed, the bootstrap standard errors and confidence intervals of QPGP parameters $(\omega, \theta, \sigma^2)$ are obtained empirically using these bootstrap estimates. 

\section{Simulation Study}\label{s5} 

In this section, we illustrate the numerical performance of the proposed estimation algorithm as discussed in section~\ref{s4}. We also examine the performance of bootstrap standard error estimates of the proposed estimator of QPGP parameters.Then, we demonstrate the faster evaluation of likelihood and prediction for the proposed QPGP data vector. Further, we compute the average lengths of bootstrap confidence intervals and coverage probabilities for both $\omega$ and $\kappa_p$. We choose standard QPGP with period $p=10$, between period correlation $\omega=0.5$, and MacKay's periodic covariance kernel $\kappa_p$ as in \eqref{McKer}  with $\theta=1$ and $\sigma^2=1$. We compare the performances for sample sizes  $n=600$, $3000$, and $10000$. The chosen sample sizes are similar to those of the real datasets analyzed in Section~\ref{s6}. An additional simulation study to illustrate the performance of the proposed estimation methodology, under the identical experimental setup, for a larger periodicity $p=100$ (which is similar to the one of the chosen real dataset) is reported in the supplementary material. Furthermore, all experiments are performed on a standard desktop with an Intel Core~i7-12700 CPU, 16~GB DDR4 RAM, and 1~TB SSD. 

\subsection{Faster likelihood and prediction evaluation }\label{faster_likelihood_nuemeric}

We begin by comparing the computational time taken to evaluate the logarithm of the likelihood using the expressions given in \eqref{likelihood_datavector} and \eqref{likelihood_complete}, and the logarithm of the reduced likelihood given in \eqref{reduced_likelihood}. We implement the fast algorithm proposed in \cite{quasiperiodic} for computing the inverse and determinant of $\boldsymbol{\Sigma}_n$ given in \eqref{Sigma_def}. In Table~\ref{tab:computational cost likeli}, we report the exact and reduced logarithm of the likelihood values and time taken for evaluation in milliseconds. Note that the exact logarithm of the likelihood value coincides for both \eqref{likelihood_datavector} and \eqref{likelihood_complete}, and the computational time for evaluating expression \eqref{likelihood_datavector} is much larger than \eqref{likelihood_complete}.
Further, the logarithm of the reduced likelihood \eqref{reduced_likelihood} value is slightly larger than that of the exact likelihood \eqref{likelihood_complete} value.  
The difference column presents the relative difference in percentage between reduced likelihood value \eqref{reduced_likelihood} and exact likelihood value \eqref{likelihood_complete}
 (which is less than 2\% for sample size $n=600$ and further reduces as $n$ increases). The computational time to evaluate \eqref{likelihood_complete} is larger than \eqref{reduced_likelihood}.

\begin{table*}[t]
\centering
\caption{Comparative Analysis of Likelihood Estimates and Computational Cost}
\label{tab:computational cost likeli}
\setlength{\tabcolsep}{8pt}
\begin{tabular}{c c c c c}
\toprule
$n$ & Metric / Method & Eq. \eqref{likelihood_datavector} & Eq. \eqref{likelihood_complete} & Eq. \eqref{reduced_likelihood} \\
\midrule
\multirow{3}{*}{600} 
 & Log-Likelihood & $-9.61\times 10^1$ & $-9.61\times 10^1$ & $-9.46\times 10^1$ \\
 & Execution Time (ms) & $0.98$ & $0.21$ & $0.13$ \\
\midrule
\multirow{3}{*}{3000} 
 & Log-Likelihood & $-5.32\times 10^2$ & $-5.32\times 10^2$ & $-5.31\times 10^2$ \\
 & Execution Time (ms) & $6.52$ & $0.54$ & $0.23$ \\
\midrule
\multirow{3}{*}{10000} 
 & Log-Likelihood & $-1.79\times 10^3$ & $-1.79\times 10^3$ & $-1.79\times 10^3$ \\
 & Execution Time (ms) & $17.20$ & $0.62$ & $0.48$ \\
\bottomrule
\end{tabular}
\end{table*}

\begin{table}[htbp]
\centering
\caption{RMSE of proposed estimator and MLE based on 1000 runs, along with respective computational cost}
\label{tab1}
\setlength{\tabcolsep}{10pt}
\begin{tabular}{c l ccc r}
\toprule
& & \multicolumn{3}{c}{RMSE} & \\
\cmidrule(lr){3-5}
$n$ & Estimator & $\omega$ & $\theta$ & $\sigma^2$ & Time / run (ms) \\
\midrule
\multirow{2}{*}{600} 
 & Proposed & $0.0639$ & $0.1185$ & $0.1251$ & $2.04$ \\
 & MLE      & $0.0366$ & $0.0192$ & $0.0961$ & $6183.23$ \\
\midrule
\multirow{2}{*}{3000} 
 & Proposed & $0.0276$ & $0.0511$ & $0.0551$ & $4.08$ \\
 & MLE      & $0.0161$ & $0.0089$ & $0.0441$ & $39504.15$ \\
\midrule
\multirow{2}{*}{10000} 
 & Proposed & $0.0148$ & $0.0274$ & $0.0313$ & $5.06$ \\
 & MLE      & $0.0088$ & $0.0056$ & $0.0267$ & $53721.56$ \\
\bottomrule
\end{tabular}
\end{table}
We now compare the time taken to compute the prediction using the expression given in \eqref{eq:pred} and \eqref{Y_pred_QPGP}. Table~\ref{tab:computational cost prediction} reports the prediction integrated prediction squared error (IPSE) defined as $\mbox{IPSE}\triangleq\frac1n\sum_{t=2}^n(\hat{Y}_t-Y_t)^2$ and time (in milliseconds) taken to compute using expressions \eqref{eq:pred} and \eqref{Y_pred_QPGP}. Similar to the likelihood values evaluations, we observe that the values of IPSE using both the expressions \eqref{eq:pred} and \eqref{Y_pred_QPGP} match. However, the computational time of IPSE based on \eqref{eq:pred} is significantly larger than that of using \eqref{Y_pred_QPGP}. A substantial computational cost of IPSE using  \eqref{eq:pred}, for $n=10000$, is due to repetitive evaluation of the computationally expensive inverse of  $\boldsymbol{\Sigma_{t-1}
}$ for large $t$. These experiments demonstrate the computational advantages of the proposed structural equation based QPGP. 

\begin{table}[h]
    \centering
    \caption{Computational time in milliseconds for IPSE}
    \label{tab:computational cost prediction}
    \begin{tabular}{c c r c r}
        \toprule
        & \multicolumn{2}{c}{Expression \eqref{eq:pred}} & \multicolumn{2}{c}{Expression \eqref{Y_pred_QPGP}} \\
        \cmidrule(lr){2-3} \cmidrule(lr){4-5}
        $n$ & IPSE & Time & IPSE & Time \\
        \midrule
        $600$   & $1.45\times 10^{-1}$ & $1.37\times 10^2$ & $1.45\times 10^{-1}$ & $0.54$ \\
        $3000$  & $1.48\times 10^{-1}$ & $2.14\times 10^4$ & $1.48\times 10^{-1}$ & $1.91$ \\
        $10000$ & $1.50\times 10^{-1}$ & $4.78\times 10^7$ & $1.50\times 10^{-1}$ & $24.83$ \\
        \bottomrule 
    \end{tabular}
\end{table}
\subsection{Finite sample performance of proposed estimator}\label{s5b}

We now present the finite sample performance of the proposed estimation methodology in terms of root mean squared error (RMSE) based on 1000 simulation runs and compare it with the maximum likelihood estimates (MLE), which are obtained by minimizing the negative logarithm of the likelihood function given in \eqref{likelihood_complete} using a grid search algorithm. For MLE, we choose the grid for $\omega\in [0,0.99] $, $\theta\in [0.5,1.5]$ and $\sigma^2\in [0.5,1.5]$ with step sizes 0.01. .  

Table~\ref{tab1} shows the RMSE of the proposed estimator and the MLE of the QPGP parameters. It also shows the computational time per run in milliseconds for the proposed estimator as well as MLE. We observe that RMSE of the proposed estimator is slightly larger than that of MLE.  
Note that the computational time of MLE is tremendously larger than the proposed estimator. We also observed that a coarser grid size reduces the computational cost of MLE; however, it leads to a larger  RMSE of the MLE in comparison to the proposed methodology. This indicates that the proposed estimation strategy has a sizable computational advantage over the grid search based MLE and exhibits a comparable accuracy in comparison to MLE. 

\subsection{Finite sample performance of bootstrap standard errors} \label{s5c}

We estimate the bootstrap standard errors of the proposed estimators $\hat{\omega}$, $\hat{\theta}$ and $\hat{\sigma}^2$ based on $M(=1000)$ bootstrap samples corresponding to each run by using the resampling steps outlined in subsection~\ref{s4.3}. We also compute the standard errors of these estimators across $1000$ independent runs. 

\begin{figure*}
\centering
\begin{tabular}{ccc}
S.E.($\hat{\omega}$) & S.E.($\hat{\theta}$) & S.E.($\hat{\sigma}^2$)\\
\includegraphics[width=0.32\textwidth, height=1.8in]{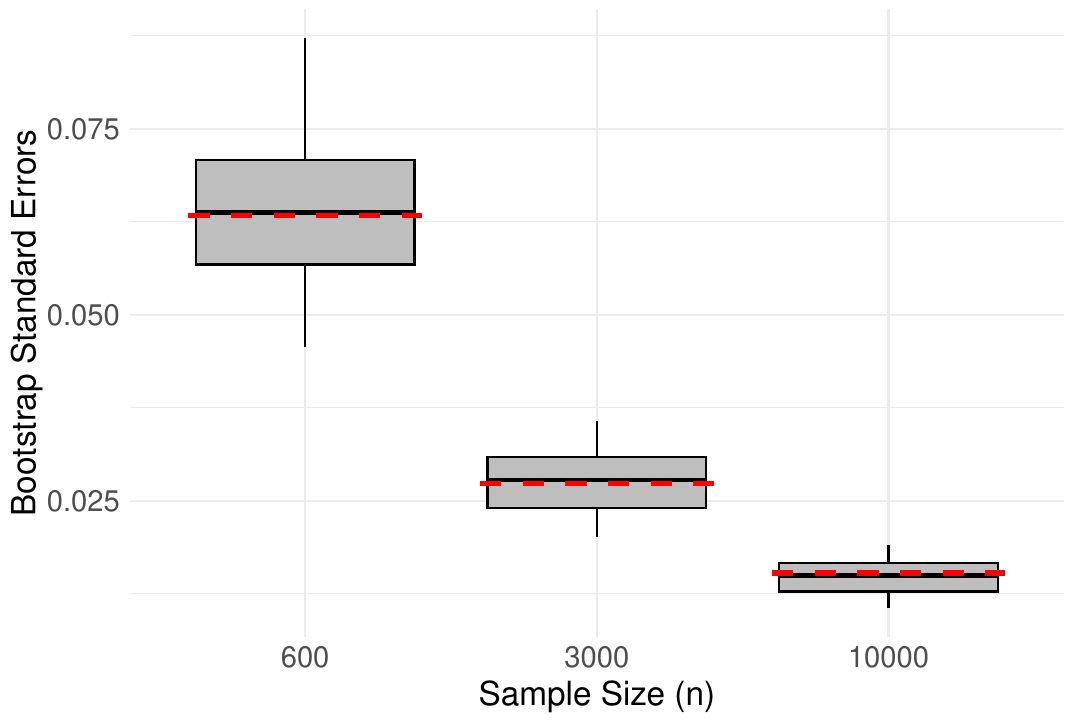} 
&
\includegraphics[width=0.32\textwidth, height=1.8in]{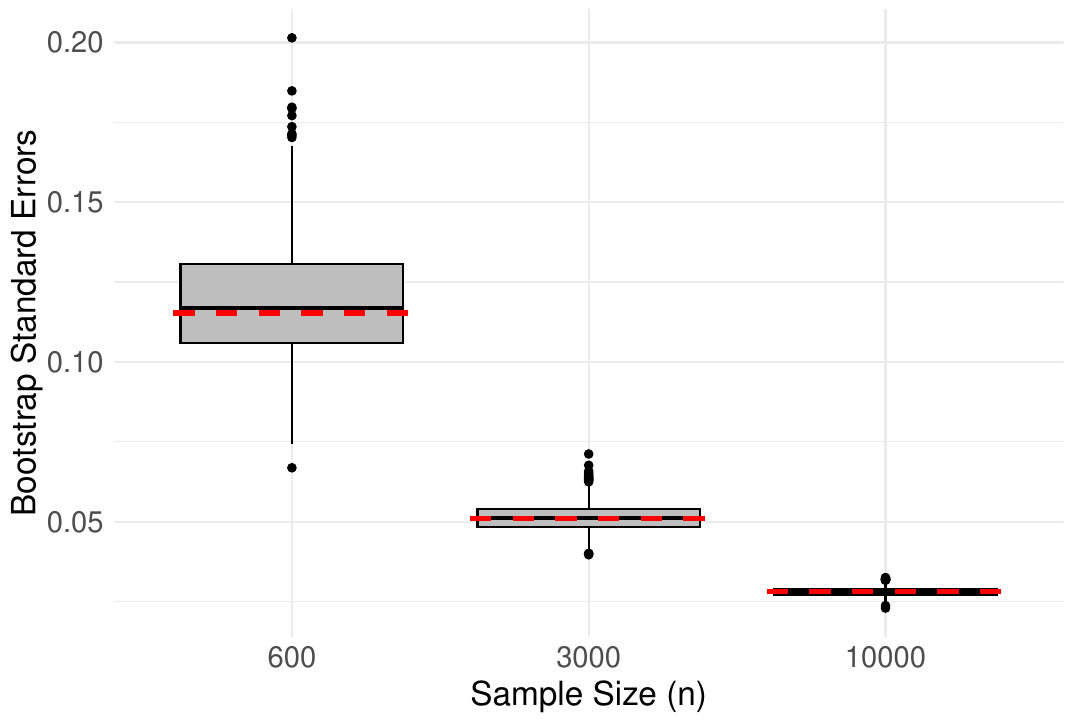}
&
\includegraphics[width=0.32\textwidth, height=1.8in]{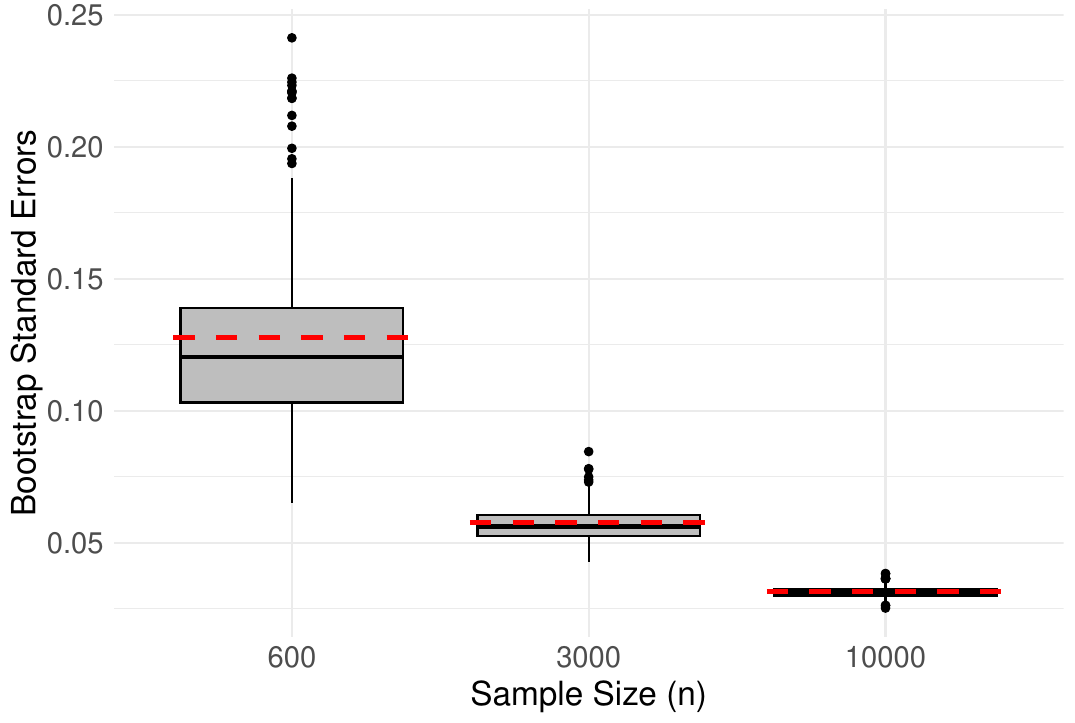}
\end{tabular}
\caption{The box plots of bootstrap standard errors (computed from $M=1000$ bootstrap samples) of $\hat{\omega}$, $\hat{\theta}$ and $\hat{\sigma}^2$, based on $1000$ simulation runs of standard QPGP with period $p=10$, $\omega=0.5$ and Mackay's periodic kernel with $( \theta=1, \sigma^2=1)$, are shown in left, center and right panel, respectively. Each panel consists of three box plots corresponding to sample sizes $n=600$, $3000$, and $10000$. The empirical standard error of estimators across simulation runs is shown as a dashed horizontal red line.}
\label{fig:bootstrap}
\end{figure*}

The left column of Figure~\ref{fig:bootstrap} shows the boxplots of bootstrap standard errors of $\hat{\omega}$ based on $1000$ independent runs for sample sizes $n=600$, $3000$ and $10000$. The decreasing spread of the boxplots as $n$ increases indicates a reduction in variability, consistent with a smaller standard deviation for larger sample sizes.
The red dashed line corresponds to the standard error of $\hat{\omega}$ across simulation runs. We observe that the gap between median bootstrap standard error and across-run standard errors reduces as the sample size increases. Further, the width of boxes also reduces as the sample size ($n$) increases. The center and right columns of Figure~\ref{fig:bootstrap} show the boxplots of bootstrap standard errors of $\hat{\theta}$ and $\hat{\sigma}^2$. We observe similar patterns as in the case of $\hat{\omega}$. This indicates that the bootstrap standard errors of the proposed estimator approximate the true standard errors reasonably well. 

\subsection{Performance of Bootstrap Confidence Intervals} \label{coverage}
We compute the average length of the 95\% bootstrap confidence intervals and their empirical coverage probabilities for the parameters ${\omega}$ and ${\kappa}_p$, based on $M(=1000)$ bootstrap samples corresponding to each run by using the resampling steps outlined in subsection \ref{s4.3}. We compute this for both estimators across 1000 independent runs.

\begin{figure}[htbp]
    \centering
    \begin{minipage}{0.48\textwidth}
        \centering
        \includegraphics[width=\linewidth]{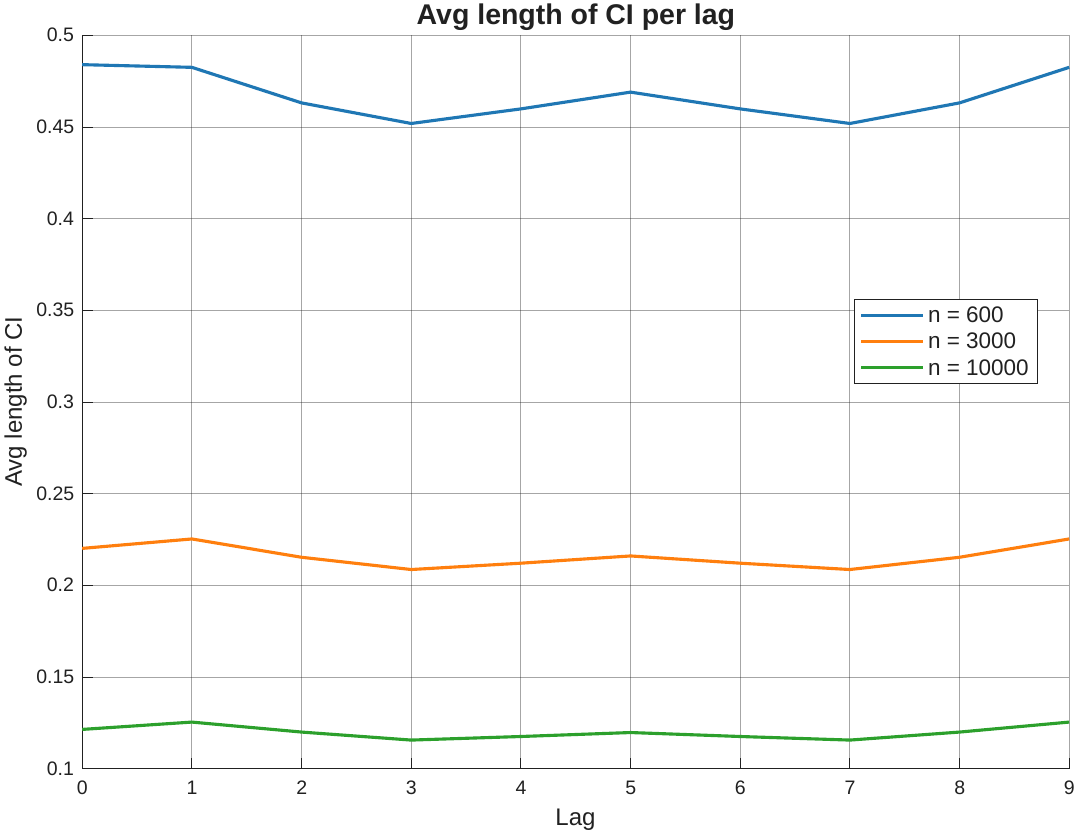}
        \caption{Average length of bootstrap  CI for ${\kappa}_p$.}
        \label{fig:elength_p10}
    \end{minipage}%
    \hfill
    \begin{minipage}{0.48\textwidth}
        \centering
        \includegraphics[width=\linewidth]{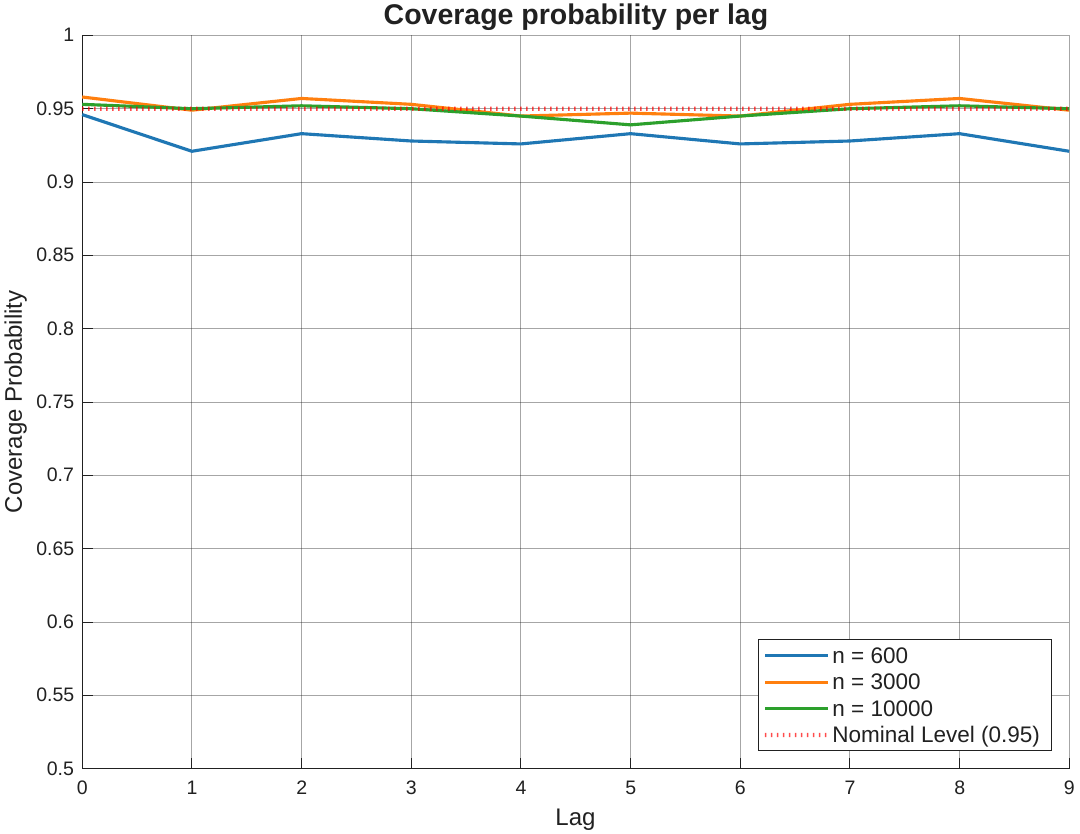}
        \caption{Empirical coverage probability of bootstrap CI  for ${\kappa}_p$.}
        \label{fig:coverage_p10}
    \end{minipage}
\end{figure}

Table \ref{coverage_omega},  and Figures \ref{fig:elength_p10} and \ref{fig:coverage_p10} show the average lengths and coverage probabilities of the 95\% bootstrap confidence intervals for $\omega$ and $\kappa_p$, respectively, computed across 1000 independent simulations for sample sizes $n =600, 3000$ and $10000$. These indicate that increasing the sample size leads to narrower confidence intervals while driving the coverage probabilities closer to the nominal level 95\%.  

\begin{table}
    \centering
    \caption{Average length and coverage probability of bootstrap  CI for $\omega$}
    \label{coverage_omega}
    \begin{tabular}{ccc}
        \toprule
        \textbf{$n$} & \textbf{Average length } & \textbf{Empirical coverage } \\
        \midrule
        600    & 0.2530  & 0.957 \\
        3000   & 0.1096  & 0.943 \\
        10000  & 0.0604  & 0.959 \\
        \bottomrule
    \end{tabular}
\end{table}

\section{Case Studies}\label{s6}
In this section, we fit the proposed QPGP model to three distinct real datasets, which are known to be quasi-periodic signals. Suppose the exact periodicity of a quasi-periodic signal is not known but believed to belong to an integer set $\mathcal{P}$. In that case, we fit standard QPGP with a general periodic kernel for each $p\in \mathcal{P}$  using Algorithm~\ref{alg1} and determine the periodicity $p$ that corresponds to the smallest reduced negative logarithm of likelihood $\tilde{\ell}_n$ given in \eqref{likelihood_with_partial} (if $n=kp$ for some positive integer $k$, then use $\tilde{\ell}_n$ as given in \eqref{reduced_likelihood}). 
To compare the performance of fitted QPGP over different periodic covariance kernels, we consider the following choices of  $\kappa_p$ kernels: 
\begin{itemize}
    \item General periodic kernel
    \item  MacKay’s kernel given in \eqref{McKer}.
    \item Periodic Mat\'ern's kernel given in  \eqref{MatKer} with $\nu = 1.5$. 
    \item Cosine kernel given in \eqref{CosKer} with $\iota=1$.
\end{itemize}

We evaluate the RMSE corresponding to all the chosen kernels and the benchmark method in \cite{quasiperiodic} and show it in Table~\ref{rmse}. 
We report here the estimates of the QPGP parameters, along with their bootstrap standard errors, corresponding to the kernel that yields the smallest RMSE among the chosen periodic kernels. The standard errors are computed based on $M=1000$ resamples. We report the same details of estimates of QPGP parameters for all the chosen periodic covariance kernels in the supplementary material. 

\subsection{Carbon Dioxide Emission Signal} 
We consider the monthly carbon dioxide emission data, measured in ppm by SIO (Scripps Institution of Oceanography, San Diego) air sampling network, from the year 1958 to 2003, publicly available at \href{https://www.osti.gov/dataexplorer/biblio/dataset/1389346}{the DOE Data Explorer}. The carbon dioxide emission signal exhibits an increasing trend over the years, along with quasi-periodic behavior \cite{rasmussen}. 
We first adjust the trend in the data by fitting a quadratic regression over time using least squares, and proceed to fit the QPGP on the trend-adjusted CO$_2$ emission signal.
The data set consists of $n=612$ time instances with six missing entries. We imputed the missing entries by using linear interpolation.
The approximate periodicity of the data appears to be $p=12$. As discussed, we fitted a general QPGP for $p \in \mathcal{P}=\{2,3,\dots,20\}$, and the smallest reduced negative likelihood $\tilde{\ell}_n$ corresponds to $p=12$. This is in line with the approximate periodicity of the data.

\begin{table*}[htbp]
\centering
\caption{RMSE corresponding to proposed QPGP vs. Benchmark \cite{quasiperiodic} with estimation time and prediction time for chosen datasets \tablefootnote{The bold entry in each cell corresponds to the lowest value, i.e., the minimum RMSE and the lowest computational time, while the underlined entry denotes the second lowest value.}}
\label{rmse}
\begin{tabular}{llccc}
\toprule
Dataset & Methods & RMSE & Est. Time (ms) & Pred. Time (ms) \\ 
\midrule

\multirow{5}{*}{\textit{CO$_2$}} 
& QPGP-General & \textbf{0.404} & \textbf{20.24} & 4.82 \\
& QPGP-Matern ($\nu=1.5$) & 0.512 & 37.75 & \textbf{3.29} \\
& QPGP-Mackay & 0.461 & 42.37 & \underline{3.31} \\
& QPGP-Cosine ($\iota=1$) & 0.700 & \underline{30.45} & 8.82 \\
& Benchmark \cite{quasiperiodic} & \underline{0.450} & 632.06 & 83.37 \\
\cmidrule(lr){1-5}

\multirow{5}{*}{\textit{Sunspot}} 
& QPGP-General & \underline{35.589} & \textbf{16.07} & \underline{2.06} \\
& QPGP-Matern ($\nu=1.5$) & \textbf{33.956} & 47.78 & \textbf{1.95} \\
& QPGP-Mackay & 37.094 & 61.81 & 2.18 \\
& QPGP-Cosine ($\iota=1$) & 41.093 & \underline{46.35} & 7.10 \\
& Benchmark \cite{quasiperiodic} & 35.824 & 274.22 & 38.67 \\
\cmidrule(lr){1-5}

\multirow{5}{*}{\textit{Water Level}} 
& QPGP-General & \underline{0.036} & \textbf{1461.06} & \underline{1257.29} \\
& QPGP-Matern ($\nu=1.5$) & \textbf{0.027} & 1520.19 & \textbf{1213.21} \\
& QPGP-Mackay & 0.110 & 1501.74 & 1994.56 \\
& QPGP-Cosine ($\iota=1$) & 0.090 & \underline{1484.46} & 1401.55 \\
& Benchmark \cite{quasiperiodic} & 0.080 & 3720.93 & 9144.02 \\
\bottomrule

\end{tabular}
\end{table*}

\begin{figure}
    \centering
\centering
    \includegraphics[width=2.3in,height=1.8in]{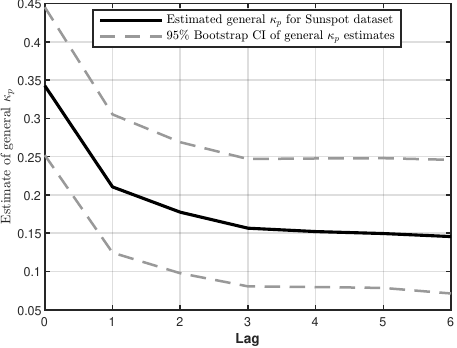}
    \caption{The plot shows the estimates of general $\kappa_p(\cdot)$ against lag in black solid line corresponding to the CO$_2$ dataset, along with the 95\% bootstrap confidence limits in grey-dashed lines.}
    \label{co2_general_kappa}
\end{figure}

\begin{figure}[htbp]
    \centering
    \includegraphics[width=\linewidth]{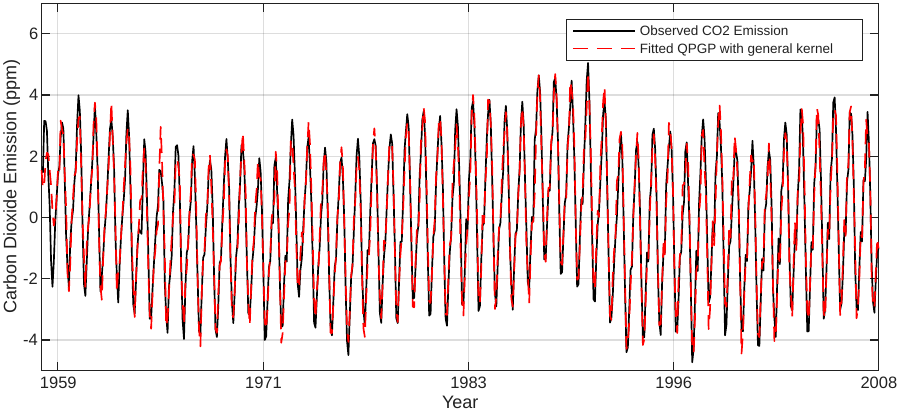}
    \caption{The plot shows the detrended carbon dioxide emission levels vs. year in black solid line together with fitting standard QPGP with $p=12$ and general kernel in dashed red line. The grey-shaded region corresponds to the estimated $95\%$ prediction intervals using plug-in estimates. } 
    \label{co2}
\end{figure}

The top section of Table~\ref{rmse} shows the RMSE values for CO$_2$ emission data corresponding to the chosen periodic covariance kernels $\kappa_p$ with $p=12$. The smallest RMSE corresponds to the general choice of covariance kernel. The estimate of $\omega$ corresponding to the general kernel turns out to be $0.9752$ with bootstrap standard error $0.0085$ and 95\% confidence interval as $(0.9705,1.0038)$. Figure~\ref{co2_general_kappa} shows the plot of estimates of the general covariance kernel against lag in a solid black line, along with $95\%$ confidence limits in the dashed black lines. We also observe that the estimated $\kappa_p(\cdot)$ values are included in $95\%$ confidence limit. Figure~\ref{co2} shows the plot of detrended CO$_2$ emission data in a black solid line and the fitted QPGP corresponding to the general kernel in a dashed red line. We observe that the proposed QPGP fits the data reasonably well. 

\subsection{Sunspot Numbers Data}
Sunspot numbers observed over the years appear to exhibit a quasi-periodic pattern. The underlying solar magnetic dynamo, which involves nonlinear and chaotic processes, reflects the irregularities in cycle timing and intensity in sunspots.  Further, the turbulence in the solar plasma also affects the periodic pattern of the sunspot numbers (\cite{sunspot_lit}). We consider the yearly sunspot numbers from $1703$ to $2025$. The dataset consists of $n=322$ samples. The data is publicly available at \href{http://www.sidc.be/silso/datafiles}{the SILSO database}.
\cite{quasiperiodic} fitted the QPGP model (that corresponds to MacKay's kernel) to the sunspot data. It is well known that the approximate periodicity of sunspot numbers is 11 years \cite{Balogh2014}.  As indicated earlier, we fitted standard QPGP for the general covariance kernel to sunspot data corresponding to  $p\in \mathcal{P}=\{2,4,\dots,20\}$. The smallest reduced negative likelihood ($\tilde{\ell}_n$) corresponds to  $p=11$. This is in tune with the well-known approximate periodicity of the sunspot numbers.   

\begin{figure}[h]
    \centering
    \includegraphics[width=\linewidth]{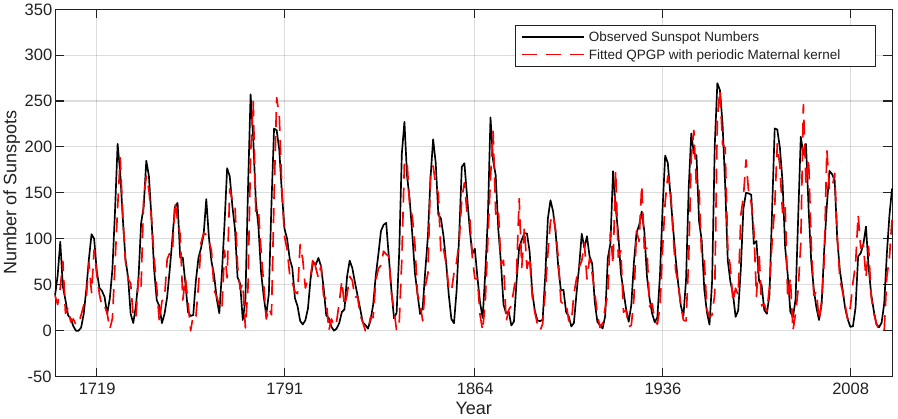}
    \caption{The plot shows sunspot numbers vs. year in a black solid line together with fitting standard QPGP with period $p=11$ and Mat\'ern kernel in a dashed red line. The grey-shaded region corresponds to the estimated $95\%$ prediction intervals using plug-in estimates. }
    \label{sunspot}
\end{figure}

The middle section of Table~\ref{rmse} shows the RMSE values corresponding to the chosen periodic covariance kernels with $p=11$.
The smallest RMSE value corresponds to the periodic Mat\'ern kernel. 
Table~\ref{Est_para_sunspot} shows the estimates of fitted QPGP parameters (with $p=11$, $\kappa_p$ as periodic Mat\'ern kernel)   along with bootstrap standard errors and 95\% confidence intervals. Note that the confidence interval of QPGP parameters does not include $0$, which indicates that the estimates of QPGP parameters are statistically significant.
Figure~\ref{sunspot} shows the plot of sunspot numbers over the years in a black solid line, along with the fitted QPGP with $p=11$ with Mat\'ern kernel in a dashed red line. We observe that the proposed QPGP mostly fits the sunspot numbers well, except for a few years when the signal appears to be relatively weak.  

\begin{table}[h]
\centering
\caption{Estimates of QPGP parameter corresponding to $p=11$ and periodic Mat\'ern kernel ($\nu=1.5$) for sunspot data}
\label{Est_para_sunspot}
\begin{tabular}{cccc}
\toprule
Parameter & Estimate & Standard Error & Confidence Interval \\
\midrule
$\omega$   & $0.7228$    & $0.06375$  & $(0.5861, 0.8429)$   \\
$\sigma^2$ & $2568.1523$ & $563.3239$ & $(1439.5634, 3699.3789)$ \\
$\theta$   & $0.7599$    & $0.0814$   & $(0.5436, 0.8539)$   \\
\bottomrule
\end{tabular}
\end{table}

\begin{figure*}
    \centering
    \includegraphics[width=5.2in]{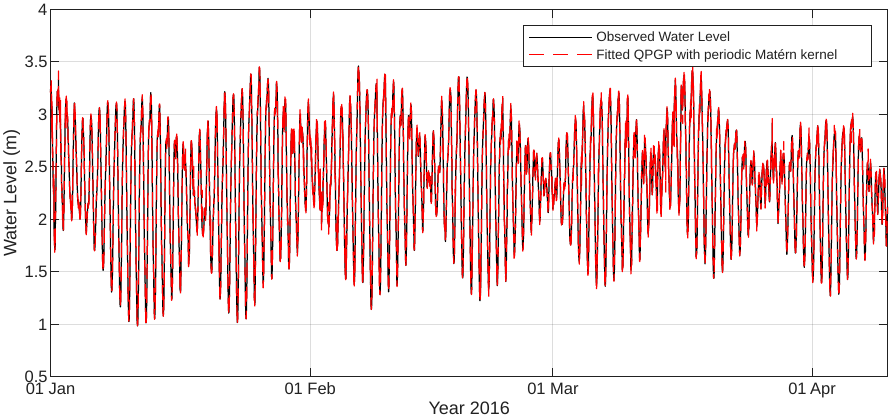}
    \caption{The plot shows the water level in a black solid line against days, together with fitted standard QPGP (with $p=148$ and $\kappa_p$ periodic Mat\'ern kernel) in a dashed red line. The grey-shaded region corresponds to the estimated $95\%$ prediction intervals using plug-in estimates.  }
    \label{tide}
\end{figure*}

\subsection{Water Level Signal} 
The water level at a specific location of the sea depends on the complex climate phenomenon and appears to be quasi-periodic due to the tide. Since tides are affected by multiple natural cycles due to the gravitational pull of the moon and the sun, these cycles of tides are a source of periodicity in the water level. The local weather conditions, sea level, and bathymetry of the location also affect the tidal measurements, which adds to the periodic variation in water levels  (see \cite{tide_qpgp}). 
We consider the water levels dataset, recorded over a uniform time interval at an automatic tide gauge on Mornington Island in Queensland\footnote{Publicly available at \href{https://www.data.qld.gov.au/dataset/mornington-island-tide-gauge-archived-interval-recordings/resource/53254b19-7aa2-4ca5-ac34-2153675e6026?view_id=d8d5d761-7e90-4588-8b92-2987336246e7}{Queensland Government’s open data portal.}}.
This data set contains the water level records measured from January 1, 2016 00:00 Hrs. to April 9, 2016 23:50 Hrs over a uniform time interval of 10 minutes and consists of $n= 14400$ observations. 
 We choose this high-frequency water level quasi-periodic dataset to illustrate that the proposed QPGP model can be fitted to long quasi-periodic signals with large sample counts efficiently, while providing uncertainty quantification of estimated parameters. 

The approximate periodicity of the water level appears to be $24$ hours. We therefore fitted standard QPGP with the general kernel for $p\in\mathcal{P}=\{132,137,\dots, 156\}$ (between 22 hours to 26 hours) to determine $p$. The smallest reduced negative likelihood value corresponds to $p=148$ (24 hours and 40 minutes). As we observed in the CO$_2$ emission and sunspot numbers, the described method to determine $p$ for water level data is also in tune with conventional observations.

The bottom section of Table~\ref{rmse} shows the RMSE values for the water level data corresponding to the chosen periodic covariance kernels $\kappa_p$ with $p=148$. The smallest RMSE corresponds to Mat\'ern periodic kernel; however, the RMSE value corresponding to the general kernel is very close to it. 

Table~\ref{Est_para_water_level} shows the estimates of fitted QPGP parameters (with $p=148$, $\kappa_p$ as periodic Mat\'ern kernel)   along with bootstrap standard errors and 95\% confidence intervals.  The confidence intervals of the QPGP parameters exclude $0$, which shows that the estimates of the parameters are statistically significant. Figure~\ref{tide} shows the plot of the water level in a black solid line, along with the fitted QPGP (with $p=148$ and $\kappa_p$ as Mat\'ern kernel) in a dashed red line. 
We observe that the fitted QPGP is very close to the data.      

\begin{table}[h]
\centering
\caption{Estimates of QPGP parameter corresponding to $p=148$ and periodic Mat\'ern kernel ($\nu=1.5$) for water level data}
\label{Est_para_water_level}
\begin{tabular}{c r r r}
\toprule
Parameter & Estimate & Standard Error & Confidence Interval \\
\midrule
$\omega$   & $0.9673$ & $0.0102$ & $(0.9432, 0.9824)$ \\
$\sigma^2$ & $0.0358$ & $0.0872$ & $(0.0237, 0.3508)$ \\
$\theta$   & $0.8338$ & $0.4892$ & $(0.5649, 2.1742)$ \\
\bottomrule
\end{tabular}
\end{table}

\section{Conclusion}\label{s7}
In this article, we develop a novel family of Quasi-Periodic Gaussian Processes by using a system of structural equations which provide a flexible framework for modeling the within period correlation of the QPGP. We show that the structural equations simplify the likelihood function, proving more computationally efficient to evaluate than the prior work on rapid likelihood evaluation \cite{quasiperiodic}. Importantly, the proposed approach generalises to a broad class of within period kernels, both parametric and non-parametric.

Given a data vector of the proposed QPGP, the maximum likelihood estimation technique of the QPGP parameters requires optimization of the likelihood function over the family of general periodic covariance kernels, which is a non-convex set. We address this issue by developing a two stage fast estimation algorithm based on a reduced likelihood function. We establish that the proposed estimation strategy provides statistically consistent estimators. We numerically show that that the accuracy of the proposed estimator is comparable to the maximum likelihood estimator. We illustrate that the proposed estimation strategy is computationally faster than prior work \cite{quasiperiodic}. Further, the structural equations reduce the computational cost of the best linear prediction of QPGP significantly. The proposed QPGP based on structural equations enables rapid sample path generation. We exploit this to construct a bootstrap methodology for standard error estimation for the proposed QPGP parameter estimators. The proposed method trades a small, quantifiable loss of estimation accuracy for a larger computational efficiency, and separately offers kernel flexibility that can recover or exceed benchmark accuracy when a better-matched kernel exists.

The general selection of periodic kernel sets a reasonable QPGP fit for all the chosen data. The QPGP with parametric choice of periodic kernel, in particular periodic Mat\'ern kernel, competes with the general periodic kernel and exhibits superior performance for two datasets. This highlights the advantage of the proposed QPGP, which offers a flexible choice of periodic covariance kernels. 
The rapid generation of the proposed QPGP and the computationally efficient proposed parameter estimation methodology enable the uncertainty quantification in the estimates, even for a long quasi-periodic signal such as the water level signal.  

\appendix
\section{Appendix: Estimation strategy for a partial periodic block} 
\subsection{Estimation strategy for a partial periodic block} \label{incomplete_block}

In this appendix, we consider the QPGP data vector $\boldsymbol{y}=[y_1,y_2,\ldots, y_n]$ where $n=kp+l$ for some $l\in\{1,2,\ldots,p-1\}$, i.e., the data vector consists of complete observations on the periodic blocks $\boldsymbol{\mathsf{y}}_i$ for $i=1,2,\ldots,k$ and partial observation of $(k+1)^{\text{th}}$ block $\boldsymbol{\mathsf{y}}_{k+1}^{(l)}$.  For the QPGP data vector  $\boldsymbol{y}$ ,the negative logarithm of the likelihood function is given as follows, with $c=n\log(\sqrt{2\pi})$:
 \begin{align}
   \ell_n(\omega, \kappa_p)&=\underbrace{\frac{1}{2}\log|\boldsymbol{\Kappa}_{l}|+ \frac{1}{2}\left(\boldsymbol{\mathsf{y}}_{k+1}^{(l)}-\omega \boldsymbol{\mathsf{y}}_{k}^{(l)}\right)^\top\boldsymbol{\Kappa}_{l}^{-1} \left(\boldsymbol{\mathsf{y}}_{k+1}^{(l)}-\omega \boldsymbol{\mathsf{y}}_{k}^{(l)}\right)}_{\text{Contribution of partial block $\boldsymbol{\mathsf{Y}}_{k+1}^{(l)}$ } }\nonumber\\
   & + \frac{k-1}{2} \log(|\boldsymbol{\Kappa}|)
    +\underbrace{ \ \ \frac12\sum_{i=1}^{k-1}(\boldsymbol{\mathsf{y}}_{k-i+1}-\omega \boldsymbol{\mathsf{y}}_{k-i})^{\top}\boldsymbol{\Kappa}^{-1}(\boldsymbol{\mathsf{y}}_{k-i+1}-\omega \boldsymbol{\mathsf{y}}_{k-i})}_{\text{Contribution of blocks $\boldsymbol{\mathsf{Y}}_k,\ldots, \boldsymbol{\mathsf{Y}}_1$}}\nonumber\\
    &+\underbrace{ \frac{1}{2}\log\left(\Big|\frac{1}{1-\omega^2}\boldsymbol{\Kappa}\Big|\right)
    +\frac{1}{2}\boldsymbol{\mathsf{y}}_1 ^\top \left(\frac{1}{1-\omega^2}\boldsymbol{\Kappa}\right)^{-1}\boldsymbol{\mathsf{y}}_1}_{\text{Marginal contribution of block $\boldsymbol{\mathsf{Y}}_1$}} +c,\label{likelihood_with_partial}
\end{align}
As discussed in section~\ref{s4}, we ignore the marginal contribution of the initial periodic block $\boldsymbol{\mathsf{Y}}_1$ in the likelihood given in \eqref{likelihood_with_partial} in our estimation strategy. Like \eqref{reduced_likelihood}, the scaled negative logarithm of reduced likelihood function is updated as follows. 
 \begin{align}
\tilde{\ell}_n(\omega, \kappa_p)
    &=\frac{1}{k-1}\log|\boldsymbol{\Kappa}_{l}|+\frac1{k-1}\left(\boldsymbol{\mathsf{y}}_{k+1}^{(l)}-\omega \boldsymbol{\mathsf{y}}_{k}^{(l)}\right)^\top\boldsymbol{\Kappa}_{l}^{-1} \left(\boldsymbol{\mathsf{y}}_{k+1}^{(l)}-\omega \boldsymbol{\mathsf{y}}_{k}^{(l)}\right)
    \nonumber\\
    &+\log(|\boldsymbol{\Kappa}|)+\frac1{k-1}\sum_{i=1}^{k-1}(\boldsymbol{\mathsf{y}}_{k-i+1}-\omega \boldsymbol{\mathsf{y}}_{k-i})^{\top}\boldsymbol{\Kappa}^{-1}(\boldsymbol{\mathsf{y}}_{k-i+1}-\omega \boldsymbol{\mathsf{y}}_{k-i}).    \label{reduced_likelihood_with_partial}
\end{align}
Note that, by using the partition and transformation
\begin{eqnarray}\boldsymbol{\Kappa}&\triangleq&\left[ 
    \begin{array}{c:c}
             \boldsymbol{\Kappa}_{l}&\boldsymbol{\Kappa}_{l,p-l}\\
             \hdashline
        \boldsymbol{\Kappa}_{p-l,l}&\boldsymbol{\Kappa}_{p-l}
    \end{array}\right],\label{def_kappa_part_partial}\\
    \boldsymbol{\Kappa}_{(p-l).l}&\triangleq&\boldsymbol{\Kappa}_{p-l}-\boldsymbol{\Kappa}_{p-l,l} \boldsymbol{\Kappa}_{l}^{-1}\boldsymbol{\Kappa}_{l,p-l},\label{s22.1_def}
\end{eqnarray} 
we have
\begin{equation}\label{det_kappa_part}
    |\boldsymbol{\Kappa}|= |\boldsymbol{\Kappa}_l|\times |\boldsymbol{\Kappa}_{(p-l).l}|.
\end{equation}
By using the transformation and partition 
\begin{eqnarray}
    \boldsymbol{\mathfrak{Z}}_{i}(\omega)&\!\!\!=\!\!\!&\boldsymbol{\mathsf{y}}_{i}-\omega \boldsymbol{\mathsf{y}}_{i-1} \mbox{ for } i=2,\ldots,k,\label{def_zomega}\\   \boldsymbol{\mathfrak{Z}}_i(\omega)&\!\!\!=\!\!\!&
    \left[\begin{array}{c}
         
        \boldsymbol{\mathfrak{Z}}_i^{(l)}(\omega)\\
        \hdashline\boldsymbol{\mathfrak{Z}}_i^{(p-l)}(\omega)
     
    \end{array}\right]= 
    \left[\begin{array}{c}
         
        \boldsymbol{\mathsf{y}}_i^{(l)}-\omega\boldsymbol{\mathsf{y}}_{i-1}^{(l)}\\
        \hdashline\boldsymbol{\mathsf{y}}_i^{(p-l)}-\omega\boldsymbol{\mathsf{y}}_{i-1}^{(p-l)}
    \end{array}\right],\label{def_zomega_part}
\end{eqnarray}
and some algebra, the quadratic summand of the fourth term on the RHS of \eqref{reduced_likelihood_with_partial} simplifies as follows.
\begin{align}
    &(\boldsymbol{\mathsf{y}}_{k-i+1}-\omega \boldsymbol{\mathsf{y}}_{k-i})^{\top}\boldsymbol{\Kappa}^{-1}(\boldsymbol{\mathsf{y}}_{k-i+1}-\omega \boldsymbol{\mathsf{y}}_{k-i})= \boldsymbol{\mathfrak{Z}}_{k-i+1}^{(l)}(\omega)^\top \boldsymbol{\Kappa}_l^{-1} \boldsymbol{\mathfrak{Z}}_{k-i+1}^{(l)}(\omega)\nonumber\\
&+\left(\boldsymbol{\mathfrak{Z}}_{k-i+1}^{(p-l)}(\omega)-\boldsymbol{\Kappa}_{p-l,l}\boldsymbol{\Kappa}_{l}^{-1}\boldsymbol{\mathfrak{Z}}_{k-i+1}^{(l)}(\omega) \right)^\top \boldsymbol{\Kappa}_{(p-l).l}^{-1}\left(\boldsymbol{\mathfrak{Z}}_{k-i+1}^{(p-l)}(\omega)-\boldsymbol{\Kappa}_{p-l,l}\boldsymbol{\Kappa}_{l}^{-1}\boldsymbol{\mathfrak{Z}}_{k-i+1}^{(l)}(\omega) \right).
    \label{forthterm_tilde_ell_partial}
\end{align}
By using \eqref{det_kappa_part} and \eqref{forthterm_tilde_ell_partial},  $\tilde{\ell}_n$, given in \eqref{reduced_likelihood_with_partial}, can be viewed as a twice differentiable function of $\omega$, $\boldsymbol{\Kappa}_l$, $\boldsymbol{\Kappa}_{p-l,l}\boldsymbol{\Kappa}_l^{-1}$ and $\boldsymbol{\Kappa}_{(p-l).l}$. By equating the first derivatives of $\tilde{\ell}_n$ with respect to $\omega$, $\boldsymbol{\Kappa}_l$, $\boldsymbol{\Kappa}_{p-l,l}\boldsymbol{\Kappa}_l^{-1}$ and $\boldsymbol{\Kappa}_{(p-l).l}$ to $0$, the stationary points of $\tilde{\ell}_n$ satisfy the following relations.
\begin{align}
\omega=&\frac{\sum_{i=1}^{k-1} \boldsymbol{\mathsf{y}}_{k-i}^\top  \boldsymbol{\Kappa}^{-1} \boldsymbol{\mathsf{y}}_{k-i+1}+\boldsymbol{\mathsf{y}}_{k}^{(l)\top}  \boldsymbol{\Kappa}_l^{-1} \boldsymbol{\mathsf{y}}_{k+1}^{(l)}}{\sum_{i=1}^{k-1} \boldsymbol{\mathsf{y}}_{k-i}^\top  \boldsymbol{\Kappa}^{-1} \boldsymbol{\mathsf{y}}_{k-i}+\boldsymbol{\mathsf{y}}_{k+1}^{(l)\top}  \boldsymbol{\Kappa}_{l}^{-1} \boldsymbol{\mathsf{y}}_{k+1}^{(l)}} \label{d_omega}
\end{align}
\begin{align}
\boldsymbol{\Kappa}_l=&\frac{1}{k}\sum_{i=0}^{k-1}\boldsymbol{\mathfrak{Z}}_{k-i+1}^{(l)}(\omega)\boldsymbol{\mathfrak{Z}}_{k-i+1}^{(l)}(\omega)^\top \label{d_K11}
\end{align}
\begin{align}
\boldsymbol{\Kappa}_{p-l,l} = \left(\frac1{k-1}\sum_{i=1}^{k-1}\boldsymbol{\mathfrak{Z}}_{k-i+1}^{(p-l)}(\omega)\boldsymbol{\mathfrak{Z}}_{k-i+1}^{(l)}(\omega)^\top \right) \left(\frac1{k-1}\sum_{i=1}^{k-1}\boldsymbol{\mathfrak{Z}}_{k-i+1}^{(l)}(\omega)\boldsymbol{\mathfrak{Z}}_{k-i+1}^{(l)}(\omega)^\top\right)^{-1} \boldsymbol{\Kappa}_l\label{d_B}
\end{align}
\begin{align}
    \boldsymbol{\Kappa}_{(p-l).l}=&\frac{1}{k-1}\sum_{i=1}^{k-1}\left(\boldsymbol{\mathfrak{Z}}_{k-i+1}^{(p-l)}(\omega)-\boldsymbol{\Kappa}_{p-l,l}\boldsymbol{\Kappa}_{l}^{-1}\boldsymbol{\mathfrak{Z}}_{k-i+1}^{(l)}(\omega) \right)\nonumber\\
&\hskip-10pt\times \left(\boldsymbol{\mathfrak{Z}}_{k-i+1}^{(p-l)}(\omega)-\boldsymbol{\Kappa}_{p-l,l}\boldsymbol{\Kappa}_{l}^{-1}\boldsymbol{\mathfrak{Z}}_{k-i+1}^{(l)}(\omega) \right)^\top. \label{dK22.1}
\end{align}
Given three out of four variables among $\omega$, $\boldsymbol{\Kappa}_l$, $\boldsymbol{\Kappa}_{p-l,l}\boldsymbol{\Kappa}_l^{-1}$ and $\boldsymbol{\Kappa}_{(p-l).l}$, the second derivative of $\tilde{\ell}_n$ is positive definite at the stationary points. Therefore, using a similar argument as in subsection~\ref{s4.1}, iterative alternate minimization converges. 

We now describe the modified  iterative steps of stage~I of Algorithm~\ref{alg1} to accommodate the observed partial block $\boldsymbol{\mathsf{y}}_{k+1}^{(l)}$ in the following manner to get $\tilde{\omega}_n$ and $\tilde{\boldsymbol{\Kappa}}_n$. Given $\tilde{\boldsymbol{\Kappa}}_{(m-1)}$ with initial $\tilde{\boldsymbol{\Kappa}}_{(0)}=\boldsymbol{I}_p$ for $m\ge 1$, we set  
\begin{equation*}
    \tilde{\omega}_{(m)}=\frac{\sum_{i=1}^{k-1} \boldsymbol{\mathsf{y}}_{k-i}^\top  \tilde{\boldsymbol{\Kappa}}_{(m-1)}^{-1} \boldsymbol{\mathsf{y}}_{k-i+1}+\boldsymbol{\mathsf{y}}_{k}^{(l)\top}  \tilde{\boldsymbol{\Kappa}}_{l_{(m-1)}}^{-1} \boldsymbol{\mathsf{y}}_{k+1}^{(l)}}{\sum_{i=1}^{k-1} \boldsymbol{\mathsf{y}}_{k-i}^\top  \tilde{\boldsymbol{\Kappa}}_{(m-1)}^{-1} \boldsymbol{\mathsf{y}}_{k-i}+\boldsymbol{\mathsf{y}}_{k}^{(l)\top}  \tilde{\boldsymbol{\Kappa}}_{l_{(m-1)}}^{-1} \boldsymbol{\mathsf{y}}_{k}^{(l)}}. \label{w_incomplete}
\end{equation*}
Further, given $\tilde{\omega}_{(m)}$ for $m\ge 1$, we first replace $\omega$ by $\tilde{\omega}_{(m)}$ in equations \eqref{def_zomega} and \eqref{def_zomega_part} to get $\boldsymbol{\mathfrak{Z}}_{\cdot}^{(l)}(\tilde{\omega}_{(m)})$ and $\boldsymbol{\mathfrak{Z}}_{\cdot}^{(p-l)}(\tilde{\omega}_{(m)})$. Subsequently, replace $\boldsymbol{\mathfrak{Z}}_{\cdot}^{(l)}(\omega)$ by $\boldsymbol{\mathfrak{Z}}_{\cdot}^{(l)}(\tilde{\omega}_{(m)})$ and $\boldsymbol{\mathfrak{Z}}_{\cdot}^{(p-l)}(\omega)$ by $\boldsymbol{\mathfrak{Z}}_{\cdot}^{(p-l)}(\tilde{\omega}_{(m)})$ on the RHS of \eqref{d_K11}, \eqref{d_B} and \eqref{dK22.1} and set them as $\tilde{\boldsymbol{\Kappa}}_{l_{(m)}}$, $\tilde{\boldsymbol{\Kappa}}_{p-l,l_{(m)}}$ and $\tilde{\boldsymbol{\Kappa}}_{(p-l).l_{(m)}}$, respectively. Now, by using \eqref{s22.1_def} and \eqref{def_kappa_part_partial}, we set 
\begin{align*}
\tilde{\boldsymbol{\Kappa}}_{p-l_{(m)}}&\triangleq\tilde{\boldsymbol{\Kappa}}_{(p-l).l_{(m)}}+\tilde{\boldsymbol{\Kappa}}_{p-l,l_{(m)}} \tilde{\boldsymbol{\Kappa}}_{l_{(m)}}^{-1}\tilde{\boldsymbol{\Kappa}}_{l,p-l_{(m)}}\\
\tilde{\boldsymbol{\Kappa}}_{(m)}&\triangleq\left[ 
    \begin{array}{c:c}
             \tilde{\boldsymbol{\Kappa}}_{l_{(m)}}&\tilde{\boldsymbol{\Kappa}}_{l,p-l_{(m)}}\\
             \hdashline
        \tilde{\boldsymbol{\Kappa}}_{p-l,l_{(m)}}&\tilde{\boldsymbol{\Kappa}}_{p-l_{(m)}}
    \end{array}\right].
\end{align*}
As described in stage~I of Algorithm~\ref{alg1}, we terminate these iterative steps when the first derivative of $\tilde{\ell}_n$ evaluated at $(\tilde{\omega}_{(m)}, \tilde{\boldsymbol{\Kappa}}_{(m)})$ is less than prespecified threshold $\delta$ simultaneously. Since the steps of stage~II of Algorithm~\ref{alg1}  are based on $\tilde{\omega}_n$ and $\tilde{\boldsymbol{\Kappa}}_n$, Algorithm~\ref{alg1} shows how to get $\hat{\omega}$ and $\hat{\kappa}_p$.   

Since the contribution of the partial periodic block $\boldsymbol{\mathsf{Y}}_{k+1}^{(l)}$ in the likelihood from \eqref{likelihood_with_partial} is bounded, the consistency of $\hat{\omega}$ and $\hat{\kappa}_p$ derived in Theorem~\ref{thm5} holds good in the presence of partial periodic block data. If significant partial block information is available for inference, we recommend implementing the modified stage~I of Algorithm~\ref{alg1} instead of ignoring this.

\section{Supplementary Material}
\subsection{Covariance Function Approximation Error}

\label{approx_error}
The exact covariance function, given on the RHS of \eqref{quasi_cov} of the manuscript, depends upon the covariance of the initial building block vector $\boldsymbol{\mathsf{Y}}_1$. Further, the approximate covariance function, given on the RHS of \eqref{quasi_cov_approx} of the manuscript, eliminates the effect of the initial vector $\boldsymbol{\mathsf{Y}}_1$. Let $s\le t$. By using the periodicity of $\kappa_p(\cdot)$, the ratio between the exact (RHS of \eqref{quasi_cov})  and approximate (RHS of \eqref{quasi_cov_approx}) covariance is given by 
\begin{equation}\label{Ratio_cov_exact_approximate}
    R=1 + \omega^{2\lfloor s/p \rfloor} \left[ \frac{Cov(Y_{l( t)}, Y_{l( s)})(1-\omega^2)}{\kappa_p( l(t)- l(s))} - 1 \right],
\end{equation}
where $l( t), l( s)\in \{1,2,\ldots,p\}$(see definition of $l(\cdot)$ in Theorem 1). Since $\omega\in (-1,1)$, the second term on the RHS of \eqref{Ratio_cov_exact_approximate} converges to $0$ as $s\to\infty$. Thus, if $s$ is large, then the ratio $R$ can be arbitrarily close to $1$. In particular if $Cov(Y_{l(t),Y_{l(s)}})=\frac{\kappa_p(l(t)-l(s))}{1-\omega^2}$, then $R=1$. However, if $Cov(Y_{l(t),Y_{l(s)}})\ne\frac{\kappa_p(l(t)-l(s))}{1-\omega^2}$, then, given a specified approximation tolerance $\epsilon>0$,  $|R-1|\le \epsilon $ when
{\begin{equation*}
    s >
p \times \underbrace{\max_{l(t),l(s)\in\{1,\ldots,p\}}{\left\lceil\frac{
\left(
\ln(\epsilon) -
\ln\!\left|
\frac{Cov\!\left(Y_{l(t)},Y_{l(s)}\right)(1-\omega^2)}{\kappa_p\!\left(l(t)-l(s)\right)}
- 1
\right|\right)}{
2\ln|\omega|}
\right\rceil}}_{L(\epsilon)}.
\end{equation*}}
Note that if one burns $L(\epsilon)$ initial periodic blocks, then the ratio between the exact and approximate covariance is close to $1$ with specified approximation tolerance $\epsilon$ uniformly. 

\begin{table}[hbt]
    \centering
    \caption{Relative Frobenius distance between the exact and approximated covariance matrices.}
    \label{table_FD_exact_approximate}
    \begin{tabular}{c r r}
        \toprule
        $n$ & Periodic Initialization & Cold Initialization \\
        \midrule
        $600$   & $3.02 \times 10^{-2}$ & $1.21\times 10^{-1}$ \\
        $3000$  & $1.34\times 10^{-2}$ & $5.38 \times 10^{-2}$ \\
        $10000$ & $7.35\times 10^{-3}$ & $2.94\times 10^{-2}$ \\
        \bottomrule
    \end{tabular}
\end{table}

As discussed above, the exact covariance depends upon the covariance of the initial building block $\boldsymbol{\mathsf{Y}}_1$ used for initialisation. We consider the following two covariance matrices of $\boldsymbol{\mathsf{Y}}_1$ for numerical quantification of approximation error.
\begin{itemize}
    \item Periodic init. ($Var(\boldsymbol{\mathsf{Y}}_1)= \boldsymbol{\Kappa}$); $L(\epsilon)= \left\lceil\frac{\ln(\epsilon)}{2\ln(|\omega|)}-1\right\rceil$ 
    \item Cold init. ($Var(\boldsymbol{\mathsf{Y}}_1) = \boldsymbol{0}$); $L(\epsilon)= \left\lceil\frac{\ln(\epsilon)}{2\ln(|\omega|)}\right\rceil$.
\end{itemize}
Table~\ref{table_FD_exact_approximate} shows the relative Frobenius distance between the exact and approximate covariance matrix under the simulation setup chosen in Section~V of the manuscript, i.e., the process $\{Y_t\}_{t=1}^n$ is a QPGP with $p=10, \ \omega=0.5$ and $\kappa_p$ as MacKay's kernel with $\theta=1$ and $\sigma^2=1$. This indicates that the approximation error is very small even with a smaller sample size. The relative Frobenius distance corresponding to periodic initialization is smaller than that of cold initialization.  
Different QPGP initializations may lead to different degrees of approximation error, so this numerical illustration is specific to the chosen covariance structure of the initial periodic block. 

\subsection{Hyperparameter Estimation} \label{hyperparameter}{An additional optimization step in \eqref{parameteric_kappa} is used if necessary to enforce a parametric structure of the covariance kernel. 
 When the within-period correlation structure is known, incorporating this information through parametric optimization improves estimation accuracy for both the parameters $\omega$ and $\boldsymbol{\Kappa}$.  We conducted the estimation and prediction comparison in 1000 simulated samples with $p=10$, $\omega=0.5$, and within-period covariance as MacKay's kernel with $\sigma^2=1$ and $\theta=1$. Table \ref{comment3_rmse} shows that, for the parameter $\omega$, the MSE of estimator \eqref{parameteric_kappa} is smaller  than \eqref{tilde_k_p}. The average (over 1000 simulation runs) Frobenius distance between $\boldsymbol{\Kappa}$ and the estimate of $\boldsymbol{\Kappa}$ obtained using \eqref{parameteric_kappa} is also smaller than \eqref{tilde_k_p}.   
 The Frobenius distance is also viewed as element-wise integrated MSE of the elements of $\boldsymbol{\Kappa}$.
 This structural refinement leads to a more robust model, resulting in lower prediction MSE (Table~\ref{comment3_rmse}). }

As discussed in subsection \ref{s4.1} of the main manuscript, the hyperparameters can also be estimated in Stage~I using grid search (SIGS). Table~\ref{comment3_sigs} shows that SIGS estimates have a smaller MSE (especially for $\boldsymbol{\Kappa}$) at a significantly higher computational cost.

\subsection{Proof of Theoretical results}
\label{thm_proofs}
{\it Proof of Theorem \ref{thm:cov_QPGP}:} For $t\in\mathbb{Z}^+$, define $i(t)\triangleq \lfloor t/p\rfloor$. Note that $i(t)\in \mathbb{Z}^+\cup \{0\} $. Since $t = i(t)p+l(t)$, $Y_t$ belongs to the $(i(t)+1)^{\text{th}}$ periodic block (i.e., $\boldsymbol{\mathsf{Y}}_{i(t)+1}$) at its $l(t)^{\text{th}}$ coordinate. Similarly, $s= i(s)p+l(s)$and $Y_s\in \boldsymbol{\mathsf{Y}}_{i(s)+1}$.  

For $s\le t$,  $i(t)$ and $i(s)$ must satisfy either of the following two scenarios: (a)  $i(s)< i(t)$, (b) $i(s)=i(t)$ with $l(s) \leq l(t)$. 
By using the recursive equation \eqref{eq:recursion} repeatedly, we have
\begin{align}
   \hskip-10pt Y_t= \begin{cases}
    \sum_{m=0}^{i(t)-i(s)-1}\omega^m \mathsf{Z}_{i(t)+1
    -m,l(t)}+\omega^{i(t)-i(s)} Y_{i(s)p+l(t)}&\mbox{if } i(t)>i(s)\\ 
     Y_{i(s)p+l(t)} & \mbox{if } i(t)=i(s) 
    \end{cases},\label{sameblock}
\end{align}  
where $\boldsymbol{\mathsf{Z}}_i\triangleq [\mathsf{Z}_{i,1},\ldots, \mathsf{Z}_{i,p}]^\top$.
By using the independence of periodic block $\boldsymbol{\mathsf{Y}}_{i(s)+1}$ and periodic building blocks  ($\boldsymbol{\mathsf{Z}}_{i(s)+2},\boldsymbol{\mathsf{Z}}_{i(s)+3},\dots, \boldsymbol{\mathsf{Z}}_{i(t)+1}$) and \eqref{sameblock}, we have
\begin{align}
E(Y_tY_s )
\!=\!
\begin{cases}
\omega^{i(t)-i(s)}E(Y_{i(s)p+l(t)}Y_{i(s)p+l(s)})& \hskip-8pt\mbox{if } i(t)>i(s)
\\  E(Y_{i(s)p+l(t)}Y_{i(s)p+l(s)})&  \hskip-8pt\mbox{if } i(t)=i(s)
  \end{cases}.   \label{1.1}
\end{align}
Again by using \eqref{eq:recursion} repeatedly, for $l\in \{1,2,\ldots,p\}$, we have
\begin{align} Y_{i(s)p+l}=\begin{cases}
    \sum_{m=0}^{i(s)-1}\omega^m\mathsf{Z}_{i(s)+1-m,l} +\omega^{i(s)}Y_l & \mbox{if } i(s)\ge 1\\
    Y_l & \mbox{if } i(s)=0
\end{cases}.\label{firstblock}
\end{align}
Now by using the independence of initial periodic block $\boldsymbol{\mathsf{Y}}_1$ and building-blocks $\{\boldsymbol{\mathsf{Z}}_{i(s)}\}_{i(s)>1}$ and \eqref{firstblock}, we have
\begin{align}
E(Y_{i(s)p+l(t)}Y_{j(s)p+l(s)})&=E\left(\sum_{m=0}^{i(s)-1}\omega^m\mathsf{Z}_{i(s)-m,l(t)} \cdot\sum_{m=0}^{i(s)-1}\omega^m \mathsf{Z}_{i(s)-m,l(s)}\right)\nonumber\\ &\ \ \ \ \ +\omega^{2i(s)} E\left(Y_{l(t)}Y_{l(s)}\right)\nonumber\\
&\hskip-10pt=\left[\frac{1-\omega^{2i(s)}}{1-\omega^2}\right]\kappa_p(l(t)-l(s))+ \omega^{2i(s)} E\left(Y_{l(t)}Y_{l(s)}\right).\label{Subsequent_37}
\end{align}
The proof is completed by plugging in the expression on the RHS of  \eqref{Subsequent_37} in that of \eqref{1.1} and using the periodicity of the covariance kernel $\kappa_p$.~\hfill{$\blacksquare$}

{\it Proof of Proposition \ref{prop:quasi_cov_simple}:}  The proof follows by plugging in $E\left(Y_{l(t)}, Y_{l(s)}\right)=\frac{\kappa_p(l(t)-l(s))}{1-\omega^2}$ on the RHS of \eqref{Subsequent_37}.~\hfill{$\blacksquare$}

{\it Proof of Theorem \ref{thm5}} ({\it part (1)}).
Since $\boldsymbol{Y}_n=[Y_1,Y_2,\ldots,Y_n]$ is a QPGP vector with period $p$ and parameters $\omega_0$ and $\kappa_{0p}$, by using \eqref{eq:recursion} and the following identity
\begin{align*}
    \boldsymbol{\mathsf{Y}}_{i+1}-\omega \boldsymbol{\mathsf{Y}}_{i}
=\boldsymbol{\mathsf{Z}}_{i+1}-(\omega-\omega_0)\boldsymbol{\mathsf{Y}}_i,
\end{align*}
for all $i=1,2,\ldots, k-1$, the reduced likelihood function as defined in \eqref{reduced_likelihood} can be expressed as follows.
\begin{align}
     \tilde{\ell}_n(\omega, \boldsymbol{\Kappa})=&\log(|\boldsymbol{\Kappa}|)
    +\frac1{k-1}\sum_{i=1}^{k-1}\boldsymbol{\mathsf{Z}}_{k-i+1}^{\top}\boldsymbol{\Kappa}^{-1}\boldsymbol{\mathsf{Z}}_{k-i+1}\nonumber\\
    &-2(\omega-\omega_0)\frac1{k-1}\sum_{i=1}^{k-1}\boldsymbol{\mathsf{Z}}_{k-i+1}^{\top}\boldsymbol{\Kappa}^{-1}\boldsymbol{\mathsf{Y}}_{k-i}\nonumber\\
    &+(\omega-\omega_0)^2\frac1{k-1}\sum_{i=1}^{k-1}\boldsymbol{\mathsf{Y}}_{k-i}^{\top}\boldsymbol{\Kappa}^{-1}\boldsymbol{\mathsf{Y}}_{k-i}.\label{decomposed_reduced_likelihood}
\end{align}
Note that the first term on the RHS of \eqref{decomposed_reduced_likelihood} does not depend on $n$ whereas the other terms depend on $n$ via $k=n/p$. The limit of $\tilde{\ell}_n$ over $n\to \infty$ is equivalent to the convergence of the RHS of \eqref{decomposed_reduced_likelihood} as $k\to\infty$ since $p$ is fixed. 
We establish the probability converges of the second, third and fourth term on the RHS of \eqref{decomposed_reduced_likelihood} by showing the convergence of mean and variances of these terms continuously. 

We begin with the second term on the RHS of \eqref{decomposed_reduced_likelihood}: 
\begin{align}
    &\hskip-10pt E\left[\frac{1}{k-1} \sum_{i=1}^{k-1}\boldsymbol{\mathsf{Z}}_{k-i+1}^\top \boldsymbol{\Kappa}^{-1}\boldsymbol{\mathsf{Z}}_{k-i+1}\right]\nonumber\\ 
    \hskip-10pt=&\frac{1}{k-1} \sum_{i=1}^{k-1}E[tr(  \boldsymbol{\Kappa}^{-1}\boldsymbol{\mathsf{Z}}_{k-i+1}\boldsymbol{\mathsf{Z}}_{k-i+1}^\top)]=tr(\boldsymbol{\Kappa}^{-1}\boldsymbol{\Kappa}_0).\label{second_term_mean} 
\end{align} 
Using the independence of building blocks $\{\boldsymbol{\mathsf{Z}}_i,\ i\ge 1\}$ and variance of random quadratic form (see equation~(50) on pp.~77 for the formula of variance of random quadratic forms in \cite{searle}):
\begin{align}
    &\hskip-10pt \text{Var}\left[\frac{1}{k-1} \sum_{i=1}^{k-1}\boldsymbol{\mathsf{Z}}_{k-i+1}^\top \boldsymbol{\Kappa}^{-1}\boldsymbol{\mathsf{Z}}_{k-i+1}\right]\nonumber\\ =&\frac{1}{(k-1)^2} \sum_{i=1}^{k-1}\text{Var}( \boldsymbol{\mathsf{Z}}_{k-i+1}^\top \boldsymbol{\Kappa}^{-1}\boldsymbol{\mathsf{Z}}_{k-i+1})=\frac{2tr((\boldsymbol{\Kappa}^{-1}\boldsymbol{\Kappa}_0)^2)}{k-1}. \label{second_term_var}
\end{align}
By using \eqref{second_term_mean}, the expectation of the second term on the RHS of \eqref{decomposed_reduced_likelihood} converges to $tr(\boldsymbol{\Kappa}^{-1}\boldsymbol{\Kappa}_0)$ continuously over $\boldsymbol{\Kappa}\in\mathfrak{K}$. Similarly, by using \eqref{second_term_var}, the variance of the second term on the RHS of \eqref{decomposed_reduced_likelihood} converges to $0$ continuously over $\boldsymbol{\Kappa}\in\mathfrak{K}$ as $k\to\infty$.
Therefore, the second term on the RHS of \eqref{decomposed_reduced_likelihood} converges continuously in probability, i.e., as $ k \to \infty$,  
\begin{equation}
\frac{1}{k-1} \sum_{i=1}^{k-1}\boldsymbol{\mathsf{Z}}_{k-i+1}^\top \boldsymbol{\Kappa}^{-1}\boldsymbol{\mathsf{Z}}_{k-i+1} \xrightarrow{P} tr(\boldsymbol{\Kappa}^{-1}\boldsymbol{\Kappa}_0) \label{R1}  
\end{equation}  
Using a similar argument, Lemmas~\ref{lemma1} and \ref{lemma2} show that the 3rd and 4th terms on the RHS of \eqref{decomposed_reduced_likelihood} converge continuously over $\omega\in[-1,1]$ and $\boldsymbol{\Kappa}\in\mathfrak{K}$. By using Lemmas~\ref{lemma1} and \ref{lemma2} as $k\to\infty$:
\begin{align}
   &\frac{(\omega-\omega_0)}{k-1} \sum_{i=1}^{k-1}\boldsymbol{\mathsf{Z}}_{k-i+1}^\top\boldsymbol{\Kappa}^{-1}\boldsymbol{\mathsf{Y}}_{k-i}\xrightarrow{P}0  \label{R2}\\
   \hskip-20pt &\frac{(\omega-\omega_0)^2}{k-1} \sum_{i=1}^{k-1}\boldsymbol{\mathsf{Y}}_{k-i}^\top\boldsymbol{\Kappa}^{-1}\boldsymbol{\mathsf{Y}}_{k-i}\xrightarrow{P}
    \frac{(\omega-\omega_0)^2}{1-\omega_0^2} tr(\boldsymbol{\Kappa}^{-1}\boldsymbol{\Kappa}_0). \ \ \quad\label{R3}
\end{align}
Now by using \eqref{decomposed_reduced_likelihood}, \eqref{R1}, \eqref{R2} and \eqref{R3}, $\tilde{\ell}_n \xrightarrow{P} \tilde{\ell}$ as $n\to\infty$ continuously over $(\omega,\boldsymbol{\Kappa})$. This completes the proof of part (1).

\textit{part (2):} By using \eqref{limit_tilde_l}, the first order derivative of the limiting function $\tilde{\ell}$ with respect to $\omega$  and $\boldsymbol{\Kappa}$ is given as follows:
\begin{align}
   & \frac{\partial \tilde{\ell}}{\partial \omega}=\frac{2(\omega-\omega_0)}{1-\omega_0^2}tr(\boldsymbol{\Kappa}^{-1}\boldsymbol{\Kappa}_0)\label{tilde_l_deriv_omega}\\
   & \frac{\partial \tilde{\ell}}{\partial \boldsymbol{\Kappa}}=\boldsymbol{\Kappa}^{-1}-\left(1+\frac{(\omega-\omega_0)^2}{1-\omega^2}\right)\boldsymbol{\Kappa}^{-1}\boldsymbol{\Kappa}_0 \boldsymbol{\Kappa}^{-1}\label{tilde_l_deriv_Kappa}.
\end{align}
Note that, by using \eqref{tilde_l_deriv_omega} and \eqref{tilde_l_deriv_Kappa}, $\omega_0$ and $\boldsymbol{\Kappa}_0$ is a stationary point of $\tilde{\ell}$. Now, the second derivative of $\tilde{\ell}$ with respect to $\omega$ and $\boldsymbol{\Kappa}$ is given as follows.
\begin{align*}
    &\frac{\partial^2 \tilde{\ell} }{\partial \omega^2}=\frac{tr(\boldsymbol{\Kappa}^{-1}\boldsymbol{\Kappa}_0)}{1-\omega_0^2}\\
    &\frac{\partial^2 \tilde{\ell}}{\partial \boldsymbol{\Kappa}^2}=-\boldsymbol{\Kappa}^{-1}\otimes\boldsymbol{\Kappa}^{-1}+\left(1+\frac{(\omega-\omega_0)^2}{1-\omega^2}\right) \\&\hskip30pt \times (\boldsymbol{\Kappa}^{-1}\otimes (\boldsymbol{\Kappa}^{-1}\boldsymbol{\Kappa}_0 \boldsymbol{\Kappa}^{-1})+(\boldsymbol{\Kappa}^{-1}\boldsymbol{\Kappa}_0\boldsymbol{\Kappa}^{-1})\otimes \boldsymbol{\Kappa}^{-1}),
\end{align*}where $\otimes$ denotes the Kronecker product between matrices. We refer to \cite{matrix_cookbook} for differentiation formulae with respect to matrices (in particular, see equations (57), (60) and (61) on pp. 9--10 of \cite{matrix_cookbook}). Further, note that the Hessian matrix of $\tilde{\ell}$ at the stationary point $(\omega_0,\boldsymbol{\Kappa}_0)$ simplifies to \begin{align*}
H &=\begin{bmatrix}
    \frac{p}{1-\omega_0^2}& 0\\
    0&\boldsymbol{\Kappa}_0^{-1}\otimes \boldsymbol{\Kappa}_0^{-1}
\end{bmatrix},
\end{align*}
which is positive-definite. {
For $\omega\in[-1,1]$ and $\boldsymbol{\Kappa}\in\mathfrak{K}$, define
\[
\Delta(\omega, \boldsymbol{\Kappa})
:= \tilde{\ell}(\omega, \boldsymbol{\Kappa})
- \tilde{\ell}(\omega_0, \boldsymbol{\Kappa}_0),
\]
which simplifies to  
\begin{align*}
\Delta(\omega, \boldsymbol{\Kappa})
=&  \underbrace{- \log |\boldsymbol{\Kappa^{-1}\Kappa}_0|
+ \operatorname{tr}(\boldsymbol{\Kappa}^{-1}\boldsymbol{\Kappa}_0) - p}_{T_1}
+ \underbrace{\operatorname{tr}(\boldsymbol{\Kappa}^{-1}\boldsymbol{\Kappa}_0)
\frac{(\omega - \omega_0)^2}{1 - \omega_0^2}}_{T_2}.    
\end{align*}

Since $\boldsymbol{\Kappa}$ and $ \boldsymbol{\Kappa}_0$ are positive-definite, the eigenvalues (say $\lambda_1,\dots,\lambda_p$) of $\boldsymbol{\Kappa}^{-1}\boldsymbol{\Kappa}_0$ are positive. Thus, $T_1=\sum_{i=1}^p \bigl(\lambda_i - \log \lambda_i - 1\bigr)\ge 0$ and $T_1=0$ if and only if $\lambda_1=\lambda_2=\cdots=\lambda_p=1$ which happens if only if $\boldsymbol{\Kappa}=\boldsymbol{\Kappa}_0$. Further, note that $T_2=(\sum_{i=1}^p \lambda_i)(\omega-\omega_0)^2/(1-\omega_0^2)\ge 0$ and it attains zero value if and only if $\omega=\omega_0$.
Thus,
\[
\Delta(\omega, \boldsymbol{\Kappa})=T_1+T_2 \ge 0,
\]
which attains the value $0$ if and only if $(\omega, \boldsymbol{\Kappa}) = (\omega_0, \boldsymbol{\Kappa}_0)$. This completes the proof of part (2).} 

\textcolor{black}{\textit{part (3): } Since the set $[-1,1]\times \mathfrak{K}$ is a closed bounded subset of $\mathbb{R}^{p^2+1}$, the sequence of continuous functions $\tilde{\ell}_n$  are uniformly continuous. Let a non-random sequence $(\omega_n,\boldsymbol{\Kappa}_n)\to (\omega, \boldsymbol{\Kappa})$ as $n\to\infty$. Then, we have
\begin{align}
|\tilde{\ell}_n(\omega, \boldsymbol{\Kappa})-\tilde{\ell}(\omega, \boldsymbol{\Kappa})|\le &
|\tilde{\ell}_n(\omega, \boldsymbol{\Kappa})-\tilde{\ell}_n(\omega_n, \boldsymbol{\Kappa}_n)|+|\tilde{\ell}_n(\omega_n, \boldsymbol{\Kappa_n})-\tilde{\ell}(\omega, \boldsymbol{\Kappa})|.\label{sup_breakup}
\end{align}
By using the uniform continuity of $\tilde{\ell}_n$, we have
\begin{align}
    |\tilde{\ell}_n(\omega, \boldsymbol{\Kappa})-\tilde{\ell}_n(\omega_n, \boldsymbol{\Kappa}_n)|\le \sup_{(\omega, \boldsymbol{\Kappa})\in [-1,1]\times \mathfrak{K}} \tilde{\ell}_n(\omega, \boldsymbol{\Kappa}) \times d((\omega_n, \boldsymbol{\Kappa}_n), (\omega, \boldsymbol{\Kappa})),
\end{align}
where $d(\cdot,\cdot)$ is the Euclidean distance between $(p^2+1)$ dimensional real vectors. By using part~(1), and continuity of $\tilde{\ell}$, and compactness of set $[-1,1]\times \mathfrak{K}$, $\tilde{\ell}_n(\omega,\boldsymbol{\Kappa})$ is uniformly bounded in probability. By using the convergence of $(\omega_n, \boldsymbol{\Kappa}_n)$ and uniform boundedness of $\tilde{\ell}_n$, 
the first term on the RHS of \eqref{sup_breakup} converges to $0$ with probability $1$ uniformly over $(\omega, \boldsymbol{\Kappa})\in [-1,1]\times \mathfrak{K}$. By using part (1) of Theorem~\ref{thm5}, the second term on the RHS of \eqref{sup_breakup} converges to $0$ in probability as $n\to\infty$. Thus, we have
\begin{align}\label{sup_conv}
   \hskip-10pt \sup_{(\omega, \boldsymbol{\Kappa})\in [-1,1]\times \mathfrak{K}}|\tilde{\ell}_n(\omega, \boldsymbol{\Kappa})-\tilde{\ell}(\omega, \boldsymbol{\Kappa})| \xrightarrow{P}0 \mbox{ as } n\to\infty.
\end{align}
Since the stage~I estimator $(\tilde{\omega}_n, \tilde{\boldsymbol{\Kappa}}_n)$ is the minimizer of $\tilde{\ell}_n$, by using part(1) of Theorem~\ref{thm5}, we have 
\begin{equation}
    \tilde{\ell}_n(\tilde{\omega}_n, \tilde{\boldsymbol{\Kappa}}_n)\le \tilde{\ell}_n(\omega_0,\boldsymbol{\Kappa}_0) = \tilde{\ell}(\omega_0, \boldsymbol{\Kappa}_0)+o_P(1) \label{step3}
\end{equation}
Since $(\omega_0, \boldsymbol{\Kappa}_0)$ is the unique minima of $\tilde{\ell}$, by using \eqref{step3},
\begin{align}
    \hskip-20pt0&\le \tilde{\ell}(\tilde{\omega}_n, \tilde{\boldsymbol{\Kappa}}_n)-\tilde{\ell}(\omega_0, \boldsymbol{\Kappa}_0)\nonumber\\&\le \tilde{\ell}(\tilde{\omega}_n, \tilde{\boldsymbol{\Kappa}}_n)-\tilde{\ell}_n(\tilde{\omega}_n, \tilde{\boldsymbol{\Kappa}}_n) +o_P(1)\nonumber\\
    &\le \sup_{(\omega, \boldsymbol{\Kappa})\in [-1,1]\times \mathfrak{K}}|\tilde{\ell}(\omega, \boldsymbol{\Kappa})-\tilde{\ell}_n(\omega, \boldsymbol{\Kappa})| +o_P(1).\label{Gap_between_tilde_l_estimate_minima}
\end{align}
Now by using \eqref{sup_conv} and \eqref{Gap_between_tilde_l_estimate_minima}, we have
\begin{eqnarray}
    \hskip-20pt0&\le& \tilde{\ell}(\tilde{\omega}_n, \tilde{\boldsymbol{\Kappa}}_n)-\tilde{\ell}(\omega_0, \boldsymbol{\Kappa}_0) \xrightarrow[]{P}0 \mbox{ as } n\to\infty.\label{conv_tilde_l_minimizer}
\end{eqnarray}
Now, suppose the stage~I estimator $(\tilde{\omega}_n,\tilde{\boldsymbol{\Kappa}}_n)\not\xrightarrow[]{P} (\omega_0, \boldsymbol{\Kappa}_0)$, then there exists $\epsilon>0$ and $\delta>0$ such that
\begin{equation}
    P(d((\tilde{\omega}_n, \tilde{\boldsymbol{\Kappa}}_n),(\omega_0, \boldsymbol{\Kappa}_0))\ge \varepsilon) > \delta \ \forall \ n.\label{not_prob_conv}
\end{equation}
Since $(\omega_0, \boldsymbol{\Kappa}_0)$ is the unique minima of $\tilde{\ell}$, $\forall \epsilon>0$, there exists $\eta>0$ such that
\begin{eqnarray}
    \inf_{\{(\omega,\boldsymbol{\Kappa}): d((\omega,\boldsymbol{\Kappa}),(\omega_0,\boldsymbol{\Kappa}_0))\ge\epsilon\}} \tilde{\ell}(\omega,\boldsymbol{\Kappa})> \tilde{\ell}(\omega_0,\boldsymbol{\Kappa}_0)+\eta.\label{unique_minima_def}
\end{eqnarray}
Now, \eqref{unique_minima_def} and \eqref{not_prob_conv} contradicts \eqref{conv_tilde_l_minimizer}. Thus, we have 
\begin{equation}
(\tilde{\omega}_n,\tilde{\boldsymbol{\Kappa}}_n)\xrightarrow[]{P} (\omega_0, \boldsymbol{\Kappa}_0) \mbox{ as } n\to\infty. \label{tilde_est_con}   
\end{equation}}

We now turn to establish the convergence of stage II estimator $(\hat{\omega},\hat{\boldsymbol{\Kappa}})$ of Algorithm~\ref{alg1}. Note that, by using \eqref{tilde_k_p}, \eqref{tilde_est_con} and \eqref{def_bloc_cov}, for fixed $p$ and $|t|<p$, we have, as $n \to \infty$
\begin{align}
    \tilde{\kappa}_p(t)\xrightarrow[]{P}\frac{1}{p-|t|}\sum_{j=1}^{p-|t|}\boldsymbol{\Kappa}_0(j,j+|t|)=\kappa_{0p}(t) .\label{tilde_k_p_conv}
\end{align}
By using \eqref{tilde_k_p_conv} and given fixed $p$, the spectrum of $\tilde{\kappa}_p$ defined as $\tilde{f}(\lambda)$ (for $\lambda\in[-\pi,\pi]$) in \eqref{tilde_f_def} converges as follows.
\begin{eqnarray}
    \tilde{f}(\lambda)\xrightarrow[]{P} f_0(\lambda)\triangleq \frac1{2\pi} \sum_{|t|<p}\kappa_{0p}(t )e^{-i t \lambda } \mbox{ as } n\to\infty.\label{tilde_f_conv}
\end{eqnarray}
Since $f_0(\lambda)\ge 0$ for all $\lambda\in[-\pi,\pi]$, by using \eqref{tilde_f_conv}, we have
\begin{eqnarray}
    \max(\tilde{f}(\lambda),0)\xrightarrow[]{P} f_0(\lambda)\mbox{ as } n\to\infty.\label{max_tilde_f_0_conv}
\end{eqnarray}
Now, using \eqref{hat_k_p}, \eqref{tilde_f_def}, \eqref{tilde_f_conv}, and \eqref{max_tilde_f_0_conv}, for $|t|<p$, we have
\begin{eqnarray}
\hat{\kappa}_p(t)\xrightarrow[]{P}  \int_{-\pi}^\pi e^{it\lambda} f_0(\lambda) \ d\lambda =\kappa_{0p}(t) \mbox{ as } n\to\infty. \label{hat_k_p_conv}
\end{eqnarray}
We now establish the convergence of $\hat{\omega}$ proposed in \eqref{hat_w_def} to complete the proof. By using \eqref{eq:recursion} and \eqref{hat_w_def}, we have
\begin{align}
        \hat{\omega} &= \omega_0+\dfrac{ tr(\hat{\boldsymbol{\Kappa}}^{-1}\frac1{k-1}\sum_{i=1}^{k-1} \boldsymbol{\mathsf{Z}}_{i+1}  \boldsymbol{\mathsf{Y}}_{i}^\top)}{tr(\hat{\boldsymbol{\Kappa}}^{-1}\frac1{k-1}\sum_{i=1}^{k-1} \boldsymbol{\mathsf{Y}}_{i}  \boldsymbol{\mathsf{Y}}_{i}^\top)}.\label{hat_w_decompose}
\end{align}
A similar argument to the proof of Lemma~\ref{lemma1} and \ref{lemma2} gives
\begin{align}
    \frac1{k-1}\sum_{i=1}^{k-1} \boldsymbol{\mathsf{Z}}_{i+1}  \boldsymbol{\mathsf{Y}}_{i}^\top &\xrightarrow[]{P}\boldsymbol{0} \mbox{ as } n\to\infty,\label{num1_conv}\\
    \frac1{k-1}\sum_{i=1}^{k-1} \boldsymbol{\mathsf{Y}}_{i}  \boldsymbol{\mathsf{Y}}_{i}^\top&\xrightarrow[]{P}\frac{1}{1-\omega_0^2}\boldsymbol{\Kappa}_0 \mbox{ as } n\to\infty.\label{den1_conv}
\end{align}
By using \eqref{hat_k_p_conv}, \eqref{hat_w_decompose}, \eqref{num1_conv} and \eqref{den1_conv}, we have
\begin{equation}
    \hat{\omega}\xrightarrow[]{P}\omega_0 \mbox{ as } n\to\infty.
\end{equation}
This completes the proof.~\hfill{$\blacksquare$}

\begin{lemma} Let $\boldsymbol{Y}_n=[Y_1,\dots, Y_n]^\top$ be a QPGP vector with period $p$ and parameters $\omega_0$ and $\kappa_{p0}$. Suppose $n=kp$ and $\boldsymbol{\Kappa}\in \mathfrak{K}$. Then, as $k\to \infty$, we have 
\begin{equation}\label{cross_quad}
\frac{(\omega-\omega_0)}{k-1}\sum_{i=1}^{k-1}\boldsymbol{\mathsf{Y}}_{i}^\top\boldsymbol{\Kappa}^{-1}\boldsymbol{\mathsf{Z}}_{i+1} \xrightarrow{P}0,    
\end{equation}
continuously over $\omega\in[-1,1]$ and $\boldsymbol{\Kappa}\in\mathfrak{K}$.
 \label{lemma1}
\end{lemma}
\textit{Proof of Lemma~\ref{lemma1}:}  Note that
\begin{align}
     E\left(\frac{(\omega-\omega_0)}{k-1}\sum_{i=1}^{k-1}\boldsymbol{\mathsf{Y}}_{i}^\top\boldsymbol{\Kappa}^{-1}\boldsymbol{\mathsf{Z}}_{i+1}\right)=\frac{(\omega-\omega_0)}{k-1}\sum_{i=1}^{k-1} E[ E(\boldsymbol{\mathsf{Y}}_{i}^\top\boldsymbol{\Kappa}^{-1}\boldsymbol{\mathsf{Z}}_{i+1}|\boldsymbol{\mathsf{Y}}_{i})]=0. \label{lemma_exp}
\end{align}
Further,
\begin{align}
\text{Var}\left(\frac{1}{k-1}\sum_{i=1}^{k-1}\boldsymbol{\mathsf{Y}}_{i}^\top\boldsymbol{\Kappa}^{-1}\boldsymbol{\mathsf{Z}}_{i+1}\right) 
    &=\frac{1}{(k-1)^2}\sum_{i=1}^{k-1}\text{Var}\left(\boldsymbol{\mathsf{Y}}_{i}^\top\boldsymbol{\Kappa}^{-1}\boldsymbol{\mathsf{Z}}_{i+1}\right)\nonumber\\
    &\hskip-30pt+\frac{2}{(k-1)^2}\sum_{i=2}^{k-1}\sum_{j=1}^{i-1} \text{Cov}(\boldsymbol{\mathsf{Y}}_i^\top \boldsymbol{\Kappa}^{-1}\boldsymbol{\mathsf{Z}}_{i+1}, \boldsymbol{\mathsf{Y}}_j^\top \boldsymbol{\Kappa}^{-1}\boldsymbol{\mathsf{Z}}_{j+1}) \label{lemma1_var}
\end{align}
By using \eqref{lemma_exp} and Definition~\ref{def:QPGP}, the summands of the first term on the RHS of \eqref{lemma1_var} simplify for $i=1,2,\ldots,k-1$:
 \begin{align}
     \text{Var}[\boldsymbol{\mathsf{Y}}_{i}^\top\boldsymbol{\Kappa}^{-1}\boldsymbol{\mathsf{Z}}_{i+1}]&=E(\boldsymbol{\mathsf{Y}}_{i}^\top\boldsymbol{\Kappa}^{-1}\boldsymbol{\mathsf{Z}}_{i+1})^2 \nonumber \\
    &=E[E(tr(\boldsymbol{\mathsf{Z}}_{i+1}\boldsymbol{\mathsf{Z}}_{i+1}^\top  \boldsymbol{\Kappa}^{-1} \boldsymbol{\mathsf{Y}}_{i} \boldsymbol{\mathsf{Y}}_{i}^\top\boldsymbol{\Kappa}^{-1})|\boldsymbol{\mathsf{Y}}_{i})] \nonumber\\
    &=E[tr(\boldsymbol{\Kappa_0}  \boldsymbol{\Kappa}^{-1} \boldsymbol{\mathsf{Y}}_{i} \boldsymbol{\mathsf{Y}}_{i}^\top\boldsymbol{\Kappa}^{-1}))]=\frac{1}{1-\omega_0^2}tr((\boldsymbol{\Kappa}_0\boldsymbol{\Kappa}^{-1})^2).\label{lemma1_var_term2}
    \end{align}

Similarly by using \eqref{eq:recursion} and \eqref{lemma_exp}, the summands of the second term on the RHS of \eqref{lemma1_var}  is simplified for $j< i$:
 \begin{align}
     &\text{Cov}(\boldsymbol{\mathsf{Y}}_i^\top \boldsymbol{\Kappa}^{-1}\boldsymbol{\mathsf{Z}}_{i+1}, \boldsymbol{\mathsf{Y}}_j^\top \boldsymbol{\Kappa}^{-1}\boldsymbol{\mathsf{Z}}_{j+1})\nonumber\\
     &=E \left(\left(\omega_0^{i-j}\boldsymbol{\mathsf{Y}}_j+\sum_{m=0}^{i-j-1}\omega_0^m \boldsymbol{\mathsf{Z}}_{i-m}\right)^\top\boldsymbol{\Kappa}^{-1}\boldsymbol{\mathsf{Z}}_{i+1} \boldsymbol{\mathsf{Y}}_j^\top \boldsymbol{\Kappa}^{-1}\boldsymbol{\mathsf{Z}}_{j+1}\right)\nonumber\\
     &=\omega_0^{i-j} E (\boldsymbol{\mathsf{Y}}_j^\top \boldsymbol{\Kappa}^{-1}\boldsymbol{\mathsf{Z}}_{i+1}\boldsymbol{\mathsf{Y}}_j^\top \boldsymbol{\Kappa}^{-1}\boldsymbol{\mathsf{Z}}_{j+1})+\sum_{m=0}^{i-j-1} \omega_0^m E(\boldsymbol{\mathsf{Z}}_{i-m}^\top\boldsymbol{\Kappa}^{-1}\boldsymbol{\mathsf{Z}}_{i+1} \boldsymbol{\mathsf{Y}}_j^\top \boldsymbol{\Kappa}^{-1}\boldsymbol{\mathsf{Z}}_{j+1}) \label{lemma1_cov}
 \end{align}
Since $j<i$ for each summand in the second term on the RHS of \eqref{lemma1_var}, by using the  independence between  $\boldsymbol{\mathsf{Z}}_{i+1}$ and $ (\boldsymbol{\mathsf{Y}}_j,  \boldsymbol{\mathsf{Z}}_{j+1}, \ldots, \boldsymbol{\mathsf{Z}}_{i})$ and conditional expectation arguments both the terms on the RHS of \eqref{lemma1_cov} vanishes. Therefore, by using \eqref{lemma1_var} and \eqref{lemma1_var_term2}, we have
\begin{align}
&\text{Var} \left(\frac{(\omega-\omega_0)}{k-1} \sum_{i=1}^{k-1}\boldsymbol{\mathsf{Y}}_{i}^\top\boldsymbol{\Kappa}^{-1}\boldsymbol{\mathsf{Z}}_{i+1}\right)=\frac{(\omega-\omega_0)^2tr(\boldsymbol{\boldsymbol{\Kappa}_0\Kappa}^{-1})^2}{(k-1)(1-\omega_0^2)}. \label{var_thirdterm_decomp} 
\end{align}
Since the RHS of \eqref{var_thirdterm_decomp} converges to $0$ continuously over $\omega\in[-1,1]$ and $\boldsymbol{\Kappa}\in\mathfrak{K}$ as $k\to\infty$, this completes the proof.~\hfill{$\blacksquare$}

\begin{lemma}
Let $\boldsymbol{Y}_n=[Y_1,\dots, Y_n]^\top$ be a QPGP vector with period $p$ and parameters $\omega_0$ and $\kappa_{p0}$. Suppose $n=kp$ and $\boldsymbol{\Kappa}\in\mathfrak{K}$. Then, as $k\to\infty$, we have
\begin{equation}
\frac{(\omega-\omega_0)^2}{k-1}\sum_{i=1}^{k-1}\boldsymbol{\mathsf{Y}}_{i}^\top\boldsymbol{\Kappa}^{-1}\boldsymbol{\mathsf{Y}}_{i} \xrightarrow[]{P}\frac{(\omega-\omega_0)^2tr(\boldsymbol{\Kappa}^{-1}\boldsymbol{\Kappa}_0)}{1-\omega_0^2}, \end{equation}
continuously over $\omega\in[-1,1]$ and $\boldsymbol{\Kappa}\in\mathfrak{K}$. 
\label{lemma2}
\end{lemma}
\textit{Proof of Lemma~\ref{lemma2}:} By using Proposition~\ref{prop:quasi_cov_simple}, we have 
\begin{align}
    &E\left(\frac{(\omega-\omega_0)^2}{k-1}\sum_{i=1}^{k-1}\boldsymbol{\mathsf{Y}}_{i}^\top\boldsymbol{\Kappa}^{-1}\boldsymbol{\mathsf{Y}}_{i}\right)\nonumber\\&=\frac{(\omega-\omega_0)^2}{k-1}\sum_{i=1}^{k-1}tr(\boldsymbol{\Kappa}^{-1}E(\boldsymbol{\mathsf{Y}}_{i}\boldsymbol{\mathsf{Y}}_{i}^\top))=\frac{(\omega-\omega_0)^2tr(\boldsymbol{\Kappa}^{-1}\boldsymbol{\Kappa}_0)}{1-\omega_0^2}.\label{lemma2_exp}
\end{align}\vskip-5mm
Note that 
\begin{align}
    &\hskip-10pt \text{Var} \left(\frac{1}{k-1} \sum_{i=1}^{k-1}\boldsymbol{\mathsf{Y}}_{i}^\top\boldsymbol{\Kappa}^{-1}\boldsymbol{\mathsf{Y}}_{i}\right)=\frac{1}{(k-1)^2}  \sum_{i=1}^{k-1}\text{Var}(\boldsymbol{\mathsf{Y}}_{i}^\top\boldsymbol{\Kappa}^{-1}\boldsymbol{\mathsf{Y}}_{i})\nonumber\\
    &\ \ +\frac{2}{(k-1)^2}\sum_{i =2}^{k-1}\sum_{j=1}^{i-1}\text{Cov}(\boldsymbol{\mathsf{Y}}_{i}^\top\boldsymbol{\Kappa}^{-1}\boldsymbol{\mathsf{Y}}_{i}, \boldsymbol{\mathsf{Y}}_{j}^\top\boldsymbol{\Kappa}^{-1}\boldsymbol{\mathsf{Y}}_{j}) \label{lemma2_var_supp}
\end{align}
By using Proposition~\ref{prop:quasi_cov_simple} and and a similar argument as used in \eqref{second_term_var}, the variance of summands of the first term on the RHS of \eqref{lemma2_var_supp} is given as follows.
\begin{align}
\text{Var}(\boldsymbol{\mathsf{Y}}_{i}^\top\boldsymbol{\Kappa}^{-1}\boldsymbol{\mathsf{Y}}_{i})&=\frac{2tr(\boldsymbol{\Kappa}^{-1} \boldsymbol{\Kappa}_0)^2}{(1-\omega_0^2)^2} \label{lemma2_var_term}
\end{align}
for $1\le i\le k-1$. We now turn to simplify the second term on the RHS of \eqref{lemma2_var_supp}. By using Definition~1 and \eqref{eq:recursion}, we have 
\begin{align}
    \boldsymbol{\mathsf{Y}}_i=\omega_0^{i-j}\boldsymbol{\mathsf{Y}}_j+\sum_{m=0}^{i-j-1}\omega_0^{m} \boldsymbol{\mathsf{Z}}_{i-m} \mbox{ for } i>j.\label{Yi_in_terms_of_Yj}
\end{align}
Therefore, by using \eqref{Yi_in_terms_of_Yj}, we have 
\begin{align}
   \boldsymbol{\mathsf{Y}}_i^\top \boldsymbol{\Kappa}^{-1}\boldsymbol{\mathsf{Y}}_i &= \omega_0^{2(i-j)}\boldsymbol{\mathsf{Y}}_j^\top \boldsymbol{\Kappa}^{-1}\boldsymbol{\mathsf{Y}}_j+2  \sum_{m=0}^{i-j-1}\omega_0^{i-j+m}\boldsymbol{\mathsf{Y}}_j^\top \boldsymbol{\Kappa}^{-1}\boldsymbol{\mathsf{Z}}_{i-m}\nonumber\\
   &\quad+\sum_{m=0}^{i-j-1}\sum_{m'=0}^{i-j-1}\omega_0^{m+m'}\boldsymbol{\mathsf{Z}}_{i-m}^\top \boldsymbol{\Kappa}^{-1}{\boldsymbol{\mathsf{Z}}}_{i-m'}. \label{Yi_quad_decompose}
\end{align}
By using \eqref{Yi_quad_decompose}, the summand of the second term on the RHS of \eqref{lemma2_var_supp} is simplified as follows for $i>j$.
\begin{align}
&\hskip-5pt\text{Cov}(\boldsymbol{\mathsf{Y}}_{i}^\top\boldsymbol{\Kappa}^{-1}\boldsymbol{\mathsf{Y}}_{i}, \boldsymbol{\mathsf{Y}}_{j}^\top\boldsymbol{\Kappa}^{-1}\boldsymbol{\mathsf{Y}}_{j})
=\omega_0 ^{2(i-j)}\text{Var}(\boldsymbol{\mathsf{Y}}_{j}^\top\boldsymbol{\Kappa}^{-1}\boldsymbol{\mathsf{Y}}_{j})\nonumber\\
&
+2\sum_{m=0}^{i-j-1}\omega_0^{i-j+m}\text{Cov}\left( \boldsymbol{\mathsf{Y}}_{j}^\top \boldsymbol{\Kappa}^{-1} \boldsymbol{\mathsf{Z}}_{i-m},\boldsymbol{\mathsf{Y}}_{j}^\top\boldsymbol{\Kappa}^{-1}\boldsymbol{\mathsf{Y}}_{j}\right)\nonumber\\
&+\sum_{m=0}^{i-j-1} \sum_{m'=0}^{i-j-1}\omega_0^{m+m'}\text{Cov}\left( \boldsymbol{\mathsf{Z}}_{i-m}^\top \boldsymbol{\Kappa}^{-1} \boldsymbol{\mathsf{Z}}_{i-m'},\boldsymbol{\mathsf{Y}}_{j}^\top\boldsymbol{\Kappa}^{-1}\boldsymbol{\mathsf{Y}}_{j}\right)\label{lemma2_cov}
\end{align}\vskip-10pt
By using the similar argument as in \eqref{lemma2_var_term}, the first term on the RHS of \eqref{lemma2_cov} is simplified as follows.
\begin{align}
    &\omega_0^{2(i-j)} Var( \boldsymbol{\mathsf{Y}}_{j}^\top\boldsymbol{\Kappa}^{-1}\boldsymbol{\mathsf{Y}}_{j})=\frac{2\omega_0^{2(i-j)}}{(1-\omega_0^2)^2}tr(\boldsymbol{\Kappa}^{-1}\boldsymbol{\Kappa}_0)^2. \label{lemma2_cov1}
\end{align}
Since $\boldsymbol{\mathsf{Y}}_j$ and $(\boldsymbol{\mathsf{Z}}_{j+1},\ldots, \boldsymbol{\mathsf{Z}}_{i})$ are independent for $j<i$, by using conditional expectation arguments, the second term on the RHS of \eqref{lemma2_cov} vanishes. By using the similar argument and the independence between $\boldsymbol{\mathsf{Y}}_j$ and $\boldsymbol{\mathsf{Z}}_{i-m}$ for $j<i$ and $m=0,1,\ldots,i-j-1$, the third term on the RHS of \eqref{lemma2_cov} also vanishes. Therefore, using \eqref{lemma2_var_supp}, \eqref{lemma2_var_term} and \eqref{lemma2_cov1}:
\begin{align}
    &\text{Var} \left(\frac{(\omega-\omega_0)^2}{k-1} \sum_{i=1}^{k-1}\boldsymbol{\mathsf{Y}}_{i}^\top\boldsymbol{\Kappa}^{-1}\boldsymbol{\mathsf{Y}}_{i}\right)\nonumber\\
    &= \frac{(\omega-\omega_0)^4}{k-1}\frac{2tr(\boldsymbol{\Kappa}^{-1}\boldsymbol{\Kappa}_0)^2}{(1-\omega_0^2)^2}\left(1+\frac{2}{(k-1)}\sum_{i =2}^{k-1}\sum_{j=1}^{i-1}
    \omega_0^{2(i-j)}\right)\nonumber\\
    &=\frac{(\omega-\omega_0)^4}{k-1}\frac{2tr(\boldsymbol{\Kappa}^{-1}\boldsymbol{\Kappa}_0)^2}{(1-\omega_0^2)^2} \Bigg[1+\frac{2}{k-1}\frac{[(k-2)\omega_0^2-(k-1)\omega_0^4-\omega_0^{2k}]}{(1-\omega_0^2)^2}\Bigg].\label{lemma2_var}
    \end{align}
Since the RHS of \eqref{lemma2_exp} converges to  $\frac{(\omega-\omega_0)^2tr(\boldsymbol{\Kappa}^{-1}\boldsymbol{\Kappa}_0)}{1-\omega_0^2}$ and the RHS of \eqref{lemma2_var} converges to $0 $ as $k\to\infty$ continuously over $\omega\in[-1,1]$ and $\boldsymbol{\Kappa}\in\mathfrak{K}$, this completes the proof.~\hfill{$\blacksquare$}

\textit{Proof of Theorem \ref{thm2}:} If $t\le p$, then $Y_t\in\boldsymbol{\mathsf{Y}_1}$ and  $\boldsymbol{Y_t}(=\boldsymbol{\mathsf{Y}}_1^{(t)})$ is a zero mean Gaussian vector with covariance matrix $\boldsymbol{\Kappa}_t\triangleq \frac{1}{1-\omega^2}(k_p(i-j))_{1\le i,j\le t}$. Thus, by using the conditional expectation of jointly Gaussian vectors (see \eqref{eq:pred}), we have 
\begin{align}
    \hat{Y}_t&=E(Y_t|\boldsymbol{Y_{t-1}})
    =\frac{1}{1-\omega^2}\boldsymbol{\Kappa}_{1,l(t)-1}\left(\frac{1}{1-\omega^2}\boldsymbol{\Kappa}_{l(t)-1}^{-1}\right)\boldsymbol{\mathsf{Y}}_1^{(t-1)}\nonumber\\&=\boldsymbol{\Kappa}_{1,l(t)-1}\boldsymbol{\Kappa}_{l(t)-1}^{-1}\boldsymbol{\mathsf{Y}}_1^{(t-1)}.\label{Yt_firstblock_pred}
\end{align}
If $t>p$. then $i(t)\ge 1$ and $Y_t=Y_{i(t)p+l(t)}$. Thus,
\begin{equation}
    \hat{Y}_t=E({Y}_{t}|\boldsymbol{{Y}_{t-1}})=E[{Y}_{i(t)p+l(t)}|\boldsymbol{\mathsf{Y}}_{i(t)+1}^{(l(t)-1)}, \boldsymbol{\mathsf{Y}}_{i(t)}, \dots, \boldsymbol{\mathsf{Y}}_{1}]. \label{Yt_interms of Yt_1}
\end{equation}
Recall the structural equations given in \eqref{eq:recursion}, we have
\begin{equation}\label{eq:56}
    \boldsymbol{\mathsf{Y}}_{i(t)+1}^{(l(t))}=\omega \boldsymbol{\mathsf{Y}}_{i(t)}^{(l(t))} + \boldsymbol{\mathsf{Z}}_{i(t)+1}^{(l(t))}. 
\end{equation}
By using \eqref{eq:56}, we have
\begin{align}
& \hskip-10ptE[{Y}_{i(t)p+l(t)}|\boldsymbol{\mathsf{Y}}_{i(t)+1}^{(l(t)-1)}, \boldsymbol{\mathsf{Y}}_{i(t)}, \dots, \boldsymbol{\mathsf{Y}}_{1}]\nonumber\\=&\ \omega Y_{(i(t)-1)p+l(t)}+E[\mathsf{Z}_{i(t)+1,l(t)}|\boldsymbol{\mathsf{Y}}_{i(t)+1}^{(l(t)-1)}, \boldsymbol{\mathsf{Y}}_{i(t)}, \dots, \boldsymbol{\mathsf{Y}}_{1}].
\label{eq:57}
\end{align}\vskip-10pt
By using \eqref{eq:56} and the independence between $\boldsymbol{\mathsf{Z}}_{i(t)+1}$ and $ (\boldsymbol{\mathsf{Y}}_{i(t)}, \dots, \boldsymbol{\mathsf{Y}}_{1})$, we have
\begin{align}
&\hskip-10ptE[\mathsf{Z}_{i(t)+1,l(t)}|\boldsymbol{\mathsf{Y}}_{i(t)+1}^{(l(t)-1)}=\boldsymbol{\mathsf{y}}_{i(t)+1}^{(l(t)-1)}, \boldsymbol{\mathsf{Y}}_{i(t)}=\boldsymbol{\mathsf{y}}_{i(t)}, \dots, \boldsymbol{\mathsf{Y}}_{1}=\boldsymbol{\mathsf{y}}_{1}]\nonumber \\=&E\Big[\mathsf{Z}_{i(t)+1,l(t)}\Big| \boldsymbol{\mathsf{Z}}_{i(t)+1}^{(l(t)-1)}=\boldsymbol{\mathsf{y}}_{i(t)+1}^{(l(t)-1)}-\omega \boldsymbol{\mathsf{Y}}_{i(t)}^{(l(t)-1)},\boldsymbol{\mathsf{Y}}_{i(t)}=\boldsymbol{\mathsf{y}}_{i(t)}, \dots, \boldsymbol{\mathsf{Y}}_{1}=\boldsymbol{\mathsf{y}}_{1}\Big]\nonumber\\=&E\Big[\mathsf{Z}_{i(t)+1,l(t)}\Big| \boldsymbol{\mathsf{Z}}_{i(t)+1}^{(l(t)-1)}=\boldsymbol{\mathsf{y}}_{i(t)+1}^{(l(t)-1)}-\omega \boldsymbol{\mathsf{y}}_{i(t)}^{(l(t)-1)}\Big]\nonumber\\
=&\boldsymbol{\Kappa}_{1,l(t)-1} \boldsymbol{\Kappa}_{l(t)-1}^{-1}\left({\boldsymbol{\mathsf{Y}}_{i(t)+1}^{l(t)-1}}-\omega \boldsymbol{\mathsf{Y}}_{i(t)}^{l(t)-1}\right).\label{pred_t_ge_p}
\end{align}
By using \eqref{eq:57} and \eqref{pred_t_ge_p}, the proof of \eqref{Y_pred_QPGP} is completed. Now consider the variance of $\hat{Y}_t$. If $t\le p$, then by using \eqref{Y_pred_QPGP}:
\begin{align}\label{var_pred_yt_tlessp}
\text{Var}(\hat{Y}_t)&=\frac{1}{1-\omega^2}\boldsymbol{\Kappa}_{1,l(t)-1}\boldsymbol{\Kappa}_{l(t)-1}^{-1}\boldsymbol{\Kappa}_{l(t)-1,1}.
\end{align}
For $t> p$, by using \eqref{Y_pred_QPGP}, \eqref{eq:56}, and independence between $\boldsymbol{\mathsf{Y}}_{i(t)}$ and $\boldsymbol{\mathsf{Z}}_{i(t)+1}$, we have 
\begin{align}
   &\text{Var}{(\hat{Y}_t)}=\  \omega^2\text{Var}\left(Y_{t-p}\right) +\text{Var}\left(\boldsymbol{\Kappa}_{1,l(t)-1} \boldsymbol{\Kappa}_{l(t)-1}^{-1} \left({\boldsymbol{\mathsf{Y}}_{i(t)+1}^{l(t)-1}}-\omega \boldsymbol{\mathsf{Y}}_{i(t)}^{l(t)-1}\right) \right).\label{var_Yt_pred_tLargerp}
    \end{align}
Now by using \eqref{eq:56}, the second term on the RHS  of \eqref{var_Yt_pred_tLargerp} simplifies as follows.
\begin{align}
    \text{Var}\left(\boldsymbol{\Kappa}_{1,l(t)-1} \boldsymbol{\Kappa}_{l(t)-1}^{-1} \left({\boldsymbol{\mathsf{Y}}_{i(t)+1}^{l(t)-1}}-\omega \boldsymbol{\mathsf{Y}}_{i(t)}^{l(t)-1}\right) \right)=\boldsymbol{\Kappa}_{1,l(t)-1}\boldsymbol{\Kappa}_{l(t)-1}^{-1}\boldsymbol{\Kappa}_{l(t)-1,1}.\label{var_pred_Yt_secondterm}
\end{align}
This completes the proof.
\vspace{-10pt}

\subsection{Additional simulations and experiments}\label{additional_sim}
Additional estimation performance simulations are included below. The experimental setup is as described in Section~\ref{s5} of the main paper, with the exception of the period, $p=100$. 

Table~\ref{s5b_supp} 
shows the RMSE of the proposed estimators and MLE based on grid search from $1000$ simulation runs, along with computational time per simulation run. The grid search space remains the same as in subsection \ref{s5b} of the main paper. For $p=100$, we observe similar findings to the $p=10$ case in terms of accuracy as well as computational cost. 

\begin{table*}[t]
\centering
\caption{Estimation accuracy, prediction performance, and computational cost (per iteration) for \eqref{tilde_k_p}, \eqref{parameteric_kappa} \& SIGS.}
\label{comment3_sigs}
\setlength{\tabcolsep}{10pt}
\begin{tabular}{c l c c c}
\toprule
$n$ & Metric & \eqref{tilde_k_p} & \eqref{parameteric_kappa} & SIGS \\
\midrule
\multirow{4}{*}{600} 
 & MSE for $\omega$ & $0.0499$ & $0.0300$ & $0.0296$ \\
 & Integrated MSE for $\boldsymbol{\Kappa}$ & $1.0233$ & $1.0037$ & $0.3346$ \\
 & Prediction MSE & $0.1797$ & $0.1492$ & $0.1484$ \\
 & Time (ms) & $1.084$ & $2.360$ & $10.215$ \\
\midrule
\multirow{4}{*}{3000} 
 & MSE for $\omega$ & $0.0210$ & $0.0123$ & $0.0122$ \\
 & Integrated MSE for $\boldsymbol{\Kappa}$ & $0.4601$ & $0.4510$ & $0.1386$ \\
 & Prediction MSE & $0.1692$ & $0.1491$ & $0.1489$ \\
 & Time (ms) & $1.954$ & $3.031$ & $13.641$ \\
\midrule
\multirow{4}{*}{10000} 
 & MSE for $\omega$ & $0.0113$ & $0.0067$ & $0.0067$ \\
 & Integrated MSE for $\boldsymbol{\Kappa}$ & $0.2567$ & $0.2518$ & $0.0775$ \\
 & Prediction MSE & $0.1624$ & $0.1491$ & $0.1491$ \\
 & Time (ms) & $5.417$ & $4.848$ & $17.273$ \\
\bottomrule
\end{tabular}
\end{table*}
\begin{table*}[t]
\centering
\caption{RMSE of proposed estimator and MLE based on 1000 runs, along with respective computational cost}
\label{s5b_supp}
\setlength{\tabcolsep}{10pt}
\begin{tabular}{c l ccc r}
\toprule
& & \multicolumn{3}{c}{RMSE} & \\
\cmidrule(lr){3-5}
$n$ & Estimator & $\omega$ & $\theta$ & $\sigma^2$ & Time / run (ms) \\
\midrule
\multirow{2}{*}{600} 
 & Proposed & $0.2267$ & $0.4989$ & $0.4481$ & $8.26$ \\
 & MLE      & $0.2201$ & $0.0245$ & $0.2734$ & $29880.29$ \\
\midrule
\multirow{2}{*}{3000} 
 & Proposed & $0.0882$ & $0.1860$ & $0.1992$ & $11.83$ \\
 & MLE      & $0.0692$ & $0.0140$ & $0.1309$ & $40576.02$ \\
\midrule
\multirow{2}{*}{10000} 
 & Proposed & $0.04574$ & $0.0943$ & $0.1057$ & $12.11$ \\
 & MLE      & $0.02833$ & $0.0107$ & $0.0847$ & $69650.91$ \\
\bottomrule
\end{tabular}
\end{table*} 

\begin{table*}[t]
\centering
\caption{Estimation accuracy, prediction performance, and computational cost (per iteration) for \eqref{tilde_k_p}, \eqref{parameteric_kappa} \& SIGS.}
\label{comment3_rmse}
\setlength{\tabcolsep}{10pt}
\begin{tabular}{c l c c c}
\toprule
$n$ & Metric & \eqref{tilde_k_p} & \eqref{parameteric_kappa} & SIGS \\
\midrule
\multirow{4}{*}{600} 
 & MSE for $\omega$ & $0.0499$ & $0.0300$ & $0.0296$ \\
 & Integrated MSE for $\boldsymbol{\Kappa}$ & $1.0233$ & $1.0037$ & $0.3346$ \\
 & Prediction MSE & $0.1797$ & $0.1492$ & $0.1484$ \\
 & Time (ms) & $1.084$ & $2.360$ & $10.215$ \\
\midrule
\multirow{4}{*}{3000} 
 & MSE for $\omega$ & $0.0210$ & $0.0123$ & $0.0122$ \\
 & Integrated MSE for $\boldsymbol{\Kappa}$ & $0.4601$ & $0.4510$ & $0.1386$ \\
 & Prediction MSE & $0.1692$ & $0.1491$ & $0.1489$ \\
 & Time (ms) & $1.954$ & $3.031$ & $13.641$ \\
\midrule
\multirow{4}{*}{10000} 
 & MSE for $\omega$ & $0.0113$ & $0.0067$ & $0.0067$ \\
 & Integrated MSE for $\boldsymbol{\Kappa}$ & $0.2567$ & $0.2518$ & $0.0775$ \\
 & Prediction MSE & $0.1624$ & $0.1491$ & $0.1491$ \\
 & Time (ms) & $5.417$ & $4.848$ & $17.273$ \\
\bottomrule
\end{tabular}
\end{table*}

\begin{figure*}
\centering
\small{
\begin{tabular}{ccc}
S.E.($\hat{\omega}$) & S.E.($\hat{\theta}$) & S.E.($\hat{\sigma}^2$)\\
\includegraphics[width=0.28\textwidth]{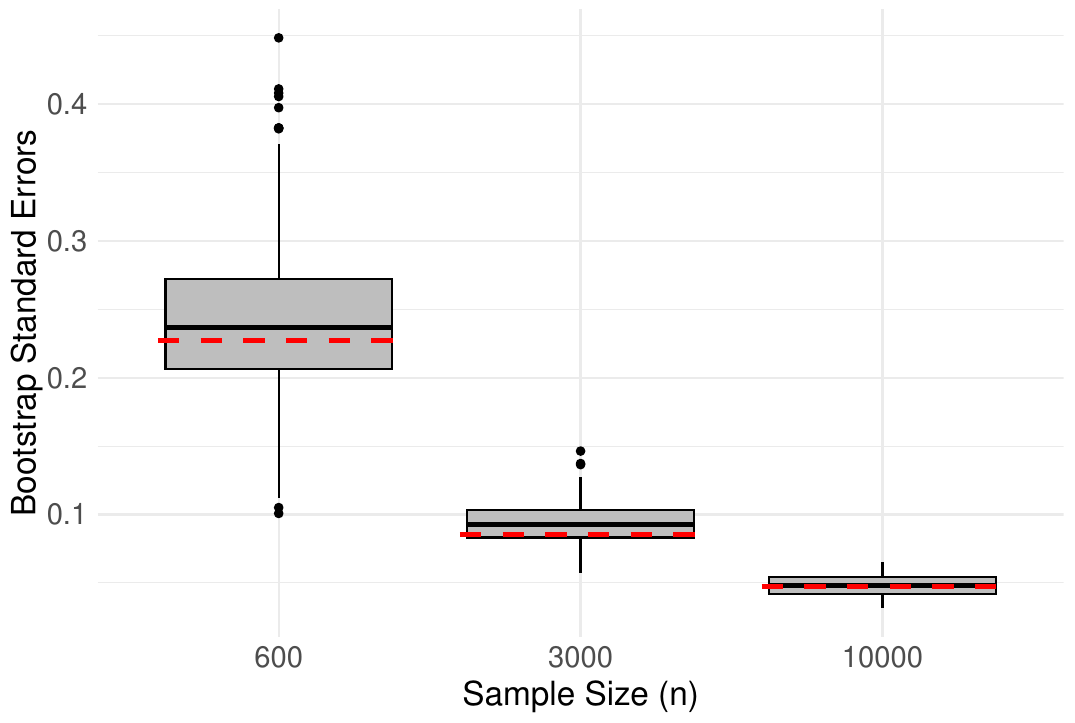} 
&
\includegraphics[width=0.28\textwidth]{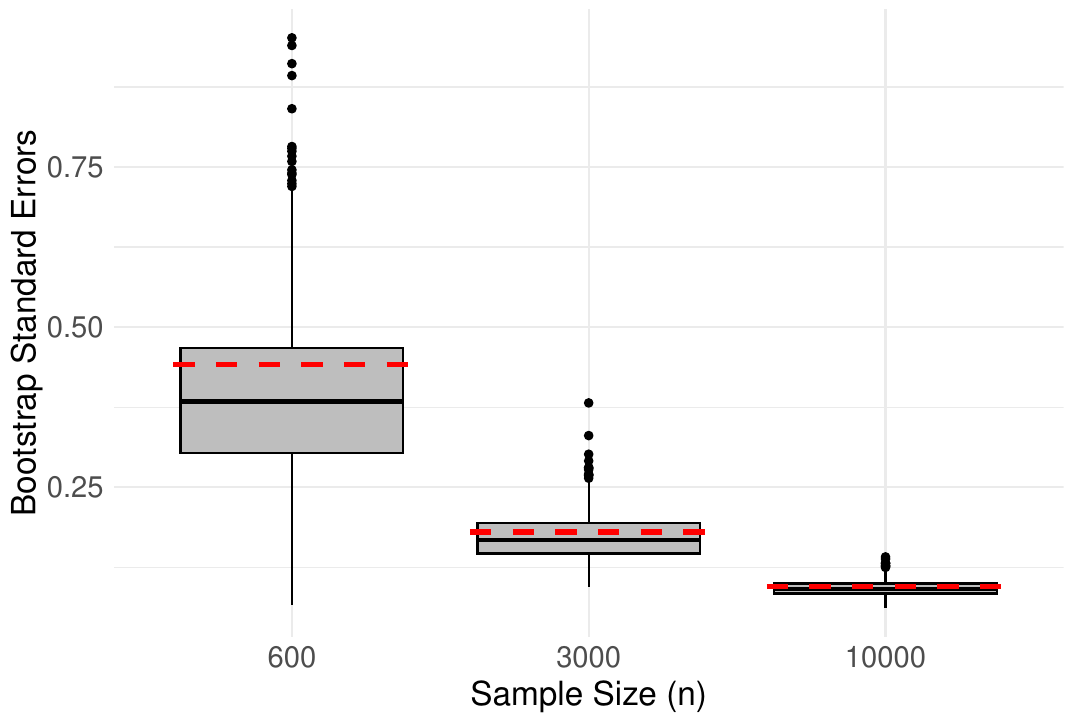}
&
\includegraphics[width=0.28\textwidth]{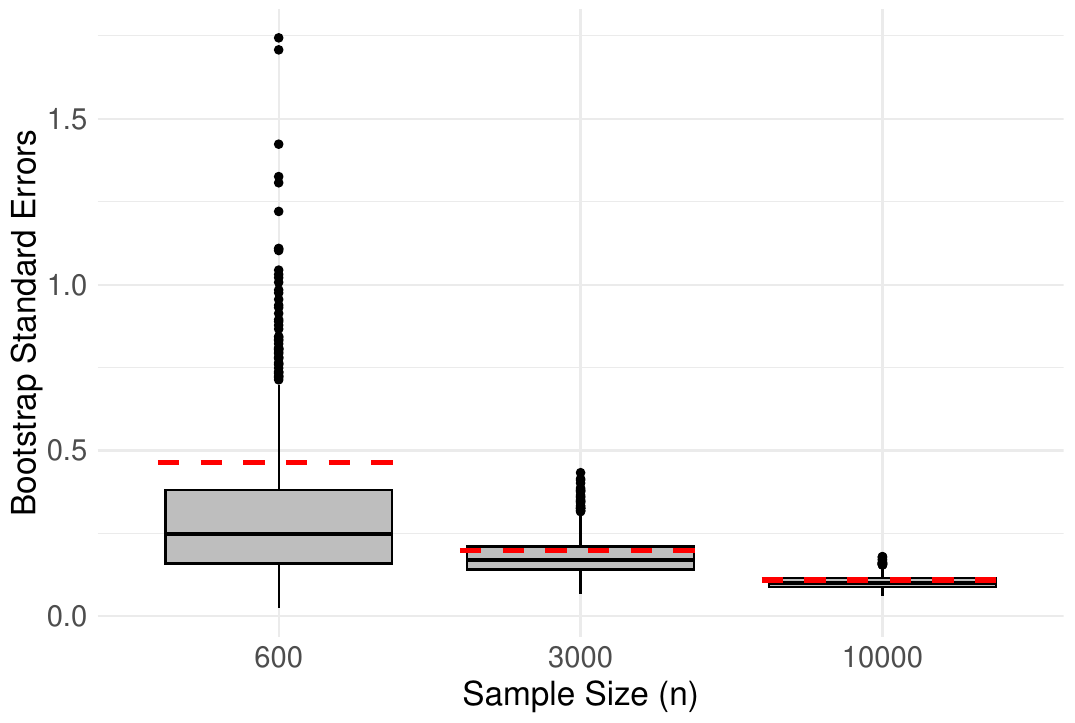}
\end{tabular}
\caption{Box plots of bootstrap standard errors ($M=1000$ bootstrap samples) of $\hat{\omega}$, $\hat{\theta}$ and $\hat{\sigma}^2$ for standard QPGP, are shown in left, center and right panel, respectively. Each panel consists of three box plots corresponding to sample sizes $600$, $3000$ and $10000$. The empirical standard error of estimators across simulation runs are shown in dashed horizontal maroon line.}
\label{fig:bootstrap_supp}
}
\end{figure*}

Fig. \ref{fig:bootstrap_supp}
shows the boxplots of bootstrap standard errors corresponding to QPGP parameters. The width of the boxes is larger for $p=100$ in comparison to $p=10$, but we observe a similar pattern to subsection~\ref{s5c} of the main paper.

\begin{figure}[htbp]
    \centering
    \begin{minipage}{0.48\textwidth}
        \centering
        \includegraphics[width=\linewidth]{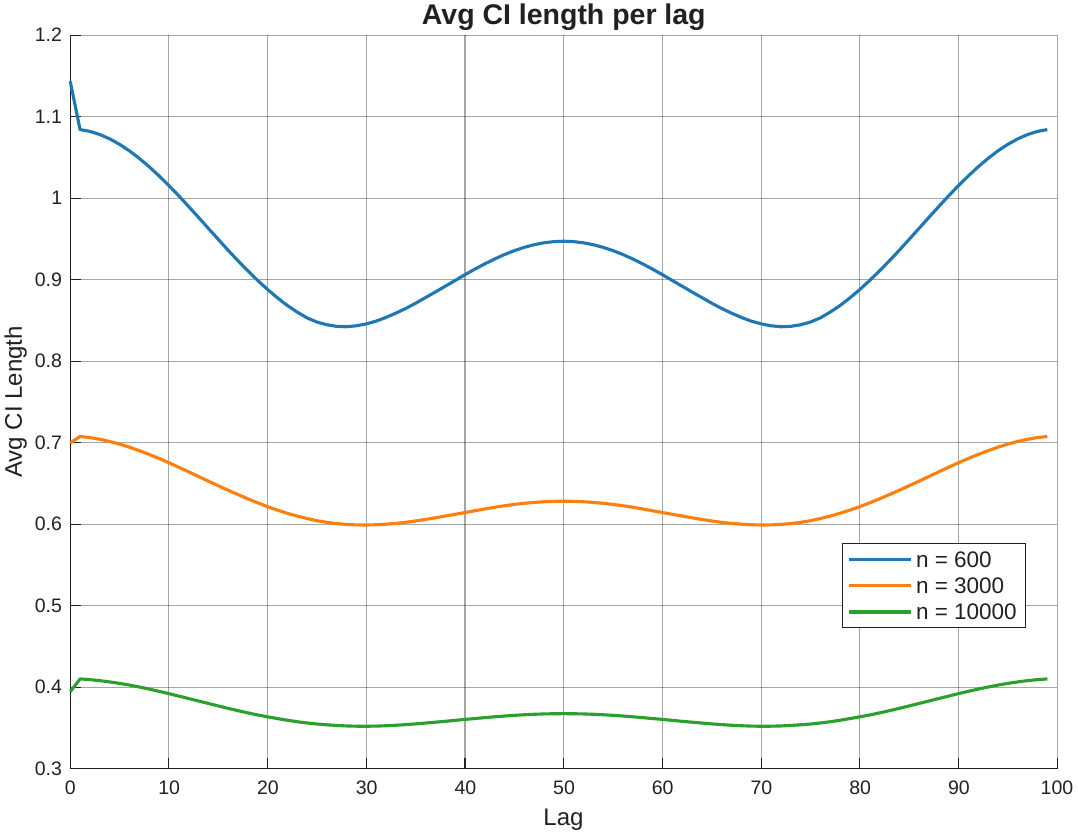}
        \caption{Avg CI length for ${\kappa}_p$.}
        \label{fig:elength_p100}
    \end{minipage}%
    \hfill
    \begin{minipage}{0.48\textwidth}
        \centering
        \includegraphics[width=\linewidth]{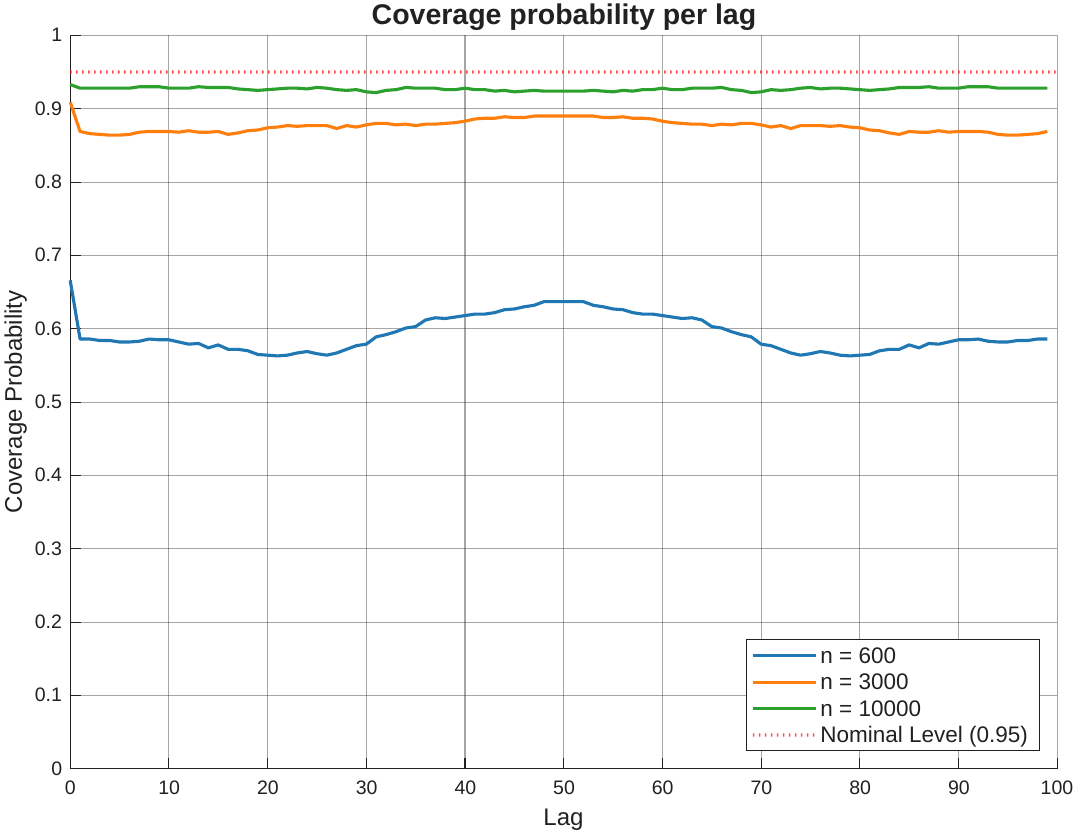}
        \caption{Coverage probability for ${\kappa}_p$.}
        \label{fig:coverage_p100}
    \end{minipage}
\end{figure}

Table \ref{tab:coverage_omega_p100} and Figs. \ref{fig:elength_p100} and \ref{fig:coverage_p100} show the average lengths and coverage probability of bootstrap confidence intervals for ${\omega}$ and ${\kappa}_p$, respectively, for $p=100$. The average lengths of the bootstrap confidence intervals are larger in comparison to those for $p=10$ due to a smaller number of blocks $k$. However, we observe a similar narrowing trend in the average length of CIs as $n$ increases, as in subsection \ref{coverage} of the main paper. We also observe that the empirical coverage probability approaches the nominal level $95\%$.

\begin{table}[htbp]
    \centering
    \caption{Average bootstrap  CI length with coverage probability for $\omega$}
    \label{tab:coverage_omega_p100}
    \begin{tabular}{ccc}
        \toprule
        \textbf{$n$} & \textbf{Avg CI Length} & \textbf{Coverage Probability} \\
        \midrule
        600    & 0.9432 & 0.975 \\
        3000   & 0.3702 & 0.967 \\
        10000  & 0.1916 & 0.969 \\
        \bottomrule
    \end{tabular}
\end{table}

Below, we provide the estimates of QPGP parameters corresponding to the chosen periodic kernels for the real data case studies discussed in section~\ref{s6} of the main paper. Recall, we chose the following periodic covariance kernels: (a) the general kernel, (b) MacKay’s kernel, (c) the periodic Matérn kernel with $\nu=1.5$, and (d) the cosine kernel with $\iota=1$. 

\begin{figure}[!t]

\centering
\setlength{\tabcolsep}{1.5pt} 
\renewcommand{\arraystretch}{1.1}

\begin{tabular}{>{\raggedright\arraybackslash}m{2cm} p{0.4\linewidth} p{0.4\linewidth}} 

& \parbox[t]{\linewidth}{\centering \footnotesize \textbf{Estimate of general $\kappa_p$}} 
& \parbox[t]{\linewidth}{\centering \footnotesize \textbf{SE of general $\kappa_p$ estimates}} \\

\vspace{-2cm}Carbon Dioxide &
\parbox[t]{\linewidth}{\centering \includegraphics[width=0.95\linewidth]{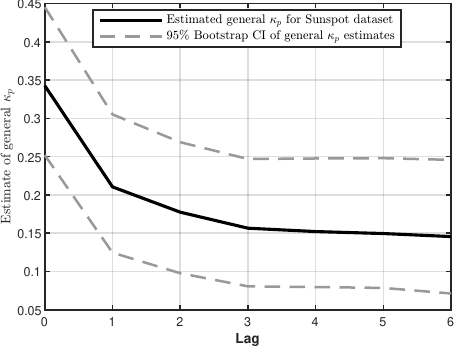}} &
\parbox[t]{\linewidth}{\centering \includegraphics[width=0.95\linewidth]{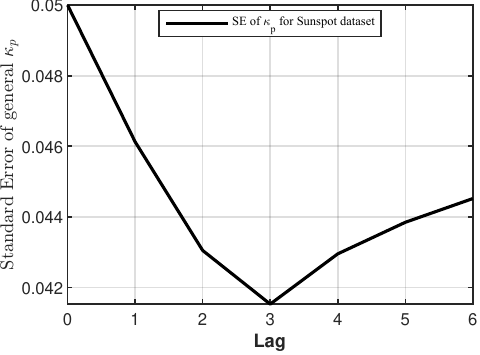}} \\

\vspace{-2cm}Sunspot Numbers &
\parbox[t]{\linewidth}{\centering \includegraphics[width=0.95\linewidth]{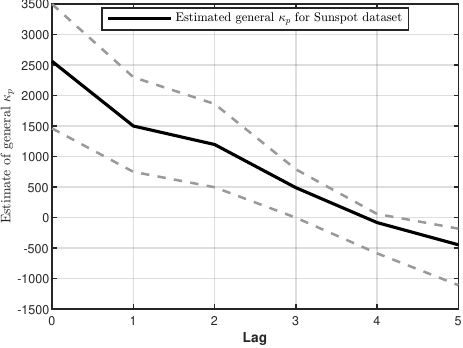}} &
\parbox[t]{\linewidth}{\centering \includegraphics[width=0.95\linewidth]{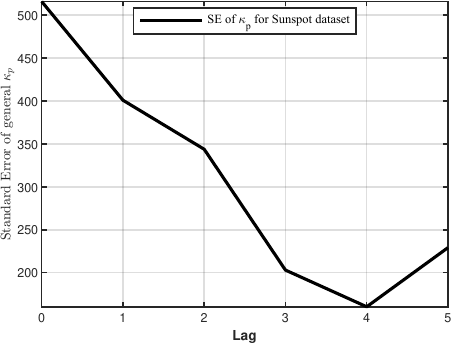}} \\

\vspace{-2cm}Water Level &
\parbox[t]{\linewidth}{\centering \includegraphics[width=0.95\linewidth]{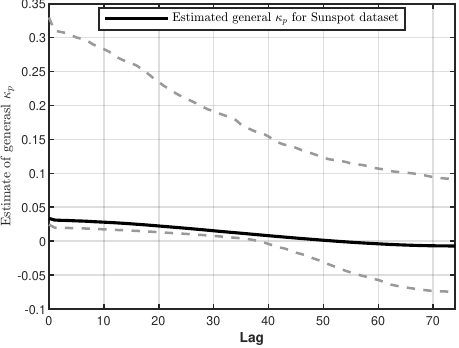}} &
\parbox[t]{\linewidth}{\centering \includegraphics[width=0.95\linewidth]{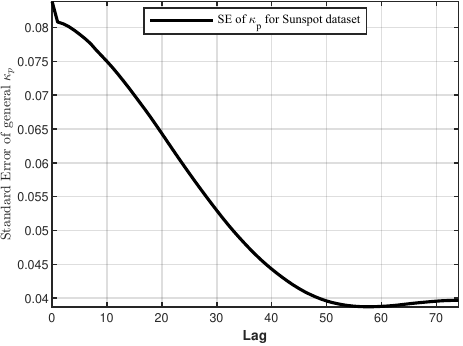}} \\

\end{tabular}

\caption{Left: estimates of general $\kappa_p$ (solid black line) with $95\%$ confidence limits (dashed black line). Right: corresponding bootstrap standard errors.}

\label{fig:dataset_kappa_table}
\end{figure}

Tables \ref{co2_all_kernels}, \ref{sunspot_all_kernels}, and \ref{tide_all_kernels} present the estimates of QPGP parameters corresponding to the selected periodic covariance kernels for the Carbon Dioxide ($p=12$), Sunspot ($p=11$), and Water Level ($p=148$) datasets, respectively, together with their bootstrap standard errors and 95\% confidence intervals. Fig.~\ref{fig:dataset_kappa_table} shows the estimates of the general covariance kernel $\kappa_p$ against lag for all three datasets, along with the associated 95\% confidence limits and bootstrap standard errors.

\begin{table}[h]
\centering
\caption{Estimates of QPGP (with $p=12$) parameters for CO$_2$ data for different $\kappa_p$}
\label{co2_all_kernels}
\begin{tabular}{l c c c c}
\toprule
Kernel & Parameter & Estimate & Std. Error & Confidence Interval \\
\midrule
General kernel 
& $\omega$ & $0.9752$ & $0.0085$ & $(0.9705, 1.0038)$ \\ 
\cmidrule(lr){1-5}

\multirow{3}{*}{\makecell[l]{MacKay \\ \eqref{McKer}}} 
& $\omega$   & $0.9752$ & $0.0085$ & $(0.9705, 1.0038)$ \\
& $\sigma^2$ & $0.2464$ & $0.0437$ & $(0.1666, 0.3329)$ \\
& $\theta$   & $0.8188$ & $0.1251$ & $(0.5975, 1.0798)$ \\
\cmidrule(lr){1-5}

\multirow{3}{*}{\makecell[l]{Mat\'ern \\ ($\nu=1.5$) \eqref{MatKer}}} 
& $\omega$   & $0.9752$ & $0.0085$ & $(0.9705, 1.0038)$ \\
& $\sigma^2$ & $0.2610$ & $0.0429$ & $(0.1826, 0.3464)$ \\
& $\theta$   & $1.9950$ & $0.4053$ & $(1.3569, 2.9443)$ \\
\cmidrule(lr){1-5}

\multirow{2}{*}{\makecell[l]{Cosine \\ \eqref{CosKer}}} 
& $\omega$   & $0.9752$ & $0.0085$ & $(0.9705, 1.0038)$ \\
& $\sigma^2$ & $0.0546$ & $0.0086$ & $(0.0355, 0.0687)$ \\
\bottomrule
\end{tabular}
\end{table}

\begin{table}[h]
\centering
\caption{Estimates of QPGP ($p=11$) parameters for Sunspot data for different $\kappa_p$}
\label{sunspot_all_kernels}
\begin{tabular}{l c c c c}
\toprule
Kernel & Parameter & Estimate & Std. Error & Confidence Interval \\
\midrule
General Kernel 
& $\omega$ & $0.7228$ & $0.06375$ & $(0.5861, 0.8429)$ \\
\cmidrule(lr){1-5}

\multirow{3}{*}{\makecell[l]{MacKay \\ \eqref{McKer}}} 
& $\omega$   & $0.7228$ & $0.06375$ & $(0.5861, 0.8429)$ \\
& $\sigma^2$ & $2335.9067$ & $519.06165$ & $(1337.9731, 3422.8676)$ \\
& $\theta$   & $1.8401$ & $0.2085$ & $(1.6809, 2.4982)$ \\
\cmidrule(lr){1-5}

\multirow{3}{*}{\makecell[l]{Mat\'ern \\ ($\nu=1.5$) \eqref{MatKer}}} 
& $\omega$   & $0.7228$ & $0.06375$ & $(0.5861, 0.8429)$ \\
& $\sigma^2$ & $2568.1523$ & $563.3239$ & $(1439.5634, 3699.3789)$ \\
& $\theta$   & $0.7599$ & $0.0814$ & $(0.5436, 0.8539)$ \\
\cmidrule(lr){1-5}

\multirow{2}{*}{\makecell[l]{Cosine \\ \eqref{CosKer}}} 
& $\omega$   & $0.7228$ & $0.06375$ & $(0.5861, 0.8429)$ \\
& $\sigma^2$ & $1254.7285$ & $323.1657$ & $(709.9667, 2002.5801)$ \\
\bottomrule
\end{tabular}
\end{table}

\begin{table}[h]
\centering
\caption{Estimates of QPGP (with $p=148$) parameters for Water Level data for different $\kappa_p$}
\label{tide_all_kernels}
\begin{tabular}{l c c c c}
\toprule
Kernel & Parameter & Estimate & Std. Error & Confidence Interval \\
\midrule
General kernel 
& $\omega$ & $0.9673$ & $0.0102$ & $(0.9432, 0.9824)$ \\
\cmidrule(lr){1-5}

\multirow{3}{*}{\makecell[l]{MacKay \\ \eqref{McKer}}} 
& $\omega$   & $0.9673$ & $0.0102$ & $(0.9432, 0.9824)$ \\
& $\sigma^2$ & $0.0334$ & $0.0827$ & $(0.0217, 0.3315)$ \\
& $\theta$   & $1.7398$ & $1.7659$ & $(0.8033, 2.3211)$ \\
\cmidrule(lr){1-5}

\multirow{3}{*}{\makecell[l]{Mat\'ern \\ ($\nu=1.5$) \eqref{MatKer}}} 
& $\omega$   & $0.9673$ & $0.0102$ & $(0.9432, 0.9824)$ \\
& $\sigma^2$ & $0.0358$ & $0.0872$ & $(0.0237, 0.3508)$ \\
& $\theta$   & $0.8338$ & $0.4892$ & $(0.5649, 2.1742)$ \\
\cmidrule(lr){1-5}

\multirow{2}{*}{\makecell[l]{Cosine \\ \eqref{CosKer}}} 
& $\omega$   & $0.9673$ & $0.0102$ & $(0.9432, 0.9824)$ \\
& $\sigma^2$ & $0.0183$ & $0.0376$ & $(0.0119, 0.1531)$ \\
\bottomrule
\end{tabular}
\end{table}

\bibliographystyle{plain}
\bibliography{refs}
\end{document}